\documentclass[11pt]{article}
\pdfoutput=1
\usepackage{jheppub}
\usepackage{graphicx}
\usepackage{color, bm, enumitem}

\newcommand{\pd}{{\partial}}
\newcommand{\ol}{\overline}
\newcommand{\wt}{\widetilde}
\newcommand{\mb}{\mathbf}

\newcommand{\Tr}{{\text{Tr}}}

\newcommand{\II}{{I\!\!I\!}}

\newcommand{\be}{\begin{equation}}
\newcommand{\ee}{\end{equation}}
\newcommand{\bp}{\begin{pmatrix}}
\newcommand{\ep}{\end{pmatrix}}
\newcommand{\bsp}{\left(\begin{smallmatrix}}
\newcommand{\esp}{\end{smallmatrix}\right)}

\newcommand{\R}{{\mathbb R}}

\newcommand{\C}{{\mathbb C}}
\newcommand{\Z}{{\mathbb Z}}

\newcommand{\CA}{{\mathcal A}}
\newcommand{\CB}{{\mathcal B}}
\newcommand{\CD}{{\mathcal D}}

\newcommand{\CF}{{\mathcal F}}

\newcommand{\CI}{{\mathcal I}}

\newcommand{\CL}{{\mathcal L}}
\newcommand{\CN}{{\mathcal N}}
\newcommand{\CO}{{\mathcal O}}
\newcommand{\CP}{{\mathcal P}}
\newcommand{\CS}{{\mathcal S}}
\newcommand{\CT}{{\mathcal T}}
\newcommand{\CV}{{\mathcal V}}
\newcommand{\CW}{{\mathcal W}}

\title{Dual Boundary Conditions in 3d SCFT's}
\author[1]{Tudor Dimofte}
\author[2]{Davide Gaiotto}
\author[3]{Natalie M. Paquette}

\affiliation[1]{Department of Mathematics and Center for Quantum Mathematics and Physics (QMAP), University of California, Davis, CA 95616, USA}
\affiliation[2]{Perimeter Institute for Theoretical Physics, 31 Caroline St.\,N., Waterloo, ON N2L 2Y5, Canada}
\affiliation[3]{Walter Burke Institute for Theoretical Physics, California Institute of Technology,
Pasadena, CA 91125, USA}

\dedicated{CALT-TH-2017-065}

\abstract{We propose matching pairs of half-BPS boundary conditions related by IR dualities of 3d $\CN=2$ gauge theories. From these matching pairs we construct duality interfaces.
We test our proposals by anomaly matching and the computation of supersymmetric indices. Examples include basic abelian dualities, level-rank dualities, and Aharony dualities.}

\begin{document}
\today
\maketitle


\section{Introduction}
\label{sec:intro}

Three-dimensional $\CN=2$ supersymmetric gauge theories have a vast, intricate network of infrared dualities. Some of the first dualities were discovered in \cite{dBHOY-N2, dBHO-N2, AHISS, Aharony, Karch, EGKRS, DoreyTong}, motivated in part by 3d $\CN=4$ mirror symmetry \cite{IS, dBHO, dBHOY}, and later extended to included level-rank-like or Seiberg-like dualities of 3d $\CN=2$ theories, with ``chiral'' matter content and nontrivial Chern-Simons couplings \cite{GiveonKutasov, Kapustin-SO, WY-Seiberg, BCC-Seiberg, 4d3d-dualities, AharonyFleischer}.
Dualities of 3d $\CN=2$ theories played a central role in the 3d-3d correspondence \cite{YamazakiTerashima, DGG, CCV, DGV-hybrid}, \emph{i.e.} in compactifications of M5 branes on 3-manifolds, and are intimately connected to three-dimensional geometry and topology. They have also recently been used \cite{Aharony-LR, JMY-LR, GY-LR, KMTW-bos, KMTW-bos2} to motivate \emph{non-supersymmetric} dualities of 3d Chern-Simons-matter theories, notably 3d ``bosonization'' dualities and their cousins \cite{Polyakov-bos, BarkeshliMcGreevy, Giombi-bos, AGY-bos, AGY-bos2, MZ-1, MZ-2, SSWW, KarchTong, KarchTong-more, MuruganNastase, HsinSeiberg-LR, ABHS, BHS, RTT, XuYou, XuCheng, Son-fermi, WangSenthil-Dirac, MetlitskiVishwanath, Mross-Dirac},%
\footnote{These references still only represent a small sample of the recent literature in this extremely fruitful area.} %
including classic particle-vortex duality \cite{Peskin-abelian, DasguptaHalperin, FisherLee}.

Our goal in this paper is to advance the study of \emph{boundary conditions} for 3d $\CN=2$ gauge theories  (with various Chern-Simons levels, matter content, and superpotential couplings) and their network of dualities.  Specifically, we focus on half-BPS boundary conditions that preserve a 2d $\CN=(0,2)$ subalgebra of the bulk $\CN=2$ supersymmetry as well as a non-anomalous $U(1)_R$ R-symmetry.
A systematic study of $\CN=(0,2)$ boundary conditions using anomalies and gauge dynamics was pioneered in \cite{GGP-walls, GGP-fivebranes}, particularly in the context of the 4d-2d correspondence (M5 branes on four-manifolds), and has led to many new and surprising results, including the $\CN=(0,2)$ trialities of \cite{GGP-triality}. Related works on $\CN=(0,2)$ boundary conditions include \cite{OkazakiYamaguchi, SugishitaTerashima, YoshidaSugiyama}; and analogous discussions of 2d $\CN=(1,1)$ boundary conditions appeared in \cite{OkazakiYamaguchi,AprileNiarchos}.
As in \cite{GGP-walls, OkazakiYamaguchi, GGP-fivebranes}, we will define boundary conditions in the UV of the bulk 3d theory, by making some elementary choices of boundary conditions for the bulk fields, and then potentially coupling to an additional 2d $\CN=(0,2)$ theory on the boundary. 

Suppose one has a pair of dual 3d $\CN=2$ theories, \emph{i.e.} a pair of UV gauge theories $\CT,\CT^\vee$ that flow to the same superconformal infrared theory $\CT_{\rm IR}$. Given a UV boundary condition (b.c.) $\CB$ for $\CT$ that flows to a superconformal b.c. $\CB_{\rm IR}$, it may be possible to find a UV boundary condition $\CB^\vee$ for $\CT^\vee$ that flows to the same $\CB_{\rm IR}$. This is what we mean by dual boundary conditions. 
There is no \emph{a priori} reason why such a  $\CB^\vee$ should exist for any $\CB$, unless $\CT^\vee$ is free. Nevertheless, we will see in numerous examples that dual pairs $(\CB,\CB^\vee)$ can be explicitly (and fairly simply) constructed.

\begin{figure}[htb]
\centering
\hspace{.5in}\includegraphics[width=5in]{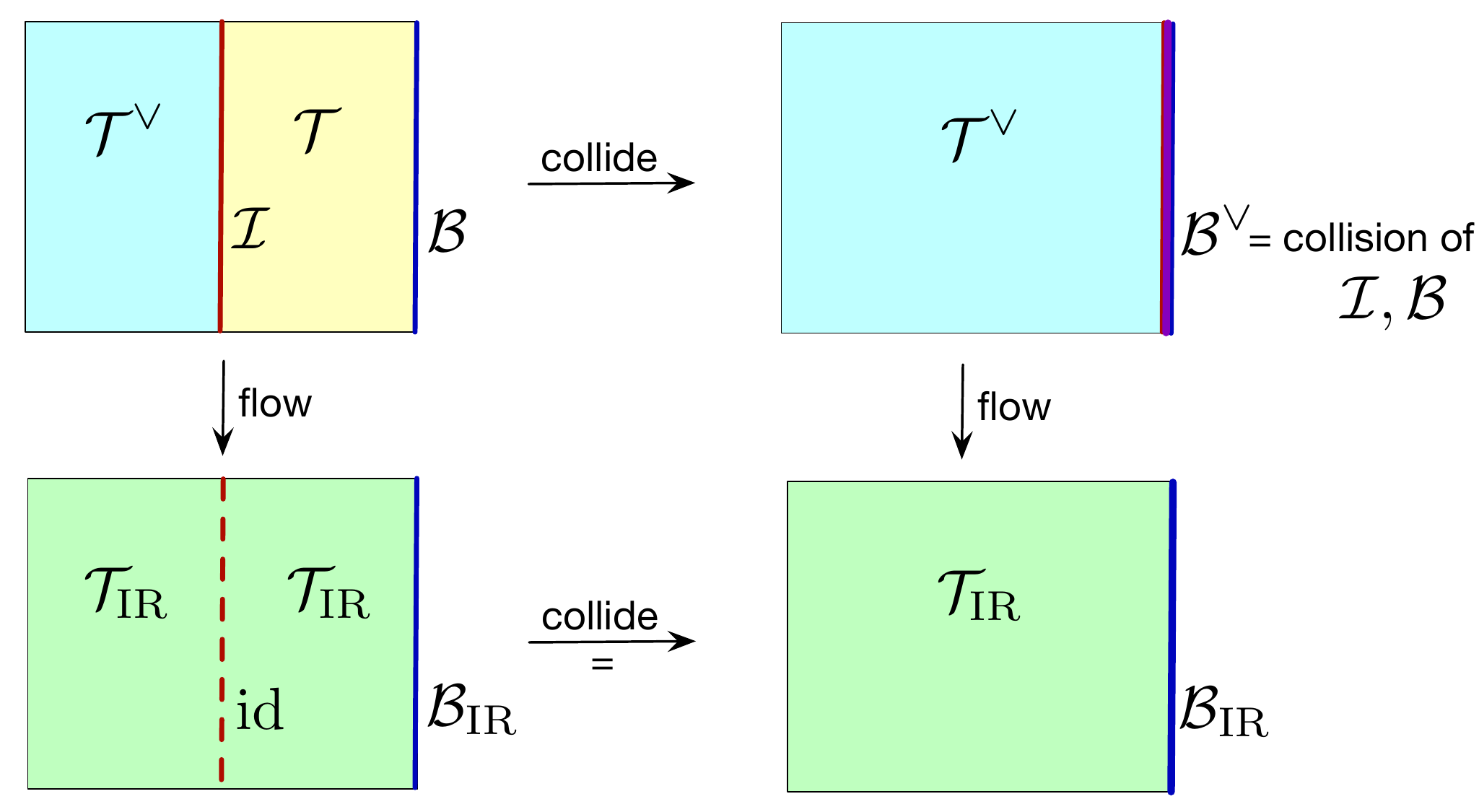}
\caption{Using collision with a duality interface to generate dual boundary conditions. Note that both collisions and the bulk flows to the IR are RG flows that hold different parameters fixed. If the diagram commutes, then the collision of $\CI$ and $\CB$ defines a dual to $\CB$.}
\label{fig:interface}
\end{figure}

One way to facilitate the existence of a dual UV boundary condition $\CB^\vee$ for any $\CB$ is to construct a duality interface $\CI$, along the lines of \cite{GaiottoWitten-Janus, GaiottoWitten-bc, GaiottoWitten-duality, DGV-hybrid, GGP-walls, BDGH}. By definition, this is an interface between the UV theories  $\CT$ and $\CT^\vee$ that flows to the trivial/identity interface in the infrared (between $\CT_{\rm IR}$ and itself). Such a duality interface can be used to generate dual boundary conditions, by defining (say) $\CB^\vee$ to be the collision of $\CI$ and $\CB$ --- assuming that the RG flow implicit in the collision commutes with the RG flows defining the IR theories and boundary conditions, as in Figure \ref{fig:interface}. We will construct duality interfaces for all the examples in this paper.

Most of the dual 3d $\CN=2$ gauge theories that we consider fit into a large family of Seiberg-like dualities, obtained from a ``parent'' duality of Aharony \cite{Aharony} by a series of flows \cite{BCC-Seiberg, 4d3d-dualities}.
Specifically, we will propose dual boundary conditions and duality interfaces for:
\begin{itemize}[leftmargin=.5cm]
\item The basic Aharony duality \cite{Aharony} relating $U(N)$ and $U(N_f-N)$ Yang-Mills theories with $N_f$ fundamental and $N_f$ antifundamental chirals, together with additional singlets and a superpotential on one side.

\item Its precursor \cite{AHISS}, corresponding to the case $N_f=N$, which relates $U(N)$ Yang-Mills with $N$ fundamental and $N$ antifundamental chirals to a Landau-Ginzburg model. For $N_f=N=1$, this becomes the basic duality between SQED and the XYZ model, which played a prominent role in the 3d-3d correspondence.

\item Supersymmetric level-rank duality, relating $U(N)_{k+N}$ and $U(k)_{-k-N}$, as well as $U(N)_{k+N,k}$ and $SU(k)_{-k-N}$
pure Yang-Mills-Chern-Simons theories \cite{GiveonKutasov, BCC-Seiberg}, a supersymmetric generalization of classic level-rank duality \cite{NS-LR1, NS-LR2, NT-LR, CLZ-LR, NS-LR3}.

\item Level-rank duality with fundamental and/or antifundamental matter, relating $U(N)_{k+N-\frac{N_f+N_a}{2}}$ theory with $N_f$ ($N_a$) fundamental (antifundamental) chirals and $U(k)_{-k-N+\frac{N_f+N_a}{2}}$ theory with $N_a$ ($N_f$) fundamental (antifundamental) chirals \cite{GiveonKutasov, BCC-Seiberg}.

\item The basic supersymmetric particle-vortex ``triality'' relating a free chiral, $U(1)_{\frac12}$ theory with a charged chiral, and $U(1)_{-\frac12}$ theory with a charged chiral. This may be considered a special case of level-rank duality with matter, or the simplest 3d $\CN=2$ ``mirror symmetry'' \cite{dBHOY-N2}, or obtained from SQED$\leftrightarrow$XYZ duality by a real mass deformation.

\end{itemize}
We also consider $SU(2)_1$ theory with adjoint matter, which is dual to a free chiral \cite{Jafferis-appetizer}. 

In all these cases, we find that a Neumann-like b.c. for one 3d theory $\CT$ is dual to a Dirichlet-like b.c. for $\CT^\vee$. Making this precise requires a longer discussion of multiplets and supersymmetric boundary conditions.
Such a discussion was initiated in \cite{GGP-walls, GGP-fivebranes, OkazakiYamaguchi}, and we will summarize and slightly extend it in Section \ref{sec:bdy}. The basic construction of an $\CN=(0,2)$ boundary condition involves
\begin{itemize}[leftmargin=.5cm]
\item[1)] Choosing Neumann or Dirichlet b.c. for the gauge fields (the b.c. is then extended by $\CN=(0,2)$ SUSY to the entire gauge multiplet).
\item[2)] Choosing Neumann or Dirichlet b.c. for the scalars in each chiral multiplet (the b.c. is then extended by $\CN=(0,2)$ SUSY to the entire chiral multiplet).
\item[3)] Choosing additional 2d $\CN=(0,2)$ boundary matter (or a boundary CFT) with appropriate boundary couplings.
\item[4)] In the case of Neumann b.c. for the 3d gauge fields, ensuring that all boundary gauge anomalies vanish.
\item[5)] If there is a bulk superpotential, ensuring that it is properly ``factorized'' at the boundary.
\end{itemize}
Together with matching of boundary flavor symmetries and boundary 't Hooft anomalies, the constraints of anomaly cancellation and superpotential factorization turn out to be surprisingly restrictive. Just as in the initial examples of \cite{GGP-walls, GGP-fivebranes}, most of the dual pairs of boundary conditions that we propose are actually the simplest possible choices that are consistent with all these constraints.

We note that some basic dualities of boundary conditions in abelian theories, in particular particle-vortex duality and SQED$\leftrightarrow$XYZ, were already proposed in \cite{GGP-walls} and \cite{OkazakiYamaguchi}, based on holomorphic block identities of \cite{BDP-blocks} and a moduli space analysis.
An extended discussion of Neumann-like b.c. for pure Yang-Mills-Chern-Simons theory, their IR behavior, and their connection to level-rank duality appeared in \cite{GGP-fivebranes} and led to $\CN=(0,2)$ trialities \cite{GGP-triality}.
Related constructions of WZW and coset models on boundaries in supersymmetric Chern-Simons theory appeared in \cite{DHSV, ArmoniNiarchos}.

Our main tool in the analysis of boundary conditions is the 3d half-index. Indices, half-indices, and more general partition functions have been used to spectacular effect in verifying and even predicting dualities of supersymmetric theories in recent years (see the review \cite{PestunZabzine} and references therein). Examples involving dual boundary conditions and interfaces for 4d $\CN=2$ theories, closely analogous to our current topic, appeared in  \cite{DGG-defects, HLP-wall, DGG-index, Gang-halfindex, DGV-hybrid}. Early checks of bulk 3d dualities based on the 3d index and $S^3$ partition function include \cite{KWY, KWY-dualities, KWY-Seiberg, HHL, IY-index, KW-index}.

 The 3d half-index ``counts'' boundary operators in the cohomology of one of the supercharges in the $\CN=(0,2)$ superalgebra preserved by the boundary condition.%
\footnote{The twist of 3d $\CN=2$ theory with respect to this supercharge was recently studied by \cite{ACV}. The twisted theory is topological in the direction perpendicular to the boundary, and holomorphic in the plane parallel to the boundary.
Thus boundary operators counted by the index have the structure of a chiral algebra. Indeed, if the 3d bulk theory were empty, the boundary chiral algebra would precisely coincide with that appearing in the half-twist of 2d $\CN=(0,2)$ theories \cite{Witten-mirror, Witten-phases}, and the half-index would coincide with the 2d elliptic genus \cite{SW1,SW2,Witten-elliptic}. } %
The 3d half-index may equivalently be identified with the partition function of a 3d $\CN=2$ theory on a hemisphere $HS^2$ times $S^1$, with a chosen boundary condition on $\pd(HS^2\times S^1)\simeq S^1\times S^1$. From this perspective, it is clear that if the 3d bulk is trivial the partition function should simply reduce to an $S^1\times S^1$ partition function, which turns out to be the 2d elliptic genus.

A UV formula for the 3d half-index of a gauge theory with Neumann b.c. for the gauge fields was given in \cite{GGP-walls, GGP-fivebranes}, analogous to UV formulae for the 2d elliptic genus \cite{GGP-walls, GG-surface, BEHT1, BEHT2}. By ``UV formula,'' we mean a prescription for computing the half-index from the UV field content of a theory as in \cite{KMMR} --- essentially obtained by counting operators constructed from the classical fields and (in the case of Neumann b.c.) using a contour integral to project to gauge invariants. Intuitively, this is possible because the index is insensitive to RG flow, and largely insensitive to UV superpotential couplings.  \footnote{UV formulae for the index are only available if the IR $U(1)_R$ coincides with some (possibly unknown) linear combination of the UV $U(1)_R$ symmetry and other $U(1)$ flavour symmetry generators which are present in the UV. This is the main reason we focus on boundary conditions which preserve some UV $U(1)_R$ symmetry. } %
Such UV formulae for indices or half-indices are often reproduced by localization computations of partition functions.
In particular, localization of the full 3d index was performed in \cite{Kim-index, IY-index} (further analyzed, extended, and related to three-sphere partition functions \cite{KWY,HHL,HHL-squashed} in \cite{KW-index, BDP-blocks, CGV-tt*, AMRS, BeniniPeelaers, FMY}); and localization of the 3d half-index with Neumann b.c. appeared in \cite{YoshidaSugiyama}.%
\footnote{These localization computations were all largely inspired by the computation of \cite{Pestun-loc} of the 4d $S^4$ partition function. We again refer readers to \cite{PestunZabzine} for a review of localization techniques, with many further references.} %
In Section 3 we propose an extension of the half-index to Dirichlet b.c., which \emph{instead} of a projection to gauge invariants involves a sum over boundary monopole sectors.

The half-indices that we consider in this paper are closely related to the holomorphic blocks of \cite{Pasquetti-blocks, BDP-blocks} (see also \cite{HKP-blocks,Taki-blocks,HYY, AganagicOkounkov, GPPV}). As discussed in \cite{BDP-blocks}, holomorphic blocks are defined in the \emph{infrared} of a 3d $\CN=2$ theory. They are defined as partition functions of a mass-deformed theory on $\C\times S^1$, with a boundary condition near infinity on $\C$ labelled by the choice of an IR vacuum.
The most general $\CN=(0,2)$ boundary condition in the IR may be described as a superposition (direct sum) of boundary conditions labelled by vacua, each optionally decorated with a boundary $\CN=(0,2)$ theory. Correspondingly, the $\C\times S^1$ partition function of a general IR boundary condition will be a linear combination of holomorphic blocks, with coefficients valued in elliptic genera.

Since any (SUSY-preserving) UV boundary condition $\CB$ flows to \emph{some} IR boundary condition, one should expect that the UV half-index of $\CB$ will be equal to such a linear combination of holomorphic blocks. We indeed find this to be the case, though we do not analyze the phenomenon in detail here.
We show that half-indices are solutions to the same difference equations that characterize the holomorphic blocks (up to a systematic modification that we explain), which implies that the former is a linear combination of the latter. These are also the same difference equations that are obeyed by the $S^2\times S^1$ index and by the $S^3$ partition function of a 3d theory \cite{DGG, DGG-index}; in all cases, the difference equations capture relations in the category of bulk BPS line operators.

\subsection{Future directions and additional motivations}

The results of this paper can be extended in numerous orthogonal directions, some of which originally motivated this work.
\begin{enumerate}
\item The analysis of boundary gauge and 't Hooft anomalies and the matching of symmetries on the two sides of the dualities
is not specific to supersymmetric gauge theories. The same anomaly matching considerations allow one to 
formulate plausible pairs of dual boundary conditions for non-supersymmetric ``bosonization'' dualities. 
Preliminary results were announced in \cite{Natifest}. While this draft was completed, further work appeared on the subject \cite{Aitken:2017nfd}. 

\item In addition, we expect that one can recover dual pairs of boundary conditions in \cite{Aitken:2017nfd} associated with bosonization and particle-vortex duality by judiciously breaking supersymmetry in our dual pairs of boundary conditions (analogous to the `bulk' study of \cite{KMTW-bos, KMTW-bos2}).
We have performed a similar breaking 
of the 3d $\CN=4$ abelian mirror symmetry interface obtained in \cite{BDGH} to the 3d $\CN=2$ interface. We derive the latter in Section \ref{sec:int-tet} by independent means. In completely breaking SUSY, of course, one is subject to the usual caveats associated with uncontrolled RG flows.

\item The tools we employed in our analysis apply equally well to boundary conditions for 3d $\CN=4$ gauge theories, 
preserving $\CN=(2,2)$ and $\CN=(0,4)$ SUSY. The corresponding dualities of boundary conditions may also be derivable through 
brane constructions~\cite{ChungOkazaki}.

\item The $\CN=(2,2)$ boundary conditions played an important role in the physical explanation of symplectic duality \cite{BDGH}. They should also be useful in interpreting the results of~\cite{AganagicOkounkov}. 

\item On the other hand, $\CN=(0,4)$ boundary conditions for 3d ${\cal N}=4$ gauge theories play an important role in past \cite{Gaiotto-twisted} and upcoming \cite{GaiottoCostello} work on gauge theory constructions of vertex operator algebras with applications to the Geometric Langlands program. Our half-index calculations provide important checks in the form of characters of the relevant VOAs.

\item Our abelian examples are the basic building blocks for an analysis of boundary conditions and duality interfaces 
for 3d ${\cal N}=2$ theories of class R \cite{DGG, DGG-Kdec}, which correspond geometrically to ideal triangulations of 3-manifolds. They are also the building blocks for an IR description of codimension-two defects in 4d ${\cal N}=2$ theories of class S, along the lines of \cite{GGP-defects} (and related to \cite{GMN-2d4d, GGS-resolvents}).

\item Some of our proposed dual pairs of boundary conditions are mutually consistent thanks to some known dualities of 
$\CN=(0,2)$ gauge theories \cite{GGP-triality}. It would be interesting to explore the relationship further. 

\item Many 3d dualities descend from Seiberg duality for $4d$ ${\cal N}=1$ gauge theories, \emph{cf.} \cite{4d3d-dualities}. The basic $\CN=(0,2)$ boundary conditions 
in the 3d theories would arise naturally from cigar configurations with a $\CN=(0,2)$ surface defect at the tip of the cigar. 
It may be possible to lift some dualities of boundary conditions to Seiberg dualities of surface defects~(see e.g. \cite{GaiottoKim}).

\item In the opposite direction, it would be interesting to reduce some of the 3d dualities of boundary conditions considered here to two dimensions, along the lines of of \cite{AHKT,ARW}, and relate to the analysis of UV boundary conditions in 2d $\CN=(2,2)$ GLSM's carried out in \cite{SugishitaTerashima, HondaOkuda, HoriRomo}. 

\end{enumerate}

\section{2d $\CN=(0,2)$ boundary conditions}
\label{sec:bdy}

We describe several very general families of boundary conditions for 3d $\CN=2$ theories that preserve two-dimensional $\CN=(0,2)$ supersymmetry, and their associated anomalies. We also introduce some compact notation that we'll use throughout the paper.

Most aspects of boundary conditions presented below appeared in a slightly more condensed form in \cite{GGP-walls, OkazakiYamaguchi, GGP-fivebranes}. 
New features of our discussion include a refined computation of boundary anomalies, showing that they are independent of the signs of bulk real masses or the signs of bulk Chern-Simons levels --- in contrast to the well-known shifts of bulk Chern-Simons levels induced by massive fermions. 
We also initiate a study of singular boundary conditions, analogous to the Nahm-pole b.c. in 4d $\CN=4$ super-Yang-Mills \cite{GaiottoWitten-bc,GaiottoWitten-duality} and generalizing the 3d $\CN=4$ analysis of \cite{ChungOkazaki}.
\medskip

\noindent\textbf{Basic conventions:} \medskip

We identify the local neighborhood of a boundary with half of 3d Minkowski space $\R^{1,1}\times \R_{\leq 0}$. We use coordinates $x^\mu$ $(\mu=0,1)$ on $\R^{1,1}$, and a spatial coordinate $x^\perp \leq 0$ transverse to the boundary. In later sections, we will also consider Euclidean signature, with complex coordiantes $z,\bar z$ in the plane of the boundary.

The 3d $\CN=2$ SUSY algebra has four real odd generators that can be regrouped into complex spinors $Q_\alpha$, $\ol Q_\alpha$ ($\alpha=-,+$), satisfying
\be \label{3dSUSY} \begin{array}{l} \{Q_+,\ol Q_+\} = -2P_+\,,\\[.1cm] \{Q_-,\ol Q_-\} = 2P_-\,, \end{array}\qquad  \begin{array}{l} \{Q_+,\ol Q_-\}=-2i(P_\perp-iZ)\,,\\[.1cm] \{Q_-,\ol Q_+\} = 2i(P_\perp+iZ)\,, \end{array} \ee
where $P_\pm=\pm P_0+ P_1$ are the left and right translation operators in $\R^{1,1}$ and $Z$ is a real central charge.
The algebra admits a $U(1)_R$ R-symmetry with charges
\be \begin{array}{c|cccc} & Q_+ & \ol Q_+ & Q_- & \ol Q_- \\\hline U(1)_R & -1 & 1 & -1 & 1 \end{array}\,. \ee

A given boundary condition can (potentially) preserve any subalgebra of 3d $\CN=2$ that does not contain $P_\perp$. 
We are interested in the $\CN=(0,2)$ subalgebra that is generated by $Q_+$ and $\ol Q_+$, and inherits  the $U(1)_R$ symmetry of 3d $\CN=2$. We focus our attention on boundary conditions that preserve this $(0,2)$ subalgebra \emph{and} leave $U(1)_R$ unbroken.
Note that unbroken $U(1)_R$ is necessary in order to define a half-index. Superconformal boundary conditions always preserve some $U(1)_R$,
though it may be emergent in the IR. 

\subsection{Free chiral}
\label{sec:chiral}

We begin with the simplest 3d theory: that of a free chiral multiplet. Off-shell, it contains a complex boson $\phi$, a complex fermion $\psi_\alpha$ and its conjugate $\bar \psi_\alpha$, and a complex auxiliary field $F$, which can be  grouped in a superfield
\be \Phi_{3d} = \phi + \theta^+\psi_+ + \theta^-\psi_-  + \theta^+\theta^- F + ...\,,   \ee
with remaining components determined by the chirality condition $\ol D_+\Phi_{3d}=\ol D_-\Phi_{3d}=0$. 
The 3d theory has an action $\int d^3x d^4\theta\, \Phi_{3d}^\dagger\Phi_{3d}$, and the equations of motion set $F=0$.
Our conventions for 3d and 2d SUSY coincide with \cite{Witten-phases, AHISS} and with a dimensional reduction of 4d $\CN=1$ following \cite{WessBagger}, up to numerical factors that we do not carefully keep track of, as they do not affect the structure of the results.

The multiplet $\Phi_{3d}$ can be decomposed into two multiplets under the $(0,2)$ subalgebra, a $(0,2)$ chiral
\be \label{2dPhi} \Phi = \Phi_{3d}\big|_{\theta^-=\bar\theta^-=0} = \phi + \theta^+\psi_+ -i\theta^+\bar\theta^+\pd_+\phi \ee
that satisfies $\ol D_+\Phi = 0$, and a $(0,2)$ Fermi multiplet
\be\hspace{1in} \label{2dPsi} \Psi  = \bar\psi_- + \theta^+ f  -i\theta^+\bar\theta^+\pd_+\psi_-  \quad(\;= \ol D_-\Phi_{3d}^\dagger\big|_{\theta^-=\bar\theta^-=0} \; \text{on shell}) \ee
that also satisfies $\ol D_+\Psi=0$.  The auxiliary field $f$ in the Fermi multiplet is equal on-shell to $f = \pd_\perp \bar\phi$, as can be seen from 
$f\sim D_+\Psi\big|_{\theta^+=\bar\theta^+=0} = D_+\ol D_-\Phi^\dagger_{3d}\big|_{\text{all $\theta=0$}} \sim \pd_\perp \Phi_{3d}^\dagger\big|_{\text{all $\theta=0$}} = \pd_\perp\bar\phi$
In the language of \cite{Witten-phases}, one would say that $\Psi$ has a ``J-term'' 
\be J \sim \pd_\perp \Phi\,. \ee
Written in (0,2) superspace, the bulk action takes the form
\be \int d^2 x\, dx^\perp \Big[\int d\bar\theta^+ d\theta^+\hspace{-3.5ex} \underbrace{\big( \Phi^\dagger \pd_-\Phi + \Psi^\dagger\Psi\big)}_{\text{standard 2d $(0,2)$ kin. term}} \hspace{-2ex} + \int d\theta^+\, \Psi \hspace{-2.5ex} \underbrace{\pd_\perp \Phi}_{\text{(0,2) J-term}}\hspace{-2ex} + \, c.c.\Big] \label{freeaction} \ee
with a fermionic superpotential $\Psi\pd_\perp\Phi$ that sets $f\sim \pd_\perp\bar\phi$ and ultimately gives rise to the $\pd_\perp$ derivatives in the 3d kinetic term.

There are two basic  boundary conditions for a free 3d chiral. They follow most simply by observing that the action \eqref{freeaction} contains a term $\bar\psi_-\pd_\perp\psi_+$, which forces one or the other of the fermions to be zero at the boundary in order to avoid a boundary term in the equations of motion (at least in the absence of any additional boundary matter). The SUSY completion of setting $\psi_-$ to zero, which we'll call Neumann and denote as `N', sets to zero the entire (0,2) multiplet~$\Psi$:
\be \text{N b.c.}\,:\qquad \Psi\big|_\pd = 0\quad\Rightarrow\quad
  \pd_\perp\phi\big|_\pd = 0\,,\quad \psi_-\big|_\pd = 0\,. \ee
(We call this Neumann due to the induced Neumann b.c. on $\phi$.)
One SUSY completion of setting $\psi_+$ to zero, which we'll call Dirichlet and denote `D', sets to zero the entire multiplet~$\Phi$:
\be \text{D b.c.}\,:\qquad \Phi\big|_\pd = 0\quad\Rightarrow\quad
  \phi\big|_\pd = 0\,,\quad \psi_+\big|_\pd = 0\,.\ee

Alternatively, we can set $\Phi$ equal to a constant (background) chiral superfield at the boundary, \emph{i.e.} to a complex number $c$, leading to a deformation of the Dirichlet b.c. that we denote `D$_c$'
\be \text{D$_c$ b.c.}\,:\qquad \Phi\big|_\pd = c\quad\Rightarrow\quad
  \phi\big|_\pd = c\,,\quad \psi_+\big|_\pd = 0\,.\ee
For the D$_c$ b.c., we always assume that $c\neq 0$.

Clearly, a D$_c$ boundary condition for a free chiral will not flow in the IR to a superconformal boundary condition.
More generally, in order for D$_c$ to preserve $U(1)_R$, the R-charge of $\phi$ must be zero.
This precludes any straightforward RG flow to a superconformal boundary condition even in the presence of interactions, since the R-charge is below the unitarity bound.
There is, however, a useful loophole: in a gauge theory, if $\phi$ is not gauge invariant, it may have R-charge zero as long as all gauge-invariant chiral operators have an R-charge above the unitarity bound. Thus, giving charged chirals D$_c$ b.c. in a gauge theory may still be part of the UV definition of superconformal boundary conditions.  Note that 
in order to include D$_c$ b.c. in a gauge theory, the gauge group will have to be partly broken 
at the boundary.

\subsubsection{Boundary conditions encoded in the action}
\label{sec:NDaction}
Any quantum field theory, even a free one, will generally admit a very large variety of ``boundary conditions,'' 
some of which may not even admit a weakly-coupled description. A general classification of such boundary conditions 
would be at least as intricate as the classification of quantum field theories in one dimension lower. 

Even semi-classically, there may be a large variety of elliptic boundary conditions on the bulk fields. 
We will ignore for now the possibility of boundary conditions defined by a singular behaviour of the 
bulk fields near the boundary --- we will come back to that in Section \ref{sec:flow}. 

A fairly general strategy to define boundary conditions is to start from a simple ``reference'' boundary condition and 
deform it by a boundary action, possibly involving auxiliary boundary degrees of freedom. Intuitively, 
one picks conjugate variables $(u,v)$ in phase space and a boundary action $S_\partial(u)$ to deform the boundary conditions as 
\begin{equation}
v|_\partial =0 \to \left(v + \frac{\partial S_\partial(u)}{\partial u}\right)\Big|_\partial =0
\end{equation}
As a shortcut, we can think about the original and modified boundary condition as arising 
as boundary equations of motions for half-space actions
\begin{align}
&\int_{x^\perp \leq 0} v \partial_\perp u \cr
&\int_{x^\perp \leq 0} v \partial_\perp u + S_\partial(u)
\end{align}
The shortcut is correct as long as we forbid the boundary action from depending on $v$. 

In general, a boundary condition is supersymmetric if the perpendicular component of the super-current 
is a total derivative along the boundary. A simple way to obtain supersymmetric boundary conditions is to 
start from a supersymmetric reference boundary condition and add supersymmetry-preserving boundary couplings.   

We illustrate this perspective by reformulating the basic N, D, and D$_c$ b.c. for a free 3d chiral in terms of half-space actions.
Effectively, $\Phi$ and $\Psi$ are conjugate variables. The reference N and D boundary conditions are associated to bulk actions 
\be \label{SND} \begin{array}{l}
 S_N = \int d^2x\int_{x^\perp\leq 0} dx^\perp \Big[\int d\bar\theta^+ d\theta^+\, \big( \Phi^\dagger \pd_-\Phi + \Psi^\dagger\Psi\big)  + \int d\theta^+\, \Psi\,\pd_\perp \Phi +  c.c.\Big] \\[.2cm]
 S_D = \int d^2x\int_{x^\perp\leq 0} dx^\perp \Big[\int d\bar\theta^+ d\theta^+\, \big( \Phi^\dagger \pd_-\Phi + \Psi^\dagger\Psi\big)  - \int d\theta^+\, (\pd_\perp\Psi)\Phi +  c.c.\Big]
\end{array} \ee

Varying $S_N$ leads to a boundary term 
\be \delta S_N = \text{(bulk terms)} +  \int d^2 x\int d\theta^+ \Psi\delta\Phi\big|_\pd + c.c. \label{SNvar}\ee
whose associated equation of motion sets $\Psi\big|_\pd=0$, effectively imposing N b.c.  Similarly, the variation of $S_D$ has a boundary term $\int d\theta^+ \delta\Psi \Phi\big|_\pd$ that imposes D b.c.:
\be \delta S_N =0 \quad\Rightarrow\quad \text{N b.c.};\qquad \delta S_D =0 \quad\Rightarrow\quad \text{D b.c.}\,.\ee

The modification D$_c$ of the Dirichlet boundary condition can be achieved by adding a boundary superpotential $c\Psi$. Consider
\be \label{Dcaction} S_{\text{D}_c} := S_D + \int d^2 x\int d\theta^+\, c\Psi\big|_\pd +  c.c.\,.
\ee
Then $\delta S_{\text{D}_c}= \text{(bulk)} -\int d^2x \int d\theta^+ \delta\Psi(\Phi-c)\big|_\pd$,
so that
\be \delta S_{\text{D}_c} = 0 \quad\Rightarrow\quad \text{D$_c$ b.c.} \ee

\subsection{Boundary fields and boundary superpotentials}
\label{sec:2dmatter}

In this section we discuss boundary conditions for free chirals that involve boundary degrees of freedom. In order to obtain $\CN=(0,2)$ boundary conditions, we add $(0,2)$ chiral and/or Fermi multiplets on the boundary, coupled to the bulk with a fermionic superpotential. We describe some general consequences of such a modification.

Consider a 2d $\CN=(0,2)$ theory with chiral multiplets $C_\alpha$ and Fermi multiplets $\Gamma_i$. Recall that interactions in this theory may be introduced via a choice of `E' and `J' terms for the Fermi multiplets. In superspace
\be \label{EJ} E_i = \ol D_+\Gamma_i\,,\qquad J_i = \ol D_+ \Gamma_i^\dagger\quad \text{(on shell)}\,.  \ee
Both $E_i$ and $J_i$ are necessarily chiral (since $\ol D_+^2=0$), and we assume they are holomorphic functions of the $C_\alpha$. The 2d action may be written as $S_{2d} = \int d^2x d\bar\theta^+ d\theta^+(C_\alpha^\dagger \pd_- C_\alpha + \Gamma_i\Gamma_i^\dagger) +\int d^2x\,d\theta^+ J_i\Gamma_i + c.c.$. (We use canonical kinetic terms.)
The action is supersymmetric provided that%
\footnote{This standard constraint arises from the variation of the fermionic superpotential $\ol Q_+ \int d\theta^+ J\cdot \Gamma \sim \int d\theta^+ \ol D_+(J\cdot \Gamma) = \int d\theta^+ E\cdot J$. This will vanish as desired if $E\cdot J$ is constant. Further requiring that the theory admits a supersymmetric vacuum, as in \eqref{EJ-BPS}, forces $E\cdot J=0$.\label{foot:EJ}} %
\be \label{EJ0-2d} E\cdot J\, := \sum_i E_i J_i = 0\,. \ee
In this theory, the equations for a supersymmetric vacuum%
\footnote{Meaning: a combination of the BPS equations for the full $(0,2)$ algebra together with the equations of motion in the low-energy limit, so that derivatives in the $x^0,x^1$ directions drop out.} include
\be \label{EJ-BPS} \begin{array}{c} E_i(C) = J_i(C) = 0\quad \forall\,i\,, \\[.2cm]
   \pd_\alpha J \cdot \Gamma + \pd_\alpha E \cdot \Gamma^\dagger =0\quad \forall\,\alpha\,, \end{array} \ee
where $\pd_\alpha J := \pd J(C)/\pd C_\alpha$, etc.
 
Note that there is an inherent symmetry under exchange $\Gamma_i\leftrightarrow\Gamma_i^\dagger$ and $E_i\leftrightarrow J_i$. This well-known symmetry may actually be viewed as fermionic analogue of T-duality. Some details of this perspective are presented in Appendix \ref{app:T}.

Now suppose we have a 3d chiral multiplet $\Phi_{3d}=(\Phi,\Psi)$ with N b.c., encoded in the action $S_N$ from \eqref{SND}, and use a boundary $\CN=(0,2)$ theory as above to modify it. We introduce boundary chiral and Fermi multiplets $C_\alpha,\Gamma_i$, and couple them to the bulk by allowing the E and J terms to contain a holomorphic dependence on the boundary values of $\Phi$ . The modified action takes the form
\be S_N^{\rm mod} = S_N + i \int d^2x \int d\theta^+ {\textstyle \sum_i}J_i(C,\Phi|_\pd) \Gamma_i + c.c. + \text{(bdy kinetic terms)} \ee
The variation of $S_N^{\rm mod}$ with respect to the boundary fields $C,\Gamma$ gives rise, at low energy, to the usual 2d constraints \eqref{EJ-BPS}. In addition, the variation of $S_N^{\rm mod}$ with respect to the bulk field $\Phi$ takes the form
\be \delta S_N^{\rm mod} = \text{(bulk)} + i\int d^2x\int d\theta^+\big[\Psi\delta\Phi|_\pd +  \pd_\Phi J \cdot \Gamma\,\delta\Phi|_\pd\big] +  c.c. + \delta\text{(bdy kinetic)}\,. \ee
Setting this variation to zero leads to
\be   \Psi\big|_\pd = - \pd_\Phi J\cdot\Gamma - \pd_\Phi E\cdot \Gamma^\dagger\,. \label{N-mod} \ee
Thus, in the infrared, we expect the modified boundary condition to flow to one combining \eqref{N-mod} and $E=J=\pd_\alpha J\cdot \Gamma+\pd_\alpha E\cdot \Gamma^\dagger=0$ from \eqref{EJ-BPS}.

We may modify the D b.c. on a 3d chiral in a similar way. We begin with the action $S_D$ from \eqref{SND}. We add 2d multiplets $C_\alpha,\Gamma_i$ with E and J terms that depend only on $C$. Then we introduce a boundary J-term $J_\Psi(C)$ for the restriction to the boundary of the bulk multiplet $\Psi$. 
The modified action takes the form
\be S_D^{\rm mod}  = S_D+ i \int d^2x \int d\theta^+\big[ J_\Psi(C) \Psi|_\pd + {\textstyle \sum_i}J_i(C)\Gamma_i(C)\big] + c.c. + \text{(bdy kinetic)}\,. \ee
Setting to zero the boundary variation of this action leads at low energy to
\be \begin{array}{c} \Phi\big|_\pd = J_\Psi(C)\,,\qquad E_i(C)=J_i(C)=0\,,\\[.2cm]
 \pd_\alpha J_\Psi \Psi\big|_\pd + \textstyle\sum_i( \pd_\alpha J_i \Gamma_i+\pd_\alpha E_i\Gamma_i^\dagger)=0\,.
\end{array}\ee
The first equation is the deformed boundary condition, the other equations are required for a low energy supersymmetric vacuum. 

\subsubsection{Flips}
\label{sec:flip}

There are two simple, important special cases of modified boundary conditions for a 3d chiral.

Starting with N b.c., we may introduce a single boundary Fermi multiplet $\Gamma$ with $J=\Phi$. The modified action is
\be S_{N,\Gamma} = S_N + i \int d^2 x\int d\theta^+ \, \Phi|_\pd\Gamma +  c.c. + \text{($\Gamma$ kinetic)}\,,\ee
whose boundary variation (at low energy) sets
\be \Psi\big|_\pd = - \Gamma\,,\qquad  J=\Phi\big|_\pd = 0\,. \label{NmodD}\ee
This both relieves the constraint on the boundary value of $\Psi$ (imposed by pure N b.c.) and constrains $\Psi$. Indeed, \eqref{NmodD} looks just like D b.c.

Conversely, starting with D b.c., we may introduce a single boundary chiral multiplet $C$ and a J-term $J_\Psi=C$ for $\Psi$. The modified action is
\be \label{D-flip} S_{D,C}=  S_D + i \int d^2 x\int d\theta^+ \,\Psi\big|_\pd C+ c.c. + \text{($C$ kinetic)}\,, \ee
whose boundary variation (at low energy) sets
\be \Phi\big|_\pd = C\,,\qquad \pd_CJ_\Psi=\Psi\big|_\pd = 0\,.\ee
This frees up the boundary value of $\Phi$ while constraining $\Psi$, so that we effectively end up with N b.c.
(Note also that the definition of D$_c$ b.c. in \eqref{Dcaction} is a special case of \eqref{D-flip}, with $C=c$ a constant rather than a dynamical boundary field.)

We say that coupling to a boundary Fermi multiplet (and flowing to the IR) ``flips'' N to D, while coupling to a boundary chiral multiplet (and flowing to the IR) ``flips'' D to N:
\be \label{flips}\text{flips}:\quad \begin{array}{l@{\quad\leadsto\quad}c} \text{D[C]} & \text{N} \\
 \text{N[$\Gamma$]} & \text{D} \end{array}\,.
\ee
These ``flips'' are analogous to modifications of boundary conditions of 4d $\CN=2$ theories that were discussed in \cite{DGG}, inspired by the field-theoretic Fourier transform of \cite{Witten-SL2, KapustinStrassler}.

\subsubsection{Multiple chirals}

If we have multiple 3d free chiral multiplets $\big\{\Phi_{3d}^a\big\}_{a=1}^{N_f}$, we can define a basic boundary condition by choosing N, D, or D$_c$ b.c. for each $\Phi_{3d}^a$ independently. This basic combination of boundary conditions can be further modified by introducing boundary fields and couplings, in a straightforward generalization of the above analysis.

For example, we could use a modification to engineer a boundary condition that restricts the bosons $\phi^a$ to lie on an arbitrary holomorphic submanifold $S\subset \C^{N_f}$, and the fermions $\psi_-^a$ to take values in the (parity-reversed) conormal bundle of $S$:
\be (\phi,\psi_-)\in N^*S\subset T_{[1]}^*\C^{N_f}\,. \label{conormal} \ee
To do so, we begin with N b.c. for all the $\Phi_{3d}^a$, and introduce boundary Fermi multiplets $\{\Gamma_i\}_{i=1}^d$ with J-tems $J_i(\Phi^1|_\pd,...,\Phi^{N_f}|_\pd)$ and $E_i=0$. This results in a modified boundary condition
\be J_1(\Phi)\big|_\pd=...=J_d(\Phi)\big|_\pd = 0\,,\qquad \Psi_-^a\big|_\pd \sim \frac{\pd J_i(\Phi)}{\pd \Phi^a}\Gamma_i\quad \forall a\,, \ee
which is precisely of the form \eqref{conormal} for $S = \{J_1=...=J_d=0\}\subset \C^{N_f}$.
Such boundary conditions are familiar in the 2d B-model, where they given rise to coherent sheaves.

\subsection{Bulk superpotentials}
\label{sec:mx}

A standard bulk deformation of the theory of $N_f$ free chirals $\Phi_{3d}^a$ corresponds to adding a superpotential
\be  \int d^3x \int d^2\theta \,W(\Phi_{3d}) + c.c. \label{W} \ee
to the bulk action.
In the presence of a superpotential, Neumann-type boundary conditions require a modification (often a severe modification, which can make them partially Dirichlet) in order to preserve 2d $(0,2)$ supersymmetry \cite{Warner}.
The structure of the possible modifications is encapsulated by the same ``matrix factorizations'' that show up in 2d Landau-Ginzburg models. We review a few relevant details here, following \cite[Sec. 4.1]{GGP-fivebranes}.

First, let's recall why Neumann b.c. are incompatible with superpotentials.
When $W\neq 0$, each 3d chiral multiplet may still be decomposed into a 2d chiral $\Phi^a$ and a 2d Fermi multiplet $\Psi^a$, as in \eqref{2dPhi}, \eqref{2dPsi}. However, the Fermi multiplets no longer satisfy $\ol D_+\Psi=0$; rather, one finds
\be \ol D_+ \Psi^a \sim \frac{\pd W(\Phi)}{\pd \Phi^a}\,. \ee
In $\CN=(0,2)$ terminology, $\Psi^a$ has both a J-term $J^a \sim \pd_\perp \Phi^a$ and an E-term $E_a \sim \pd_aW$. 
We may now try to engineer a Neumann
boundary condition by writing down an appropriate half-space action and setting the boundary variation to zero --- just as in Section \ref{sec:NDaction}.
The action $S_N$ in \eqref{SND} is a natural choice, and certainly induces $\pd_\perp\phi^a\big|_\pd=0$. However, this action no longer preserves $(0,2)$ SUSY, because it violates the standard $E\cdot J=0$ constraint \eqref{EJ0-2d}. In the present case, we have
\be \label{EJ0} \text{``$E\cdot J$''}= \int_{x^\perp\leq 0} dx^\perp {\textstyle \sum_a E_a\cdot J^a} =  \int_{x^\perp\leq 0} dx^\perp {\textstyle \sum_a \pd_a W \pd_\perp \Phi^a} =  \int_{x^\perp\leq 0} dx^\perp W = W(\Phi)\big|_\pd\,, \ee
whence the constraint is $W(\Phi)\big|_\pd=0$.  This clearly cannot be satisfied if $W$ is nontrivial and the $\phi^a$ are left unconstrained on the boundary, as for standard N b.c.%
\footnote{From a more fundamental perspective, the incompatibility of N b.c. with bulk superpotentials may be seen as a consequence of the fact that N b.c. do not ensure that the bulk BPS equations $dW=0$ are satisfied at the boundary.}

Fortunately, the computation \eqref{EJ0} tells us how to remedy the situation. We can introduce additional boundary Fermi multiplets $\Gamma^i$ ($i=1,...d$) with their own E and J terms $E_\Gamma^i$, $J_\Gamma^i$ (which are holomorphic functions of $\Phi|_\pd$, and perhaps other boundary chirals) such that
\be \label{MF} \textstyle \sum_i E_\Gamma^i J_\Gamma^i = -W(\Phi)\big|_\pd+ \mathrm{const}\,. \ee
The factorization (up to a constant shift) of $W$ in this manner is a basic example of a matrix factorization.
The constant will have to vanish for boundary conditions that preserve $U(1)_R$. %
Note that many choices of $E_\Gamma$ and $J_\Gamma$ are possible.
The relation \eqref{MF} ensures that the \emph{total} $E\cdot J$ vanishes, so that the modified half-space action
\be \label{SNW} S_N + i \int d^2x \int d\theta^+\, \sum_i \Gamma^i J_\Gamma^i(\Phi|_\pd,...) + c.c. +\text{(bdy kinetic)} \ee
is once more supersymmetric. The actual boundary condition resulting from the variation of \eqref{SNW} now becomes (\emph{cf.} \eqref{N-mod})
\be \label{MFbc} \text{MF$_{J,E}$}:\quad \textstyle \Psi^a\big|_\pd \sim \sum_i \big[\pd_a J_\Gamma^i(\Phi|_\pd)\Gamma^i+\pd_aE_\Gamma^i(\Phi|_\pd)\Gamma^i{}^\dagger\big]\,;\qquad J_\Gamma^i(\Phi|_\pd) = E_\Gamma^i(\Phi|_\pd)= 0 \quad\forall\,i\,. \ee

The ``matrix factorization'' \eqref{MFbc} is the most general sort of boundary condition we will need in the presence of a superpotential. It may be a mild or severe modification of standard N b.c., depending on the precise choice of $J_\Gamma$ and $E_\Gamma$.
Notice that $J_\Gamma\big|_\pd = E_\Gamma\big|_\pd=0$ impose Dirichlet-like b.c. on some of the $\phi^a$. The additional relations $\Psi\big|_\pd \sim dJ\cdot \Gamma$ constrain some of the $\Psi$'s (those in the cokernel of the map $dJ$), leading to Neumann b.c. for others of the $\phi^a$.

We should also discuss the effect of a bulk superpotential on Dirichlet $\text{D}_c$ boundary conditions. 
Generic Dirichlet boundary conditions do not appear to break SUSY explicitly when the 
bulk superpotential is turned on, but they break it spontaneously if the boundary value of the fields 
is not a critical value of $W$. That means we need to impose $\partial W(c)=0$ in order to have (classically)
low energy supersymmetry. 

It is easy to reproduce this statement as a special case of \eqref{MFbc}. Adding auxiliary boundary Fermi multiplets $\Gamma^a$ 
with $J^a=\Phi^a-c^a$ to go from Neumann to general Dirichlet b.c. we have a constraint 
\be (\Phi^a-c^a)E_a(\Phi,c) = W(\Phi) -W(c)\,.\ee
that can be solved in a straightforward manner. 
Low energy SUSY requires both $\Phi^a=c^a$ and 
\be E_a(\Phi,c) = E_a(c,c) = \partial W(c)=0 \ee

In a similar way, we can argue that if we give Dirichlet b.c. to a subset $\Phi$ of the bulk fields, Neumann for the remaining $\Phi'$, 
then we will need a matrix factorization of $W(c,\Phi')$. Compatibility with the bulk SUSY vacua will further require at low energy 
$\partial_c W(c,\Phi')=0$. 

Although Dirichlet boundary conditions do not need extra conditions to preserve SUSY in the UV, 
there are constraints on boundary couplings of the form $J_\Phi(C^\alpha) \Psi|_\partial$,
in the sense that the $E.J = \mathrm{const}$ constraint is deformed to $E.J -W(J_\Phi) = \mathrm{const}$.

\subsubsection{Example: XYZ}
\label{sec:XYZmf}

A simple example of a 3d $\CN=2$ theory with a superpotential is the ``XYZ model,'' \emph{i.e.} three chiral fields $X_{3d},Y_{3d},Z_{3d}$ coupled by a superpotential $W = X_{3d}Y_{3d}Z_{3d}$. What are the basic (0,2) boundary conditions for this theory?

The preceding analysis suggests that we cannot choose N b.c. for all three chirals without adding some extra degrees of freedom to factorize $XYZ$. Setting D b.c. for a single chiral, say $X=0$, is already enough to preserve SUSY in the UV. On the other hand, if we deform the 
boundary condition for $X$ to D$_c$, we will still have to deal with the restriction $cYZ$ of the bulk superpotential. 

Altogether, we can consider the following simple boundary conditions:
\begin{itemize}
\item (D,D,D): D b.c. for all three fields. This can be deformed to a (D$_x$,D$_y$,D$_z$), but the deformation will break SUSY spontaneously 
unless $xy=xz=yz=0$. In other words, we should really only consider (D$_c$,D,D), (D,D$_c$,D) or (D,D,D$_c$) deformations. 
\item (D,D,N): Dirichlet for $X$ and $Y$, Neumann for $Z$ (or permutations thereof). We can consider deformations 
(D$_c$,D,N) or (D,D$_c$,N).
\item (D,N,N): Dirichlet for one field, Neumann for the other two.  
\end{itemize}
In Section \ref{sec:XYZ}, we will identify the duals of most of these boundary conditions in 3d $\CN=2$ SQED.

\subsection{Boundary 't Hooft anomalies}
\label{sec:anom}

Boundary conditions that preserve $\CN=(0,2)$ supersymmetry typically have 't Hooft anomalies for global symmetries. We would like to determine what they are. 

We use the following conventions:
\begin{itemize}
\item A purely two-dimensional left-handed chiral fermion (such as the leading component $\gamma_-$ in a Fermi multiplet $\Gamma = \gamma_-+...$)
contributes $+1$ to the anomaly for the $U(1)$ symmetry it is charged under. Letting $\mb f$ denote the field strength for the symmetry, the anomaly polynomial is $\CI_2(\mb f)=\mb f^2$.
\item A purely two-dimensional right-handed chiral fermion (such as the fermion $\chi_+$ in a chiral multiplet $C= c+\theta^+\chi_-+...$)  contributes $-1$ to the anomaly for its $U(1)$ symmetry, with anomaly polynomial $-\mb f^2$.
\item Three-dimensional $U(1)$ Chern-Simons theory (purely bosonic) has a level $k\in \Z$ naturally quantized to be an integer; and it induces an anomaly $+k\,\mb f^2$ on the boundary.
\end{itemize}

Now, a three-dimensional fermion $(\psi_+,\psi_-)$ has both left-handed and right-handed components with respect to the 2d Lorentz group. Typical boundary conditions set either $\psi_+$ or $\psi_-$ to zero at the boundary. Let $\mb f$ be the field strength for the $U(1)$ symmetry rotating $\psi$.
We claim that
\begin{itemize}
\item A three-dimensional fermion with b.c. $\psi_+\big|_\pd=0$ (so that $\psi_-$ survives at the boundary)  contributes $\tfrac12\mb f^2$ to the anomaly polynomial.
\item A three-dimensional fermion  with b.c. $\psi_-\big|_\pd=0$ (so that $\psi_+$ survives at the boundary)  contributes  $-\tfrac12\mb f^2$ to the anomaly polynomial.
\end{itemize}
To verify the claim, we introduce a real mass for the 3d fermion $\psi_\alpha$, flow to the IR, and match UV and IR anomalies. A real mass term $im \epsilon^{\alpha\beta}\psi_\alpha\bar\psi_\beta$
has two effects at energies below $|m|$. First, the fermion becomes fully massive in the bulk, and integrating it out at one-loop generates a (background) Chern-Simons term for its $U(1)$ symmetry, at level $\tfrac12\, \text{sign}(m)$ \cite{NiemiSemenoff, Redlich1, Redlich2} (\emph{cf.} \cite{AHISS} in the supersymmetric case). Second, since the Dirac equations take the form
\be \begin{array}{c}  (\pd_\perp + m)\psi_- + \text{($\pd_\pm$ of other fermions)} = 0\,, \\[.1cm]
(\pd_\perp - m)\psi_+ + \text{($\pd_\pm$ of other fermions)} = 0\,,
\end{array}
\ee
normalizable edge modes (\emph{i.e.} purely 2d fermions) may survive, with profiles
\be \label{edge-basic} \qquad\qquad  \begin{cases}\psi_- = a_-(x^\mu)\, e^{-mx^\perp}& \text{if $m<0$} \\
\psi_+ = a_+(x^\mu)\, e^{mx^\perp} & \text{if $m>0$} \end{cases}   \quad \text{(on $x^\perp\leq 0$)}\,. \ee
If the boundary condition is $\psi_+\big|_\pd=0$, then an edge mode of $\psi_-$ exists when $m<0$, so the total IR anomaly is
\be \frac 12\,\text{sign}(m) + \left\{\begin{array}{ll} 0 &\quad \text{if}\;\; m>0 \\
1 & \quad \text{if}\;\;m < 0 \end{array}\right\} = \frac12 \ee
On the other hand, if the boundary condition is $\psi_-\big|_\pd=0$, then an edge mode of $\psi_+$ exists when $m>0$, so the total IR anomaly is
\be  \frac 12\,\text{sign}(m) + \left\{\begin{array}{ll} -1 &\quad \text{if}\;\; m>0 \\
0 & \quad \text{if}\;\;m < 0 \end{array}\right\} = -\frac12 \ee
This substantiates our claim\footnote{The non-supersymmetric analysis of \cite{Aitken:2017nfd} proceeds in a similar spirit; we expect their results are recoverable from our analysis.}.

In the case of a free chiral multiplet $\Phi_{3d}$, there are two $U(1)$ symmetries around: a flavor symmetry $U(1)_f$ under which $\Phi_{3d}$ has charge $1$ (say), and the R-symmetry $U(1)_R$ under which $\Phi_{3d}$ has some charge $\rho$. The component fields have charges
\be \begin{array}{c|c@{\quad}c@{\quad}c}
& \Phi,\phi & \psi_+ & \Psi,\bar\psi_- \\\hline
U(1)_f & 1 & 1 & -1 \\
U(1)_R & \rho & \rho-1 & 1-\rho \end{array}
\ee
(Note that the superpotential $\Phi\pd_\perp\Psi$ has $U(1)_f$ charge 0 and $U(1)_R$ charge $1$, as required for unbroken flavor and R-symmetry.) Therefore, our basic boundary conditions come with anomalies
\be \label{ND-anom} \text{N b.c.}\,:\quad -\tfrac12(\mb f+(\rho-1)\mb r)^2\,;\qquad \text{D b.c.}\,:\quad \tfrac12(\mb f+(\rho-1)\mb r)^2\,,  \ee
encoded as polynomials $\CI_2(\mb f,\mb r)$ in the $U(1)_f$ and $U(1)_R$ curvatures.

This prescription is perfectly consistent with the ``flips'' that modify D to N or vice versa. For example, in order to flip D to N we must introduce a boundary chiral multiplet $C$. The boundary superpotential $\Psi C$ fixes the $U(1)_f$, $U(1)_R$ charges of $C$ to be $(1,\rho)$, so that the right-handed chiral fermion in $C$ contributes an anomaly $-(\mb f+(\rho-1)\mb r)^2$, which is exactly right to modify the D anomaly to the N anomaly.

It is somewhat insightful to repeat the analysis of edge modes for a free chiral in superspace. If we introduce a real mass $m$ as a background value of the scalar field in the $U(1)_f$ gauge multiplet, the real mass will enter the Lagrangian through complexified covariant $\pd_\perp$ derivatives. In other words, the superpotential $\Psi\pd_\perp \Phi$ in \eqref{freeaction} gets modified to
\be  \Psi\pd_\perp \Phi \quad\to\quad \Psi(\pd_\perp - m)\Phi\,. \label{Dm} \ee
Recalling that the real mass $m$ plays the role of a central charge $Z$, we may view \eqref{Dm} as a consequence of the fact that $Z$ complexifies $P_\perp$ in the SUSY algebra \eqref{3dSUSY}. The $\CN=(0,2)$ BPS equations now arising from the superpotential \eqref{Dm} are
\be \label{BPS-m} (\pd_\perp-m)\Phi = 0\,,\qquad (\pd_\perp+m)\Psi = 0\,,\ee
which lead directly to the edge mode profiles in \eqref{edge-basic}.

\subsubsection{Nonabelian symmetries}
\label{sec:anom-na}

The anomaly for a nonabelian symmetry group $G$ may be fixed by ensuring that the above rules/conventions are consistent with the maximal abelian torus of $G$. As might be expected, we find that 3d fermions that survive as chiral fields on the boundary contribute \emph{half} of the standard 2d anomaly.

We briefly recall the standard 2d result. Let $G$ be a simple compact group and $\mb R$ an irreducible unitary representation of $G$.
Let $T_{\mb R}$ denote the quadratic index of $\mb R$, defined so that the ratio of traces in any pair of representations obeys $\frac{\text{Tr}_{\mb R}}{\text{Tr}_{\mb R'}} = \frac{T_{\mb R}}{T_{\mb R'}}$, and normalized so that $T_{\rm adjoint} = 2 h$ where $h$ is the dual Coxeter number. For example, this means that if $G=SU(N)$ we have $T_{\rm fundamental}=1$ and $T_{\rm adjoint} = 2N=2h$. 
 In general, the index so normalized may be computed as a sum of lengths-squared of weights
\be T_{\mb R} = \frac{1}{\text{rank}\,G} \sum_{\lambda\in {\mb R}} |\!|\lambda|\!|^2 \ee
where long roots $\alpha$ have $ |\!|\alpha|\!|^2=2$.
When $G$ is $U(N)$ or $SU(N)$, we simply write `$\text{Tr}$' the usual trace (on the enveloping algebra) in the fundamental representation; and for general $G$ define $\text{Tr} := \frac{1}{2h} \text{Tr}_{\rm adjoint}$.
Let $\mb f$ denote the field strength of a $G$ connection. Then
\begin{itemize}
\item A 2d left-handed (resp., right-handed) complex fermion in representation $\mb R$ of $G$ contributes $\text{Tr}_{\mb R}(\mb f^2) =T_{\mb R} \text{Tr}(\mb f^2)$ (resp., $-T_{\mb R}\text{Tr}(\mb f^2)$) to the anomaly polynomial.

\item Correspondingly, the left-handed (resp., right-handed) component of a 3d fermion in representation $\mb R$ of $G$ that is unconstrained by a boundary condition contributes $\frac12T_{\mb R} \text{Tr}(\mb f^2)$ (resp., $-\frac12 T_{\mb R}\text{Tr}(\mb f^2)$) to the anomaly polynomial.

\item In this same normalization, 3d Chern-Simons term for $G$ at level $k$ contributes $k\,\text{Tr}(\mb f^2)$.
\end{itemize}

We may extend this analysis to a 3d chiral multiplet $\Phi_{3d}$ in representation $(\mb R,\rho)$ of $G\times U(1)_R$, where $G$ is simple and $U(1)_R$ is the R-symmetry. We find that Neumann and Dirichlet b.c. have boundary anomalies
\be \begin{array}{l}
\text{N b.c.:}\quad -\frac12 T_{\mb R} \text{Tr}(\mb f^2)-\frac12(\text{dim}\,\mb R)(\rho-1)^2\mb r^2\,, \\[.2cm]
 \text{D b.c.:}\quad \frac12 T_{\mb R} \text{Tr}(\mb f^2)+\frac12(\text{dim}\,\mb R)(\rho-1)^2\mb r^2\,.
  \end{array} \ee

If $G$ is not simple, the anomaly can be computed by directly ensuring agreement with the abelian anomaly for the maximal torus. We list a few special cases for $G=U(N)$. Let $\mb f$ denote the usual $N\times N$ field strength. A 3d chiral of R-charge $\rho$ in the fundamental, anti-fundamental, and adjoint representations of $G=U(N)$ has a boundary anomaly
\be \label{Un-anom} \begin{array}{rl}
\text{fund:}\quad &\pm \big[\frac12 \text{Tr}(\mb f^2)+(\rho-1)\mb r\,\text{Tr}\,\mb f+\frac12N(\rho-1)^2\mb r^2\big]\,,\\[.2cm]
\text{anti-fund:}\quad &\pm \big[\frac12 \text{Tr}(\mb f^2)-(\rho-1)\mb r\,\text{Tr}\,\mb f+\frac12N(\rho-1)^2\mb r^2\big]\,,\\[.2cm]
\text{adj:}\quad  &\pm [ N\text{Tr}(\mb f^2)-(\text{Tr}\,\mb f)^2+\frac12N^2(\rho-1)^2\mb r^2]\,,
\end{array}\ee
with a `-' sign for Neumann b.c. and a `+' sign for Dirichlet.

\subsection{Gauge fields}
\label{sec:bdy-gauge}

We next  extend the analysis of $\CN=(0,2)$ boundary conditions to gauge fields.

\subsubsection{Reduction of 3d gauge multiplets}
\label{sec:gauge2d}

For gauge group $G$, the 3d $\CN=2$ gauge multiplet contains the connection $A_m$, a real scalar $\sigma$, a complex fermion $\lambda_\alpha$, and (off shell) a real auxiliary field $D_{3d}$, all valued in the real Lie algebra $\mathfrak g$. These fields may be grouped in a Hermitian vector superfield
\be \hspace{.7in} V_{3d} = \theta\sigma^m\bar\theta A_m+i\theta\bar\theta\sigma -i\theta^2\bar\theta\bar\lambda-i\bar\theta^2\theta\lambda+\tfrac12\theta^2\bar\theta^2 D_{3d}\qquad\text{(WZ gauge)} \ee
or a gauge-covariant linear superfield
\be \Sigma_{3d} = -\frac{i}{2}\epsilon^{\alpha\beta}\ol D_\alpha D_\beta V_{3d}= \sigma+\bar\theta\sigma^{mn}\theta F_{mn}+i\theta\bar\theta D- \theta\bar\lambda+\bar\theta\lambda+... \ee
that contains the field strength and satisfies $D_\alpha D^\alpha\Sigma_{3d} = \ol D^\alpha \ol D_\alpha \Sigma_{3d}=0$, \emph{cf.} \cite{AHISS}. On shell, the auxiliary field $D_{3d}$ is set equal to the moment map for $G$ acting on the matter scalars (which necessarily parameterize a K\"ahler manifold)\,:
\be \label{D3d} \tfrac{1}{e^2}D_{3d} = \mu_{\rm matter}\,. \ee
We typically set the gauge coupling $e^2=1$, as it can be restored by dimensional analysis.
In the presence of an FI term $t$ and/or a supersymmetric Chern-Simons term at level $k$, the D-term is modified to%
\footnote{The relative sign of the moment map and Chern-Simons term in \eqref{D3dk} is somewhat important (in contrast to most numerical factors in the SUSY analysis). It enters in the (0,2) BPS equations (Section \ref{sec:flow}), which in turn control edge modes that can contribute to anomalies. 
The sign in \eqref{D3dk} was carefully computed, and agrees with our sign conventions for anomalies.} %
\be \label{D3dk}  \tfrac{1}{e^2} D_{3d} = \mu_{\rm matter}  - t- k \sigma\,. \ee

The 3d gauge multiplet may be decomposed into an $\CN=(0,2)$ gauge multiplet and an $\CN=(0,2)$ chiral multiplet. 
The chiral multiplet contains the complexified connection $\sigma+iA_\perp$ in the $x^\perp$ direction, together with the right-handed fermion $\lambda_+$. As an $\CN=(0,2)$ chiral superfield, it takes the form
\be S = \sigma+iA_\perp -2\theta^+\bar\lambda_++...\,, \ee
with its remaining $\theta^+\bar\theta^+$ component determined by the (gauge-covariant) chirality constraint $\ol D_+S=0$. In the presence of chiral matter, the superfield $S$ appears in complexified covariant derivatives. For example, the fermionic superpotential $\Psi\,\pd_\perp\Phi$ from \eqref{freeaction} that we encountered when writing an $\CN=(0,2)$ superspace action for a 3d chiral gets modified to
\be \label{dS} \int d\theta^+ \Psi(\pd_\perp-S)\Phi\,. \ee
Note how this is compatible with the BPS equations \eqref{BPS-m} in the presence of a background gauge multiplet (containing $m$ rather than $\sigma$), and the fact that $Z$ complexifies $P_\perp$ in the SUSY algebra \eqref{3dSUSY}.

The remaining fields of the 3d gauge multiplet go into an $\CN=(0,2)$ gauge multiplet, which comprises a pair of superfields
\be \label{AV-} \qquad A = \theta^+\bar\theta^+A_+\,,\qquad V_- =A_--2i\theta^+\bar\lambda_--2i\bar\theta^+\lambda_-+2\theta^+\bar\theta^+D\qquad\text{(in WZ gauge)}\,, \ee
where $A_\pm = A_0\pm A_1$. Here `$D$' is a new auxiliary field that is equal on shell to%
\footnote{The relative sign of $\pd_\perp\sigma$ appearing here is important, just like the sign of $k$ was in \eqref{D3dk}. It can ultimately fixed by observing that the $\CN=(0,2)$ BPS equations take the form of a gradient flow, as discussed in Section~\ref{sec:flow}.}
\be D = D_{3d} - \pd_\perp\sigma\,. \label{D2d}\ee
(The RHS of \eqref{D2d} will contain covariant $\pd_\perp$ derivatives when the gauge group is nonabelian.)
From the gauge multiplet one can also construct the fermionic field-strength superfield
\be \Upsilon = \lambda_-+\theta^+(F_{01}+iD)+...\,, \ee 
which is covariantly chiral $\ol D_+\Upsilon=0$.

The kinetic terms of the 3d gauge theory action may be written more or less as ordinary $\CN=(0,2)$ kinetic terms for a vector multiplet $(A,V_-)$ coupled to a chiral $S$. Some care must be taken to account for the fact that $S$ has an exotic gauge transformation%
\footnote{If $\Lambda$ is the chiral parameter of a super-gauge transformation, then the gauge transformation takes the from $S\to e^{-i\Lambda}Se^{i\Lambda}+e^{-i\Lambda}\CD_\perp e^{i\Lambda}$, mirroring the gauge transformation of $A_\perp$. For abelian gauge group, this looks like $S\to S+i\pd_\perp\Lambda$.}, %
but this is not important for our subsequent analysis. Additionally, an FI term for the abelian part of the gauge group appears in a superpotential
\be \int d^2xdx^\perp \int d\theta^+ t\,\text{Tr}\,\Upsilon + c.c.\,; \ee
while a 3d Chern-Simons term at level $k$ manifests as
\be k \int d^2xdx^\perp \int d\bar\theta^+d\theta^+\text{Tr}(A\pd_\perp V-V\pd_\perp A)-k \int d^2x dx^\perp\int d\theta^+ \text{Tr}(\Upsilon S) + c.c.\,, \ee
which also includes a superpotential piece.

We note that the $U(1)_R$ charges of fields in the 3d gauge multiplet are uniquely determined from the fact that $A_\mu,A_\perp,\sigma$ have R-charge zero. Explicitly, the R-charges are
\be \begin{array}{c|cccc}
 & \;S\; & \;\Upsilon\; &  \;\lambda_\pm\; & \;\bar\lambda_\pm\; \\\hline
U(1)_R & 0 & 1  & 1 & -1 \end{array} \,.   \label{RSU}
\ee

\subsubsection{Basic $\CN$, $\CD$ boundary conditions}
\label{sec:NDgauge}

In pure, non-supersymmetric gauge theory, there always exist two basic boundary conditions compatible with the Maxwell/Yang-Mills equations of motion. A Neumann boundary condition sets $F_{\perp\mu}\big|_\pd=0$, and preserves gauge symmetry on the boundary. A Dirichlet boundary condition sets $A_\mu\big|_\pd=0$, and breaks gauge symmetry to a global symmetry at the boundary (since only gauge transformations that are constant along the boundary preserve the constraint $A_\mu\big|_\pd=0$). In 3d $\CN=2$ gauge theory, there exist natural supersymmetric completions of these basic Neumann and Dirichlet b.c., which we denote $\CN$ and $\CD$, respectively. 

The $\CN=(0,2)$ supersymmetric completion of Dirichlet simply sets the superfields $A,V_-$ from \eqref{AV-} to zero on the boundary:
\be \label{Dgauge} \CD\;\text{b.c.}\;:\qquad A\big|_\pd = V_-\big|_\pd= 0\quad\Rightarrow\quad A_\pm\big|_\pd = 0\,,\quad D\big|_\pd=0\,,\quad \lambda_-\big|_\pd = 0\,.\ee
The constraint on the D-term translates on-shell to
\be \pd_\perp\sigma\big|_\pd =\big[ \mu_{\rm matter}+t+k\sigma\big]_\pd\,, \label{Ns}\ee
which is a modified Neumann b.c. for the scalar $\sigma$. (As usual, if the gauge group is nonabelian, $\pd_\perp$ should be promoted to a covariant derivative.) In a quantum theory, the usual one-loop corrections from massive fermions will modify the Chern-Simons term on the RHS of \eqref{Ns}.

The $\CN=(0,2)$ supersymmetric completion of Neumann may be understood as setting to zero the complexified (chiral) covariant derivative $\pd_\perp-S$ (as in \eqref{dS}) in a gauge-invariant way. This results in
\be \label{Ngauge} \CN\;\text{b.c.}\;: \qquad F_{\perp\pm}\big|_\pd = 0\,,\quad \sigma\big|_\pd = 0\,,\quad \lambda_+\big|_\pd = 0\,,\ee
which includes a Dirichlet b.c. for the scalar $\sigma$. In the presence of additional boundary matter charged under the bulk gauge group $G$, the constraint on the field strength is modified to $F_{\perp\pm}\big|_\pd = J^\pd_\pm$, where $J^\pd_\pm$ is the current for the boundary $G$ symmetry. Moreover, for abelian factors in the gauge group, one may introduce boundary FI parameters $t^\pd$, which modify the Neumann b.c. to
\be \sigma\big|_\pd = t^\pd\,. \ee
The FI parameters enter supersymmetrically via a boundary superpotential $\int d\theta^+\, t^\pd \Upsilon|_\pd+c.c.$

A standard generalization of the basic $\CN,\CD$ b.c. in gauge theory involves choosing a subgroup $H\subseteq G$ of the gauge group to remain unbroken on the boundary. 
Let $\mathfrak h\subseteq \mathfrak g$ denote the Lie algebra of $H$, and $\mathfrak h^\perp$ its orthogonal complement with respect to the Cartan-Killing form; and let $\pi:\mathfrak g\to \mathfrak h$ and $\pi^\perp:\mathfrak g\to \mathfrak h^\perp$ denote the corresponding orthogonal projections. 
Then we may define a hybrid boundary condition
\be \CN_H\;\text{b.c.}\;:\qquad \pi(A)\big|_\pd = \pi(V_-)\big|_\pd = 0\,,\qquad \pi^\perp(F_{\perp\pm})\big|_\pd =
\pi^\perp(\sigma)\big|_\pd= \pi^\perp(\lambda_+)\big|_\pd =0\,,
\ee
which is compatible with $H$ gauge symmetry at the boundary. In addition, there is a boundary flavor symmetry $N_G(H)/H$, where $N_G(H)$ is the normalizer of $H$ in $G$. Note that for two extreme choices of $H$ we have
\be \CN_{H=G} = \CN\,,\qquad \CN_{H=\{1\}} = \CD\,.\ee

In the presence of bulk chiral matter (and superpotentials, etc.), the basic $\CN$ and $\CD$ b.c. for the gauge multiplet may be combined with N, D, or D$_c$, or any of the more complicated choices of boundary conditions for matter fields discussed in Sections \ref{sec:chiral}--\ref{sec:mx}. We will see how this works in many examples.

\subsubsection{Dual-photon multiplet}
\label{sec:photon}

When the bulk gauge group is abelian (or has an abelian factor) one may dualize the gauge field to a circle-valued scalar, the ``dual photon'' $\gamma$, which obeys $d\gamma = *F$. Supersymmetrically, the complex combination $\sigma+i\gamma$ is the leading component of a 3d $\CN=2$ chiral multiplet \cite{AHISS}. Just as in Section \ref{sec:chiral}, this 3d chiral may be decomposed into
\begin{itemize}
\item a $\CN=(0,2)$ chiral multiplet $S^\vee$ containing $\sigma+i\gamma$ and $\lambda_+,\bar\lambda_+$, which may be identified as the two-dimensional T-dual of $S$ 
\item a $\CN=(0,2)$ Fermi multiplet containing $\lambda_-,\bar\lambda_-$ and the normal derivative $\pd_\perp S^\vee=\pd_\perp(\sigma+i\gamma)+...$, which may be identified with $\Upsilon$  (note that on-shell, $\Upsilon$ contains $F_{01}+iD\sim i\pd_\perp(\sigma+i\gamma)$)
\end{itemize}

\noindent Our basic $\CN$ and $\CD$ boundary conditions for abelian gauge theory now look just like basic D and N b.c. (respectively) for the 3d dual-photon multiplet! In particular, the $\CD$ b.c. \eqref{Dgauge} is compatible with setting $\Upsilon\big|_\pd=0$, while leaving the boundary value of $S^\vee$ unconstrained. Conversely, the $\CN$ b.c. \eqref{Ngauge} is compatible with setting $S^\vee\big|_\pd=0$. This may be deformed to
\be S^\vee\big|_\pd = t^\pd \label{St} \ee
by introducing a complex boundary FI parameter (including a real FI parameter and a theta-angle). Thus $\CN$ b.c. sets the dual photon $\gamma\big|_\pd$ equal to the boundary theta-angle. The deformation \eqref{St} is exactly of the same ``D$_c$'' type discussed in Section \ref{sec:chiral}, corresponding to the addition of the boundary superpotential $\int d\theta^+\, t^\pd \Upsilon|_\pd+c.c.$.

\subsubsection{Anomalies}
\label{sec:g-anom}

The gauginos $\lambda_+,\lambda_-$ in the 3d gauge multiplet necessarily have R-charge +1, as in \eqref{RSU}, and also transform in the adjoint representation of the gauge group $G$. These gauginos may therefore contribute to boundary anomalies.

A Neumann b.c. for the 3d gauge multiplet sets $\lambda_+|_\pd=0$ but does not constrain $\lambda_-$. Thus, our standard UV computation (Section \ref{sec:anom}) would suggest a contribution from $\lambda_-$ to the boundary anomaly
\be \label{gaugeN-anom} \text{$\CN$ b.c.}\,:\quad  h\, \text{Tr}(\mb f^2) + \tfrac12 \mb r^2 \ee
if $G$ is simple.
If $G$ is not simple, than a computation such as \eqref{Un-anom}, for the adjoint representation, gives the anomaly. If $G$ is abelian, then the anomaly is simply $\tfrac12\mb r^2$.

The prediction \eqref{gaugeN-anom} may again be compared with an IR calculation. We focus on the $G$ symmetry, and assume that $G$ is simple.
We introduce a supersymmetric Chern-Simons term at level $k$, which includes a real mass for the gauginos. Indeed, in the presence of the Chern-Simons term, the Dirac equations for gauginos take the form
\be \label{lambda-edge} (D_\perp+k)\lambda_+ = 0\,,\qquad (D_\perp-k)\lambda_-=0\ee
(assuming that $\lambda_\pm$ are constant in the $x^\mu$ directions parallel to the boundary),
from which we recognize $-k$ as the real mass. 
In the bulk, integrating out massive gauginos shifts the Chern-Simons coupling $k\to k-h\,\text{sign}(k)$. On the boundary, $\lambda_-$ has an edge mode when $k>0$, which contributes a standard 2d anomaly $2h\,\text{Tr}(\mb f^2)$. Thus, the total IR  anomaly for $G$ is
\be k-h\,\text{sign}(k) +  \left\{\begin{array}{ll} 2h\,\text{Tr}(\mb f^2) \;& k > 0 \\
  0& k < 0 \end{array}\right\}  = k+h\,\text{Tr}(\mb f^2) \,.\ee
This agrees with the UV prediction, which gives $k$ from the CS term, plus $h\,\text{Tr}(\mb f^2)$ from \eqref{gaugeN-anom}.

Note that, unlike 't Hooft anomalies for flavor symmetries, the anomaly for a $G$ gauge symmetry on a Neumann b.c. must somehow be cancelled! Either bulk our boundary matter may be added to effect the cancellation.

A Dirichlet b.c. for the 3d gauge multiplet behaves in the opposite way to Neumann. Since $\CD$ sets $\lambda_-|_\pd=0$ while leaving $\lambda_+$ unconstrained, it comes with an anomaly
\be \label{gaugeD-anom} \text{$\CD$ b.c.}\,:\quad -h\, \text{Tr}(\mb f^2) - \tfrac12 \mb r^2 \ee
for (say) a simple gauge group $G$. Again, this may be computed either in the UV or the IR. Note that \eqref{gaugeD-anom} contains the anomaly for the \emph{boundary flavor symmetry} $G$ on a Dirichlet b.c., rather than the bulk gauge symmetry. It is an ordinary 't Hooft anomaly, and does not need to be cancelled.

\subsubsection{Relations between gauge theory boundary conditions}

When describing boundary conditions for a free 3d chiral, we observed in Section \ref{sec:flip} that D (resp. N) can be ``flipped'' to N (resp. D) by coupling to boundary matter. In gauge theory, we can look for similar relations between $\CN$ and $\CD$. 

Modifying $\CD$ to $\CN$, possibly enriched by any boundary matter, is relatively easy. Let $G$ denote the bulk gauge group. Since $\CD$ has a global boundary $G$ symmetry, we can ``gauge'' the boundary symmetry by adding 2d $\CN=(0,2)$ gauge fields with gauge group $G$,
together with any other 2d $\CN=(0,2)$ degrees of freedom charged under the new $G$ gauge symmetry. 
(In doing so, we should carefully ensure that all $G$ anomalies cancel.) The resulting boundary condition should be equivalent to $\CN$,
as the 2d $G$ gauge fields are eaten up by the bulk gauge fields.

In the case of abelian $G=U(1)$ gauge theory, we saw in the previous section that the bulk gauge multiplet may be dualized to a chiral dual-photon multiplet, grouped into two $\CN=(0,2)$ superfields $\hat S$ and $\Upsilon$. The $\CD$ b.c. sets $\Upsilon|_\pd=0$, and the flip to $\CN$ introduces a boundary $U(1)$ gauge symmetry with field strength $\Upsilon^\pd$, coupled to $\hat S$ by a superpotential $\int d\theta^+ \hat S\big|_\pd \Upsilon^\pd$. Just as in Section \ref{sec:flip}, this coupling modifies the b.c. to
\be \hat S\big|_\pd= 0\,,\qquad \Upsilon\big|_\pd = \Upsilon^\pd\,, \ee
which is equivalent to $\CN$.

For general $G$, there is no systematic operation that maps a generic $\CN$ boundary condition to a $\CD$ boundary condition. 
In the case of abelian $G=U(1)$ gauge theory, it is possible to modify $\CN$ to $\CD$ by promoting the boundary FI parameter to a dynamical chiral superfield. 
Comparing to \eqref{St} we see that this is essentially a D\,$\rightarrow$\,N flip for the dual photon.

\subsection{BPS equations and singular boundary conditions}
\label{sec:flow}

So far, we have focused on UV boundary conditions defined by fixing the values or normal derivatives of various fields at the boundary. There is another possibility that we briefly discuss, though it will not play a major role in this paper: we may require the bulk fields to have a singular profile in the neighborhood of the boundary that is compatible with $\CN=(0,2)$ SUSY. 
A prototypical example of a singular half-BPS boundary condition (the Nahm pole b.c.) was introduced by \cite{GaiottoWitten-bc, GaiottoWitten-duality, DiacoNahm, CMTNahm} in the context of 4d $\CN=4$ super Yang-Mills theory, and we follow the basic logic of \cite{GaiottoWitten-bc, GaiottoWitten-duality} here.

One inspiration for considering singular boundary conditions arises from the IR behavior of D$_c$ b.c. Recall from Section \ref{sec:chiral} that D$_c$ boundary conditions for chiral multiplets may flow to useful superconformal boundary conditions in the presence of gauge fields. In fact, in non-supersymmetric theories, D$_c$-type b.c. may lead to conformal boundary conditions even in the absence of gauge fields! A notable example is a real 3d scalar with a $\phi^4$ potential, which flows to the 3d Ising CFT.
This theory admits a conformal boundary condition that flows from $\phi|_\pd = c$ in the UV, known as the ``exceptional transition'' boundary condition \cite{bcft}. In the IR, the operator $\phi$ simply diverges at the boundary, as the RG flow sends the scale $c$ to infinity. 

We thus find it natural to ask: in a given 3d $\CN=2$ gauge theory, can we define singular boundary conditions directly in the UV that flow to the same IR superconformal boundary condition as some D$_c$? The answer, in certain cases, seems to be affirmative.

In general, in a Lagrangian QFT, a disorder operator supported on a submanifold $\CS$ may be defined by finding a singular solution to the equations of motion in a neighborhood of $\CS$, and requiring fields to asymptotically approach this solution when performing the path integral. One may further require that the operator preserves certain symmetries (\emph{e.g.} flavor symmetry, supersymmetry, conformal symmetry...). These symmetries must then leave invariant the singular solution to the equations of motion.
In the case at hand, we want to define a singular boundary condition that preserves 2d $\CN=(0,2)$ supersymmetry, so we look for solutions to the equations of motion that are fixed by the $\CN=(0,2)$ subalgebra of 3d $\CN=2$; in other words, we look for solutions of the $\CN=(0,2)$ BPS equations.

In a 3d gauge theory with chiral matter, most of the BPS equations for scalar fields can be deduced by rewriting the theory in 2d $\CN=(0,2)$ superspace (as in Sections \ref{sec:chiral}, \ref{sec:2dmatter}, \ref{sec:bdy-gauge}), and requiring $E$, $J$, and $D$ terms to vanish. Let's describe these in turn.
\begin{itemize}
\item The vanishing of $E$ terms requires fields to sit at a critical point of the 3d superpotential, $dW=0$.
\item The vanishing of $J$ terms sets
\be (D_\perp-\sigma)\phi =  0  \label{JBPS} \ee
for the bosonic component $\phi$ of each 3d chiral, where $D_\perp-\sigma$ is the complexified covariant derivative perpendicular to the boundary. Here $\sigma\in \mathfrak g$ implicitly ``acts'' on $\phi$ in the appropriate representation; for example, if the gauge group is $G=U(1)$ and $\phi$ has charge $q$, the equation \eqref{JBPS} would read $(\pd_\perp-iqA_\perp-q\sigma)\phi=0$.
\item The vanishing of $D$ terms as in \eqref{D2d} schematically sets
\be D_\perp\sigma = \mu_{\rm matter}-t-k\sigma\,. \label{DBPS} \ee
\end{itemize}

We can now specialize to the definition of UV boundary conditions. In the UV, we expect CS terms to be subleading 
compared to YM gauge kinetic terms. Indeed, they will provide subleading corrections to the 
solutions we describe. 

It is natural to look for solutions where the fields have a singular power law behaviour near the boundary. In an axial gauge, 
(\ref{JBPS}) immediately tells us that 
\begin{equation}
\sigma \sim -\frac{s}{x^\perp} 
\end{equation}
with $s$ some constant element of the Lie algebra that determines the scaling behavior of all chiral fields 
near the boundary. On the other hand, (\ref{DBPS}) then requires the 
matter moment map $\mu_{\rm matter}$ to diverge as $\frac{s}{(x^\perp)^2}$ near the boundary. 

For simplicity, we can restrict ourselves to solutions where the chiral fields have the 
minimal divergence compatible with the behaviour of the moment map, i.e. 
\begin{equation}
\phi \sim \frac{u}{x^\perp} 
\end{equation}
Restoring dimensionful units, that would become $\frac{u}{e x^\perp}$, with $u$ dimensionless and $\phi$ and $e$ of dimension $\frac12$.
The differential equations thus collapse to ordinary equations:
\begin{equation} \label{nahmlike}
s \cdot u = u \qquad \qquad s = \mu_{\rm matter}(u)
\end{equation}

These solutions are a close analogue of the Nahm pole boundary conditions. Indeed, the Nahm pole boundary condition occurs 
precisely in this setting for 3d $\CN=4$ gauge theory \cite{ChungOkazaki}, where only the adjoint chiral in the $\CN=4$ gauge
multiplet acquires a singularity. Then $s$ and $u$ are generators of an $su(2)$ embedding in the gauge group. 
The resulting boundary conditions preserve $\CN=(0,4)$ SUSY in that case. 

Surprisingly, the equations \eqref{nahmlike} have solutions even for simple abelian gauge theories. For example, 
if $G = U(1)$ and $\phi$ consists of a single chiral of charge $1$, we can simply take $s=1$ and $u=1$. 
The resulting boundary condition preserves the same symmetries as a ($\CD$,D$_c$) boundary condition and 
presumably flows to the same boundary condition in the IR.

In general, these singular boundary conditions require the boundary gauge symmetry to be broken to gauge transformations that 
leave $s$ and $u$ fixed. The residual gauge group may then be broken further and/or coupled to extra boundary degrees of freedom. 

We leave a full discussion of such boundary conditions to future work.


\section{Building the half-index}
\label{sec:index}

In this section, we give the basic rules for computing the half-index of various boundary conditions. Formulae for the half-index with Neumann $(\CN)$ b.c. for gauge multiplets (and various b.c. for matter) were derived in \cite{GGP-walls, GGP-fivebranes, YoshidaSugiyama}. We review these results, and propose a generalization to Dirichlet $(\CD)$ b.c. for gauge multiplets that involves a sum over monopole sectors.

The half-index may be defined as the character of the vector space of \emph{local operators on the boundary}:  for a 3d  $\CN=2$ theory $\CT$ with a $\CN=(0,2)$ boundary condition $\CB$, the half-index is a trace%
\footnote{Often one finds the expression $\Tr (-1)^F q^{J+\frac R2}x^2$ instead of \eqref{defII}, where $F=2J$ is fermion number. As long as all R-charges can be chosen to have integral values, the two formulae are simply related by substituting $q^{\frac12}\to -q^{\frac12}$. All our examples will allow integral R-charges.}
\be \label{defII} \II_{\CT,\CB}(x;q) = \text{Tr}_{\text{Ops}_\CB}  (-1)^R q^{J+\frac R2} x^e\,, \ee
where $R$ is the R-charge operator, $J$ generates $Spin(2)\simeq U(1)_J$ rotations in the plane of the boundary, and $e$ is a flavor charge operator, measuring charges under a maximal torus of the flavor symmetry group on the boundary. The character \eqref{defII} counts operators in the cohomology of the supercharge $\ol Q^+$, which is part of the $\CN=(0,2)$ algebra preserved by a half-BPS boundary condition.
\footnote{As mentioned in the introduction, the cohomology of $\ol Q_+$ actually has the structure of a chiral algebra $\CA_\pd$, and from this perspective the half-index is the character of a vacuum module for $\CA_\pd$. The characters of other modules may be obtained by inserting line operators in the index.}

For a superconformal boundary condition, with the superconformal assignment of R-charges, the half-index should begin with `$1+...$' and thereafter only contain positive powers of $q$. 
More generally, the same properties should hold a 3d $\CN=2$ gauge theory and a 2d $\CN=(0,2)$ boundary condition $\CB$ that flows to a superconformal boundary condition in the infrared, assuming an appropriate assignment of R-charges. In the gauge theory there is typically some range of R-charge assignments that preserve the positivity of $J+\frac R2$. If one can choose integral R-charge assignments within this range (which will be true for the theories considered in this paper), then we expect
\be \II_{\CT,\CB} \in \Z[\![q^{\frac12}]\!][x,x^{-1}]\,,\ee
\emph{i.e.} the half-index is a formal Taylor series in $q$, whose coefficients are Laurent polynomials in $x$, themselves with integer coefficients.

An equivalent definition of the half-index is as a hemisphere `$HS^2\times S^1$' partition function, with the boundary condition $\CB$ placed at the equatorial boundary of the hemisphere $HS^2$ and wrapping $S^1$. This perspective is amenable to a localization computation \cite{YoshidaSugiyama}.
In this section, however, we take the perspective of counting boundary local operators seriously, and employ physical intuition to derive practical formulae for the half-index.

The general formula for the half-index of a Lagrangian 3d gauge theory, summarized in Section \ref{sec:II-sum}, takes the schematic form
\be \II_{\text{gauge multiplets}} \times \II_{\text{3d matter multiplets}} \times I_{\text{2d boundary index}}\,.
\ee
This is further projected to gauge-invariant operators (given a $\CN$ b.c. for gauge fields) or summed over monopole sectors (given a $\CD$ b.c. for gauge fields). 
We describe the various ingredients, from right to left.

\bigskip
\noindent\textbf{Notation:} In formulae for the half-index, we use the q-Pochhammer symbols
\be \begin{array}{ll} \displaystyle (x;q)_k := \prod_{n=0}^{k-1} (1-q^nx)\,,\qquad & (x;q)_\infty := \prod_{n\geq 0} (1-q^n x)\,,  \\[.5cm]
 \displaystyle  (q)_k :=(q;q)_k = \prod_{n=1}^k (1-q^n)\,,\qquad  & (q)_\infty := (q;q)_\infty = \prod_{n\geq 1}(1-q^n)\,. \end{array}
\ee
Also, for a general symmetry group $G$ we work with a vector of fugacities $x\in T_\C$ valued in the maximal torus of $G$. If $\mb R$ is a (unitary) representation of $G$, and $\lambda \in \text{wt}(\mb R) \simeq \text{Hom}(T_\C,\C^*)$ a weight of $\mb R$, then the fugacity corresponding to the $\lambda$ weight space is $\lambda(x)$, which we denote with the more common notation $x^\lambda$.

In this section, we will work in Euclidean signature and use complex coordinates $z,\bar z$ in the plane parallel to the boundary, so that the $(\pd_-,\pd_+)$ derivatives become $(\pd_z,\pd_{\bar z})$.

\subsection{2d theories}
\label{sec:2dindex}

A purely two-dimensional $\CN=(0,2)$ theory may be considered a boundary condition for a trivial three-dimensional bulk. In this case, the half-index is simply the elliptic genus of the 2d $\CN=(0,2)$ theory \cite{SW1,SW2,Witten-elliptic} --- or, more accurately, the ``flavored'' generalization of the elliptic genus discussed in \cite{GGP-walls, GG-surface, BEHT1, BEHT2}. We refer to these latter references for details, and simply summarize the main features here.

For a 2d $\CN=(0,2)$ gauge theory, the elliptic genus may be constructed as a UV index, by starting with a product of elliptic genera for matter multiplets, then projecting to gauge-invariants by doing an integration along the torus of the gauge group. Superpotentials (E and J terms) only affect the index insofar as they constrain flavor symmetries.

\subsubsection{Matter}
\label{sec:2dmatter-ind}

The index of a 2d chiral multiplet $\Phi=\phi+\theta^+\psi_++...$ of charge $+1$ under a $U(1)_x$ flavor symmetry (with fugacity `$x$') and R-charge 0 is
\be \label{IIC} \text{C}(x;q) = \prod_{n\geq 0} \frac{1}{(1-q^n x)(1-q^{n+1}x^{-1})} = \frac{1}{(x;q)_\infty(qx^{-1};q)_\infty}\,.  \ee 
On the RHS, we use q-Pochhammer notation $(x;q)_\infty:= \prod_{n\geq 0}(1-q^n x)$. This is the inverse of a Jacobi theta function, up to a $1/(q;q)_\infty$ prefactor. 
The operators in $\ol Q_+$-cohomology counted by this index are $\phi$ and its $\pd_z$ derivatives, $\pd_z\bar\phi$ and its further $\pd_z$ derivatives, and normal-ordered monomials in these basic operators. Indeed, the supersymmetry transformations take the form
\be \ol Q_+ \phi =0\,,\qquad \ol Q_+ \psi_+ \sim \pd_{\bar z}\bar\phi\,,\qquad \ol Q_+\bar\phi \sim \bar\psi_+\,,\qquad \ol Q_+\bar\psi_+ =0 \ee
The operators $\pd_z^n \phi$ ($n\geq 0$) are $\ol Q_+$-closed on the nose; while the operators $\pd_z^{n+1}\bar\phi$ ($n\geq 0$) obey $\ol Q_+\pd_z^{n+1}\bar\phi\sim \pd_z^{n+1}\bar\psi_+$ and are closed modulo the Dirac equation, which sets $\pd_z\bar\psi_+=0$. Since the flavor charges,  R-charges, and spins are
\be \begin{array}{c|cc} & \pd_z^n\phi \;&\; \pd_z^{n+1}\bar\phi \\\hline
 U(1)_x & 1 & -1 \\
 U(1)_R & 0 & 0 \\
 U(1)_J & n & n+1 \\
 \text{fugacity in index:} & q^n x & q^{n+1}x^{-1} \end{array}
\ee
and these operators are all bosonic, the index \eqref{IIC} results.

Similarly, the index of a 2d Fermi multiplet $\Gamma = \gamma_-+...$ counts the operators $\pd_z^n\gamma_-$ and $\pd_z^n\bar\gamma_-$ ($n\geq 0$) and normal-ordered polynomials thereof. If $\Gamma$ has $U(1)_x$ flavor charge $-1$ and R-charge $1$, then these operators have charges
\be \begin{array}{c|cc} & \pd_z^n\gamma_- \;&\; \pd_z^{n}\bar\gamma_- \\\hline
 U(1)_x & -1 & 1 \\
 U(1)_R & 1 & -1 \\
 U(1)_J & n+\frac12 & n+\frac12 \\
 \text{fugacity in index:} & -q^{n+1}x^{-1} & -q^n x \end{array}
\ee
and remembering that they are fermionic leads to the index
\be \label{IIF}   \text{F}(x;q) = (x;q)_\infty(qx^{-1};q)_\infty = \text{C}(x;q)^{-1}\,.\ee

The fact that $\text{F}(x;q)=\text{C}(x;q)^{-1}$ is not coincidental. It reflects the fact that a chiral multiplet and a Fermi multiplet of the charges given above can be coupled via a superpotential $\int d\theta^+ \Phi \Gamma$ that preserves flavor and R-symmetry. The superpotential makes all the fields massive, triggering a flow to a trivial theory in the infrared, whose index is trivial due to the identity $\text{F}(x;q)\text{C}(x;q) = 1$.

We also see that $\text{F}(x;q) = \text{F}(qx^{-1};q)$. This reflects the fact that a free Fermi multiplet may equivalently be encoded in a superfield $\Gamma=\gamma_-+...$ or $\tilde \Gamma = \bar\gamma_-+...$. In the presence of gauge fields and superpotentials, the fundamental ambiguity in how one treats the on-shell Fermi multiplet persists, though the relation between superfields $\Gamma$ and $\tilde \Gamma$ is somewhat more interesting: it can be interpreted as a fermionic remnant of 2d $\CN=(2,2)$ T-duality that we describe in Appendix \ref{app:T}.

\subsubsection{Gauge fields}
\label{sec:2dgauge}

In pure 2d $\CN=(0,2)$ gauge theory with compact gauge group $G$, the $G$ gauge multiplet takes the same form as \eqref{AV-}. The gauginos and their derivatives are in $\ol Q_+$-cohomology. In particular, $\ol Q_+\lambda_-=0$ and $\ol Q_+D_z\bar\lambda_- \sim D_zF_{z\bar z}=0$ by the equations of motion, so the operators contributing to the index are
\be \begin{array}{c|cc} & D_z^n\lambda_- \;&\; D_z^{n+1}\bar\lambda_- \\\hline
 G & \text{adj} & \text{adj} \\
 U(1)_R & 1 & -1 \\
 U(1)_J & n+\frac12 & n+\frac32 \\
 \text{fugacity in index:} & -q^{n+1}s^\alpha\; & -q^{n+1} s^{-\alpha} \end{array} \qquad (n\geq 0)
\ee
where $s\in T_\C$ is the fugacity for the $G$ symmetry, and $\alpha\in \text{wt(adj)}$ are the weights of the adjoint representation (including the zero weights), counted with multiplicity.
These operators provide a contribution $\prod_{\alpha\in\text{wt(adj)}} (qs^\alpha;q)_\infty(qs^{-\alpha};q)_\infty$ in the index, which can usefully be rewritten as
\be (q)_\infty^{2\,\text{rank}(G)} \prod_{\alpha\in\text{roots}(G)} (qs^\alpha;q)_\infty(qs^{-\alpha};q)_\infty = (q)_\infty^{2\,\text{rank}(G)} \prod_{\alpha\in\text{roots}(G)} \frac{\text{F}(s^\alpha;q)}{1-s^\alpha}\,. \ee
Projecting to $G$-invariants by integrating over the torus of $G$ with a Vandermonde determinant $\frac{1}{|\text{Weyl}(G)|} \prod_{\alpha\in\text{roots}(G)} (1-s^\alpha)$ then leads to an index
\be  \frac{(q)_\infty^{2\,\text{rank}(G)}}{|\text{Weyl}(G)|}  \oint \frac{ds}{2\pi i s} \prod_{\alpha\in \text{roots}(G)} \text{F}(s^\alpha;q)\,. \ee

For 2d gauge theory coupled to chiral or Fermi matter multiplets, the index is obtained by combining the gauge and matter contributions:
\be I_{2d}= \frac{(q)_\infty^{2\,\text{rank}(G)}}{|\text{Weyl}(G)|}  \oint \frac{ds}{2\pi i s} \prod_{\alpha\in \text{roots}(G)} \text{F}(s^\alpha;q) \times \text{(matter index}(s;q))\,. \label{IIG2d} \ee
For example, given a chiral multiplet (of R-charge zero) in representation $\mb R$ of $G$, and a Fermi multiplets (of R-charge 1) in representation $\mb R'$ of $G$, the matter index takes the form $\prod_{\lambda\in \text{wt}(\mb R)} \text{C}(s^\lambda;q)\prod_{\mu\in \text{wt}(\mb R')} \text{F}(s^\mu;q)$. The formula \eqref{IIG2d} combines operators from the chiral and Fermi multiplets with gauginos, and projects to $G$-invariants.

\subsection{3d chirals}
\label{sec:chiralindex}

We perform an analysis analogous to that of Section \ref{sec:2dmatter-ind} to find the half-index for 3d chiral multiplets. 

We begin with a single free 3d $\CN=2$ chiral multiplet $\Phi_{\rm 3d}$, with charge $+1$ under a $U(1)_x$ flavor symmetry (with fugacity $x$) and charge zero for $U(1)_R$. Recall from Section \ref{sec:chiral} that $\Phi_{\rm 2d}$ decomposes into a $\CN=(0,2)$ chiral multiplet $\Phi=\phi+\theta^+\psi_++...$ and a Fermi multiplet $\Psi=\bar\psi_-+...$. In the 3d bulk, the operators $\pd_z^n\phi$ and $\pd_z^n\bar\psi_-$ are in $\ol Q_+$-cohomology. However, in contrast to Section \ref{sec:2dmatter-ind}, the operators $\pd_z\bar\phi$ and $\psi_-$ (and their further derivatives) are no longer $\ol Q_+$-closed:  we have $\ol Q_+(\pd_z\bar\phi)\sim \pd_z\bar\psi_+$ and $\ol Q_+(\psi_-) \sim \pd_\perp\phi$, neither of which are set to zero by the 3d equations of motion. Thus, the bulk operators potentially contributing to an index are
\be \qquad  \begin{array}{c|cc} & \pd_z^n\phi \;&\; \pd_z^n \bar\psi_- \\\hline
U(1)_x & 1 & -1 \\
U(1)_R & 0 & 1 \\
U(1)_J & n & n+\frac12 \\
\text{fugacity in index}:\; & q^nx & -q^{n+1}x^{-1} \end{array}\qquad (n\geq 0)\,. \ee

On a Neumann (N) boundary condition, all the $\pd_z^n\bar\psi_-$ operators are killed. The bosonic $\pd_z^n\phi$ operators survive, and lead to the half-index
\be \label{IIN} \II_{\rm N}(x;q) = \prod_{n\geq 0} \frac{1}{1-q^n x} = \frac{1}{(x;q)_\infty}\,.  \ee
On a Dirichlet (D) b.c., the $\pd_z^n\phi$ operators are killed while the fermionic $\pd_z^n\bar\psi_-$ operators survive, giving a half-index
\be \label{IID} \II_{\rm D}(x;q) = \prod_{n\geq 0} (1-q^{n+1}x^{-1}) = (qx^{-1};q)_\infty\,.\ee

In a similar way, a 3d chiral multiplet of R-charge $\rho$ has half-indices $\II_{\rm N}((-q^{\frac12})^\rho x;q)$ and $\II_{\rm D}((-q^{\frac12})^\rho x;q)$, obtained by a standard shift in fugacities $x\to -q^{\frac12}x$.

\subsubsection{Flips}

We can use these basic half-indices to test the operations that ``flip'' between N and D b.c., as in \ref{sec:flip}. We expect that a 3d chiral with N b.c. coupled to a 2d boundary Fermi multiplet of opposite flavor charge is equivalent to D b.c.; and conversely that a 3d chiral with D b.c. coupled to a boundary chiral multiplet of the same flavor charge is equivalent to N b.c. These operations are reflected in the obvious identities
\be \II_{\rm N}(x;q) \text{F}(x;q) = \II_{\rm D}(x;q)\,,\qquad \II_{\rm D}(x;q) \text{C}(x;q)=\II_{\rm N}(x;q)\,.\ee

\subsubsection{D$_c$ and singular b.c.}
\label{sec:singind}

A D$_c$ boundary condition, where $\Phi\big|_\pd = c$ is set to a nonzero value at the boundary, is very similar to D b.c. The same $\pd_z^n\bar\psi_-$ operators survive. However, $U(1)_x$ flavor symmetry is broken, and $U(1)_R$ symmetry is preserved precisely if $\Phi$ has R-charge zero. Correspondingly, the half-index is
\be \II_{\text{D}_c}(q) = \II_{\rm D}(x;q)\big|_{x=1} = (q)_\infty\,.\ee

This analysis indicates how we should treat D$_c$  boundary conditions in more interesting gauge theories with chiral matter, superpotentials, etc.: the index may first be written down for D b.c., and then ``deformed'' to D$_c$ by specializing the flavor fugacities of the relevant chiral multiplets according to the broken flavor symmetry, \emph{i.e.}  $x\to 1$. The specialization is done in the entire index.

\subsection{3d gauge symmetry: $\CN$ b.c.}
\label{sec:II-N}

The 3d gauge multiplet decomposes into a 2d gauge multiplet with fermionic field strength $\Upsilon$ and a chiral multiplet $S$, as discussed in Section \ref{sec:gauge2d}. 
Given a Neumann ($\CN$) boundary condition, only the operators formed out of the fields in $\Upsilon$ survive at the boundary. Ignoring gauge invariance for the moment, the operators in $\ol Q_+$-cohomology are as follows.

In a pure gauge theory, the elementary bulk operators in $\ol Q_+$-cohomology are the gaugino $\lambda_-$ and its $D_z$-derivatives. In contrast to a purely 2d gauge theory, $D_z\ol\lambda_-$ is no longer closed, since $\ol Q_+D_z\ol \lambda_-\sim D_zF_{z\bar z}$ is no longer set to zero by the 3d equations of motion. Thus, the operators contributing to the half-index for gauge group $G$ are
\be \qquad  \begin{array}{c|c} & D_z^n\lambda_- \\\hline
G & \text{adj}  \\
U(1)_R & 1 \\
U(1)_J & n+\tfrac12 \\
\text{fugacity in index:}\, & -q^{n+1}s^\alpha  \end{array}\qquad (n\geq 0) \,. \ee
The full contribution to the index from polynomials in the $D_z^n\lambda_-$ is
\be    \prod_{\alpha\in \text{wt(adj)}} (qs^\alpha;q)_\infty = (q)_\infty^{\text{rank}(G)} \prod_{\alpha\in\text{roots}(G)} (qs^\alpha;q)_\infty\,. \ee
As in 2d, the fugacity $s$ is valued in the complexified torus $T_\C$ of $G$.
Projecting to gauge-invariants with a contour integral and a Vandermonde determinant leads to%
\begin{align} \label{II-gN}
(q)_\infty^{\text{rank}(G)} \oint \frac{ds}{2\pi is} \prod_{\alpha\in\text{roots}(G)} (1-s^\alpha)(qs^\alpha;q)_\infty 
 &= (q)_\infty^{\text{rank}(G)} \oint \frac{ds}{2\pi is} \prod_{\alpha\in\text{roots}(G)} (s^\alpha;q)_\infty  
 \end{align}

The half-index for a $\CN$ b.c. is not \emph{directly} sensitive to a bulk Chern-Simons level.
However, since gauge symmetry is preserved by a $\CN$ b.c., the boundary gauge anomaly must be cancelled. The computations of anomalies in Section \ref{sec:g-anom} show that unless additional matter is present, a $\CN$ b.c. for pure gauge theory only makes sense at Chern-Simons level $k=-h$. Thus it is only for $k=-h$ that \eqref{II-gN} computes an actual half-index.

In the presence of bulk or boundary matter, other Chern-Simons levels are possible. Matter is incorporated into the half-index in a standard way:
\begin{itemize}
\item The 2d index of boundary matter (or a boundary gauge theory, or a boundary CFT) should be inserted directly into the integrand of \eqref{II-gN}. The 2d theory will typically have a $G$ flavor symmetry that is gauged in coupling to the bulk, thus the 2d index will depend on the fugacity $s$.
\item For 3d chiral matter with N b.c., D b.c., D$_c$ b.c., or some combination thereof, the half-index is computed as if $G$ were a flavor symmetry, and then inserted into the integrand of \eqref{II-gN}. (Note that D$_c$ b.c. are not possible for 3d chirals charged under $G$, if gauge symmetry is to be preserved at the boundary.)
\end{itemize}
For example, a $G$ gauge theory with a chiral multiplet in representation $(\mb R,\rho)$ of $G\times U(1)_R$ has a half-index
\begin{align}
(\CN,\text{N})\,:\quad &  (q)_\infty^{\text{rank}(G)} \oint \frac{ds}{2\pi is} \prod_{\alpha\in\text{roots}(G)} (s^\alpha;q)_\infty\prod_{\lambda\in\text{wt}(\mb R)} \II_{\rm N}((-q^{\frac12})^\rho s^\lambda;q) \\
(\CN,\text{D})\,:\quad &  (q)_\infty^{\text{rank}(G)} \oint \frac{ds}{2\pi is} \prod_{\alpha\in\text{roots}(G)} (s^\alpha;q)_\infty\prod_{\lambda\in\text{wt}(\mb R)} \II_{\rm D}((-q^{\frac12})^\rho s^\lambda;q)
\end{align}
for N and D b.c. on the chiral, respectively. Additional fugacities may be added for flavor symmetries.
Again, cancellation of the gauge anomaly constrains the bulk CS level. Here $(\CN,\text{N})$ b.c. requires a CS level $k=-h+\frac12 T_{\mb R}$, while $(\CN,\text{D})$ requires $k=-h-\frac12 T_{\mb R}$. 

As observed in many analogous index computations, and in particular in $(0,2)$ elliptic genus calculations \cite{BEHT1,BEHT2,GGP-triality}, 
doing the integration carefully may require careful contours prescriptions. As long as the 2d index of boundary matter does not contribute poles to the integrand for 3d gauge fields, 
our naive operator-counting approach works in a straightforward way. This will be the case in many of our examples, where the boundary matter is a collection of 
2d Fermi multiplets. 

If the 2d index of boundary matter does contribute poles, then our naive approach would need to be re-considered. There is a simple strategy 
which should work: replace $\CN$ b.c. with a combination of $\CD$ b.c. and 2d gauge fields and then apply the 
known contour prescriptions for contour integrals associated to 2d gauge fields. 

\subsection{3d gauge symmetry: $\CD$ b.c. and boundary monopoles}
\label{sec:II-D}

Now consider pure 3d gauge theory in the presence of a Dirichlet $(\CD)$ boundary condition.

The $\lambda_-$ gaugino is killed, but there are other operators in $\ol Q_+$-cohomology formed from the leading component $\sigma+iA_\perp$ of $S$ and its $D_z$ derivatives. Recall that $\CD$ b.c. breaks $G$ gauge symmetry to a global $G_\pd$ symmetry at the boundary, allowing only gauge transformations that are constant along the boundary. 
Nevertheless, one may still perform ``residual'' gauge transformations that depend on $x^\perp$, given by elements $g(x^\perp)\in G$ that restrict to the identity at the boundary. The boundary operator $\sigma+iA_\perp$ is not quite invariant under these residual gauge transformations, since $(\sigma+iA_\perp)\big|_\pd \to (\sigma+iA_\perp)\big|_\pd+i g^{-1}\pd_\perp g\big|_\pd$. However, the ``field strength'' $[D_z,\pd_\perp+\sigma+iA_\perp]=D_z\sigma+iF_{z\perp}$ and its further $D_z$ derivatives \emph{are} invariant, and generate the $\ol Q_+$-cohomology. They have charges
\be \label{CDops}  \qquad  \begin{array}{c|c} & D_z^{n+1}\sigma+iD_z^nF_{z\perp} \\\hline
G_\pd & \text{adj}  \\
U(1)_R & 0 \\
U(1)_J & n+1 \\
\text{fugacity in index:}\, & q^{n+1}s^\alpha  \end{array}\qquad (n\geq 0) \,, \ee
leading to a half-index
\be \label{CD-pert} \prod_{\alpha\in\text{wt(adj)}} \frac{1}{(q s^\alpha;q)_\infty} = \frac{1}{(q)_\infty^{\text{rank}(G)}} \prod_{\alpha\in \text{roots}(G)} \frac{1}{(q s^\alpha;q)_\infty}\,.\ee
Here $s$ is a fugacity for the boundary $G_\pd$ symmetry, and there is no need for any projection to $G$-invariants.

Note that $F_{z\perp}$ is the chiral current for the $G_\pd$ flavor symmetry on the boundary. Thus, the operators in \eqref{CDops} may be interpreted as positive modes of a complexified version of this current. Alternatively, in an abelian gauge theory, the operators in \eqref{CDops} are just modes of the dual photon.

In a 3d gauge theory with $\CD$ b.c., coupled to 3d and/or 2d matter, the index \eqref{CD-pert} is simply multipled by the matter index. As usual, the matter index may depend on the fugacity $s$. For example, a 3d gauge theory with a chiral multiplet of R-charge zero in representation $\mb R$ of $G$, with $\CD$ b.c. for the gauge multiplet and (say) D b.c. for the chiral, has boundary operators counted by the index
\be \label{CD-pertD} \frac{1}{(q)_\infty^{\text{rank}(G)}} \prod_{\alpha\in \text{roots}(G)} \frac{1}{(q s^\alpha;q)_\infty} \prod_{\lambda\in \text{wt}(\mb R)} \II_{\rm D}(s^\lambda;q)\,. \ee
So far, the index is completely insensitive to a bulk Chern-Simons coupling.
There is no need to cancel the 't Hooft anomaly for $G_\pd$.

\subsubsection{Boundary monopole operators}

In truth, \eqref{CD-pert} and \eqref{CD-pertD} must be supplemented by \emph{nonperturbative} contributions from monopole operators. 
The $\CD$  b.c. for gauge fields are just right to support boundary monopoles with a conserved topological charge, which enhance the space of local operators. The basic analysis of these operators appeared in \cite{BDGH}, and we review the main ideas here.

It is well known (\emph{cf.} \cite{AHISS, IS-new}) that one can define a BPS monopole operator (that is, in particular, $\ol Q_+$-closed) in the bulk of a 3d $\CN=2$ theory as a disorder operator, by specifying a singular solution to the BPS equations
\be \label{mon-eq}  F = *D\sigma\,,\qquad D*\sigma=0\,.\ee
For gauge group $G=U(1)$, the basic solutions are Dirac monopoles: they have $\sigma=\frac{m}{2r}$, where $r$ is the radial distance from the singularity and $m$ is a constant. The condition that the flux through a 2-sphere surrounding the singularity be quantized, $\oint_{S^2} F \in 2\pi \Z$, constrains $m\in \Z$\,. For general gauge group $G$, the singular solutions to \eqref{mon-eq} come from embeddings of the basic abelian monopole $\sigma =\frac{1}{2r}$ into $G$, and are thus labelled by cocharacters
\be m \in \text{Hom}(U(1),T)\,, \label{cochar} \ee
where $T\subset G$ is the maximal torus of $G$. (More accurately, the embeddings are labelled by cocharacters modulo the action of the Weyl group.)

In the present case, we want to consider a monopole operator on the boundary, defined (say) by a singular solution to \eqref{mon-eq} on a half-space. It is easy to see that in abelian $G=U(1)$ gauge theory the basic solution for the scalar field
\be \label{mon-bdy}  \sigma = \frac{1}{\sqrt{(x^\perp)^2+|z|^2}}\,,\quad F=*D\sigma\qquad \text{on $x^\perp\leq 0$} \ee
is compatible with supersymmetric $\CD$ b.c. \eqref{Dgauge}-\eqref{Ns}, which do not constrain $\sigma\big|_\pd$ of $F_{z\perp}\big|_\pd$, but set $F_{z\bar z}\big|_\pd=0$. One can also find a gauge transformation \emph{near} the boundary so that the connection $A$ corresponding to \eqref{mon-bdy} satisfies $A\big|_\pd=0$ away from $z=\bar z=0$.

More generally, we can use a cocharacter $m$ as in \eqref{cochar} to embed the basic abelian solution \eqref{mon-bdy} into any gauge group $G$, thus defining a boundary monopole operator of ``charge'' $m$.

Geometrically, one may think of boundary monopole operators as follows. Take $G=U(1)$. Consider surrounding an operator $\CO$ at $x^\perp=z=\bar z=0$ with a hemisphere $HS^2$. The $\CD$~b.c. \emph{trivializes the principal $G$ bundle at the boundary} and sets $A_z\big|_\pd=A_{\bar z}\big|_\pd=0$ there. In particular, $\CD$ b.c. trivializes the $G$-bundle on $\pd(HS^2)=S^1$. In the interior of $HS^2$, the bundle may be topologically nontrivial, and its topological type is precisely measured by the curvature integral
\be \int_{HS^2} \frac{F}{2\pi}  = m \in \Z\,.\ee
This is the monopole charge of $\CO$.
The curvature integral gives a well-defined integer precisely because of the trivialization at the boundary, and the boundary condition $A_z\big|_\pd=A_{\bar z}\big|_\pd=0$.%
\footnote{In general, given a $U(1)$ bundle with connection on $HS^2$ that is trivialized at the boundary, Stokes' theorem reads $\int_{HS^2}F = 2\pi m + \oint_{S^1} A$.}

In the case of the full 3d index (or $S^2\times S^1$ partition function), it is well known how bulk monopole operators contribute. A localization computation \cite{Kim-index, IY-index} expresses the full 3d index as a sum over flux sectors on $S^2$, \emph{i.e.} over cocharacters of $G$. Notably, this is a sum over \emph{abelian} flux sectors, \emph{i.e.} over all cocharacters rather than their Weyl-orbits. The localization computation for the half-index with $\CD$ b.c. proceeds a similar way, summing over abelian fluxes on the hemisphere.

We are led to a complete, nonperturbative formula for the half-index of the form
\be \label{CD-full} \frac{1}{(q)_\infty^{\text{rank}(G)}} \sum_{m\in \text{cochar}(G)} \bigg[ \prod_{\alpha\in \text{roots}(G)}  \frac{1}{(q^{1+m\cdot \alpha}s^\alpha;q)_\infty}\bigg] q^{\frac12k_{\rm eff}\text{Tr}(m^2)} s^{k_{\rm eff}m} \times \text{[matter index]}(q^m s)\,. \ee
Here all the fugacities $s$ for the boundary $G_\pd$ symmetry have been shifted $s\to q^ms$, reflecting the fact that electrically charged states acquire spin in the presence of magnetic flux.
In terms of operators, one would expect that a monopole of charge $m$ dressed by operators of electric charge $\lambda\in \text{wt}(G)$ acquires spin $m\cdot \lambda$.
In addition, in the presence of a Chern-Simons coupling at level $k$, even a bare monopole operator is induced to have nontrivial electric charge $km$ and spin $\frac{k}{2}\text{Tr}(m^2)$.%
\footnote{Just as in Section~\ref{sec:anom-na}, `$\text{Tr}(m^2)$'
denotes the Cartan-Killing form, the same one appearing in the Chern-Simons action $k\,\text{Tr}[AdA+...]$,
normalized to be the usual trace in the fundamental representation of $\mathfrak{u}(N)$. Similarly, the Cartan-Killing form is implicitly being used to transform a magnetic charge $m\in \text{cochar}(G)$ to an electric charge $km\in \text{wt}(G)$, so that $s^{km}$ makes sense.} %
This leads to the extra weight $q^{\frac12k\,\text{Tr}(m^2)} s^{km}$ in the monopole sum.

We \emph{propose} that for boundary monopole operators the correct Chern-Simons level to include in the index \eqref{CD-full} is not the bare UV Chern-Simons level $k$ from the bulk, but rather an effective
\be \label{keff} k_{\rm eff} := \; \begin{array}{c}\text{CS level that captures the total boundary $G_\pd$ anomaly,} \\
 \text{including shifts from gauginos and bulk or boundary matter} \end{array} \ee
In other words, $k_{\rm eff}$ is such that the boundary anomaly $\CI_2$ is the exterior derivative of a Chern-Simons form at level $k_{\rm eff}$.
A strong argument for using this effective level comes from observing that bulk Chern-Simons terms in the presence of $\CD$ b.c. are equivalent in the IR to boundary chiral matter, and the effective level $k_{\rm eff}$ (or more accurately, the boundary anomaly) is the only quantity that consistently captures the effect of both. We will thoroughly test the proposed use of $k_{\rm eff}$ in many examples. It would be satisfying to reproduce this proposal using localization, as in the case of Neumann boundary conditions \cite{YoshidaSugiyama}.

As a simple test, consider an abelian $G=U(1)$ bulk gauge theory at level $k-\frac12$ with a 3d chiral of $U(1)_\pd$ charge $+1$ and R-charge $+1$ (for convenience). By our proposal, the half-index of $\CD$ b.c. on the gauge multiplet and N b.c. on the chiral is computed by
\be \II_{\CD,\text{N}}(s;q)= \frac{1}{(q)_\infty} \sum_{m\in \Z}  q^{\frac12 (k-1) m^2}s^{(k-1)m} \frac{1}{(-q^{\frac12+m}s;q)_\infty}\,, \ee
which reflects the boundary anomaly $\CI_2=(k-\tfrac12)\mb f^2-\tfrac12\mb f^2 = (k-1)\mb f^2$ in the presence of N b.c., so $k_{\rm eff}=k-1$. Similarly, the half-index with D b.c. on the chiral is computed by
\be \II_{\CD,\text{D}}(s;q)= \frac{1}{(q)_\infty} \sum_{m\in \Z}  q^{\frac12 k m^2}s^{km} {(-q^{\frac12-m}s^{-1};q)_\infty}\,, \ee
where now $\CI_2=(k-\tfrac12)\mb f^2+\tfrac12\mb f^2 = k\mb f^2$, so $k_{\rm eff}=k$. These two boundary conditions should be related by a flip, \emph{e.g.} ($\CD$,N) should be equivalent to ($\CD$,D) coupled to a boundary Fermi multiplet $\Gamma$ of $U(1)_\pd$ charge $-1$ and R-charge zero. Accordingly, we find
\be  \II_{\CD,\text{D}}(s;q) = \text{F}(-q^{\frac12}s^{-1};q) \II_{\CD,\text{N}}(s;q) \ee
due to the theta-function identity
\be (-q^{\frac12-m}s^{-1};q)_\infty = \frac{\text{F}(-q^{\frac12-m}s^{-1};q)}{(-q^{\frac12+m}s;q)_\infty} = q^{-\frac{m^2}{2}}s^{-m} \frac{\text{F}(-q^{\frac12}s^{-1};q)}{(-q^{\frac12+m}s;q)_\infty}\,\ee
that can be applied to the summand of $\II_{\CD,\text{D}}(s;q)$.

Formula \eqref{CD-full} has a natural generalization to a product of gauge groups, or gauge and flavor groups.
Let us denote
\be \begin{array}{ll}
G & \text{entire gauge group (possible a product of groups)} \\
\CF & \text{entire global symmetry group, including flavor \emph{and} R-symmetry} \\
\tilde G = G\times \CF\;& \text{full symmetry group} \\
\mathfrak g,\mathfrak f,\tilde{\mathfrak g} & \text{the gauge, global, and gauge$\times$global Lie algebras} \\
s,x,-q^{\frac12} & \text{gauge, flavor, and R-symmetry fugacities} \\
\tilde s=(s,x,-q^{\frac12}) & \text{joint fugacity for $\tilde G$} 
\end{array} \ee
Then, given $\CD$ b.c. for all of $G$, one sums over cocharacters $m\in  \text{cochar}(G)\subset \mathfrak g$, which by the embedding $G\subset \tilde G$ may also be thought of as cocharacters $m\in  \text{cochar}(\tilde G)\subset \tilde{\mathfrak g}$. In the monopole sum, all fugacities are shifted $\tilde s \to q^m\tilde s$ (meaning explicitly $(s,x,-q^{\frac12})\to (q^ms,x,-q^{\frac12})$), and there is an extra factor
\be \label{genk} q^{\frac12 \mb{k}_{\rm eff}[m,m]} \tilde s^{\mb{k}_{\rm eff}[m,-]}\, \ee
where
\be \mb k_{\rm eff}: \tilde{\mathfrak g}\times \tilde{\mathfrak g} \to \R \ee
is the bilinear form defined by the full boundary 't Hooft anomaly polynomial, and $s^{\mb{k}_{\rm eff}[m,-]}$ means $\text{exp}\big( \mb{k}_{\rm eff}[m,\log\tilde s]\big)$.

For example, in pure 3d $U(1)$ gauge theory at level $k$, with a $\CD$ b.c., the boundary anomaly polynomial is $k\mb f^2 + 2 \mb f \mb f_x$, where $\mb f_x$ is a field strength for the topological $U(1)_x$ flavor symmetry (the R-symmetry does not enter here). The corresponding bilinear form is $\bsp k&1\\1&0\esp$. Letting $s,x$ denote the fugacities for the boundary $U(1)_\pd$ symmetry and for $U(1)_x$, the half-index becomes
\be \frac{1}{(q)_\infty}\sum_{m\in \Z} q^{\frac12\bsp m & 0\esp\bsp k&1\\1&0\esp\bsp m\\ 0\esp}e^{\bsp m & 0\esp\bsp k&1\\1&0\esp\bsp \log s\\ \log x\esp} = \frac{1}{(q)_\infty}\sum_{m\in \Z} q^{\frac12 km^2}s^{km}x^m\,.\ee

\subsubsection{Comparison between $\CD$ and $\CN$}
We now have prescriptions for $\CD$ and $\CN$ boundary conditions.
when the $\CD$ and $\CN$ boundary conditions are both well-defined 
(with the same boundary conditions on the rest of the other super-multiplets), 
it should be also possible to obtain the half-index of $\CN$ boundary conditions 
by 2d gauging the half-index of the corresponding $\CD$ boundary conditions.

If the boundary matter does not contribute poles to the index, so that 
we do have a reliable prescription for both sides, it should be possible to match the answers of our two prescriptions.

We will not do so in detail, but the match is intuitively clear: 
in the correct circumstances the $\CN$ contour integral can be executed on $\mathbb{C}^*$, picking a semi-infinite sequence of poles.
On the other hand, the 2d contour integral will only pick the poles in (\ref{CD-full}) which lie in the fundamental region $|q|<|s|<1$.
These come again in semi-infinite sequences labeled by the magnetic charge $m$, as the $q^m$ shifts push the poles 
out of the fundamental region when $m$ is sufficiently large with the appropriate sign. 

On the other hand, as we mentioned before, when the boundary matter does contribute poles to the index, 
then we expect the 2d prescription applied to (\ref{CD-full}) to produce the correct, unknown prescription for  
the half-index of $\CN$ boundary conditions.

\subsection{Summary}
\label{sec:II-sum}

Altogether, the computation of a half-index involves of a 3d $\CN=2$ gauge theory $\CT$ with a 2d $\CN=(0,2)$ boundary condition $\CB$ involves:
\begin{enumerate}
\item Computing a 2d index $I_{2d}$, as in Section \ref{sec:2dindex}, for any 2d $\CN=(0,2)$ theory that is coupled to the bulk in defining $\CB$.
\item Multiplying by $\II_{\rm N}$ or $\II_{\rm D}$ half-indices for all 3d chiral multiplets as in Section \ref{sec:chiralindex}, depending on whether these multiplets are given N or D b.c., prior to coupling to any 2d matter.
\item Integrating over fugacities (\emph{i.e.} projecting to invariants) for the part of the bulk gauge group given $\CN$ b.c., using the measure \eqref{II-gN} from Section \ref{sec:II-N}.
\item Summing over monopole sectors for the part of the bulk gauge group given $\CD$ b.c., with the ``measure'' \eqref{CD-full} from Section \ref{sec:II-D}, along along with shifts of fugacities and the effective Chern-Simons contribution from \eqref{CD-full}.
\item For any boundary flavor symmetries broken by D$_c$ b.c. or singularities such as Nahm poles, setting the corresponding fugacities to $1$, as in Section \ref{sec:singind}.
\end{enumerate}

\subsection{Line operators}
\label{sec:line}

\begin{figure}[htb]
\centering
\includegraphics[width=1.6in]{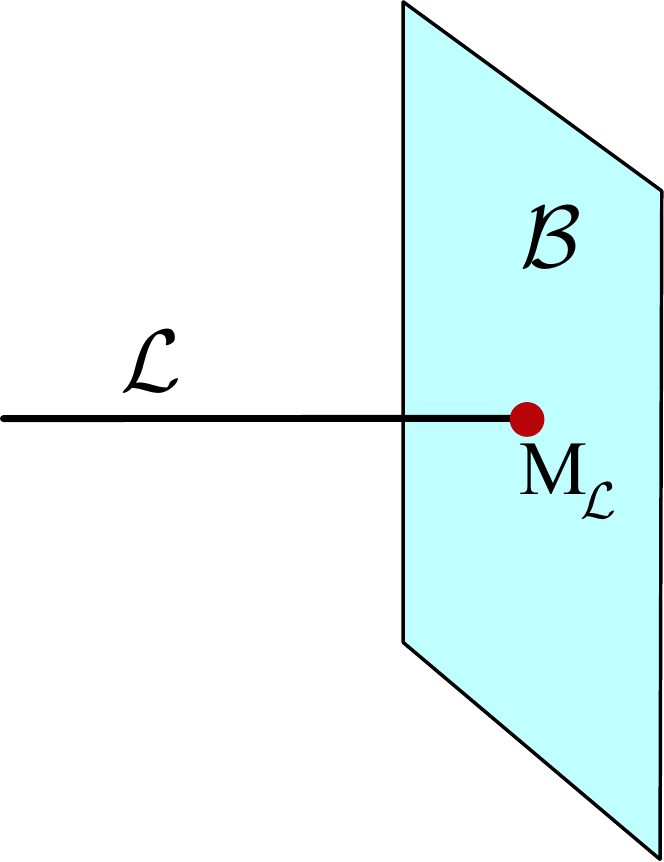}
\caption{Boundary local operators at the end of a line operator $\CL$ form a module $M_\CL$ for the boundary chiral algebra.}
\label{fig:line}
\end{figure}

A useful modification of the half-index comes from including a half-BPS line operator that preserves the $\ol Q_+$ supercharge. Here we envision a line operator $\CL$ supported on a ray perpendicular to the boundary, as in Figure \ref{fig:line}, hitting the boundary at (say) the origin $z=\bar z=0$. Such a line operator can preserve a full 1d $\CN=2$ subalgebra of 3d $\CN=2$ generated by $\ol Q_+$ and $Q_-$. The half-index can be defined to count local operators at the intersection of $\CL$ and the boundary.%
\footnote{These form a module $M_\CL$ for the usual chiral algebra of boundary local operators (since there is an OPE between generic boundary local operators and boundary local operators stuck to the endpoint of $\CL$). Thus, the half-index in the presence of $\CL$ can be interpreted as a character of the module $M_\CL$. This is analogous to the behavior of the 4d $\CN=2$ index in the presence of a surface operator \cite{GRR, GCS-surface}.}

Alternatively, under a state-operator correspondence, the half-index with a line operator $\CL$ can be interpreted as a $D^2\times S^1$ partition function with $\CL$ inserted along $\{0\}\times S^1$.

There are two simple types of line operators that we will consider here. The first is a supersymmetric Wilson line $\CW_{\mb R}$ in representation $\mb R$ of the bulk gauge group, defined as
\be \CW_{\mb R} = \CP\text{exp} \,\,i \!\!\int_{x^\perp\leq 0\atop z=\bar z=0} (A_\perp-i\sigma)dx^\perp\,.\ee
If the gauge multiplet is given Neumann $(\CN)$ b.c., the boundary operators at the end of $\CW_{\mb R}$ are no longer gauge-invariant, but rather must be in representation $\ol{\mb R}$. The half-index is accordingly modified by inserting a character
\be \CN\,:\quad  \oint \frac{ds}{2\pi is}\II_{\rm matter}(s)    \quad\overset{\CW_{\mb R}}\leadsto\quad \oint \frac{ds}{2\pi is} \text{Tr}_{\mb R}(s) \II_{\rm matter}(s)\,, \label{WN} \ee
which precisely projects to operators in representation $\ol{\mb R}$. Alternatively, if the bulk gauge multiplet is given Dirichlet $(\CD)$ b.c., the boundary operators at the end of $\CW_{\mb R}$ are simply tensored with the representation $\mb R$ of the $G_\partial$ boundary flavor symmetry. In the presence of $\CW_{\mb R}$ \emph{and} monopole flux, this can modify the spin of boundary operators by the mechanism of Section \ref{sec:II-D}; accordingly, the half-index takes the form
\be \CD\,:\quad  \sum_{m\in \text{cochar}(G)} \II_{\rm matter, CS}  \quad\overset{\CW_{\mb R}}\leadsto\quad \sum_{m\in \text{cochar}(G)} \text{Tr}_{\mb R}(q^ms)  \II_{\rm matter, CS}\,. \label{WD}\ee
More trivially, we can insert a bulk Wilson line for a flavor symmetry rather than a gauge symmetry. This again just tensors the boundary operators by representation $\mb R$ for the flavor symmetry, and multiplies the entire half-index by $\text{Tr}_{\mb R}(x)$, where $x$ is the flavor fugacity.

A second simple type of line operator is a vortex line $\CV_n$ for an \emph{abelian} gauge or flavor symmetry, which can be understood as an insertion of $n$ units of flux
\be F_{z\bar z} \sim 2\pi n\, \delta^{(2)}(z,\bar z) \ee
through the $z,\bar z$ plane.  More precisely, $\CV_n$ is a disorder operator that requires the connection to attain the singular profile $A\sim n \,d\theta$ near $z=\bar z=0$. Working in holomorphic gauge $A_{\bar z}=0$, this profile looks like
\be A_{z} \sim \frac{n}{z}\,,\qquad A_{\bar z}=0\,, \ee
and can reached by applying a singular complex gauge transformation $g(z)=z^{n}$ to a smooth configuration.
This forces charged matter fields to have a zero or pole of order $\sim n$ at the location of the vortex.
The insertion of a $\CV_n$ vortex line for an abelian gauge or flavor symmetry $G$ shifts the spin of all charged operators by $-n$ units (times the operator's charge), and thus acts on the index by shifting the corresponding fugacity
\be s\overset{\CV_n}\longrightarrow q^{-n} s\,\quad\text{or}\quad x\overset{\CV_n}\longrightarrow q^{-n} x\,. \label{Vn} \ee

Vortex lines for a dynamical abelian gauge symmetry are relatively boring, as they can be screened in the bulk (in other words, the gauge transformation $g(z)=z^{n}$ can be undone). With $\CN$ b.c. for the gauge symmetry, they can also be screened on the boundary, so the $\CN$ half-index is insensitive to a $\CV_n$ insertion. This is manifest in \eqref{Vn}, since the shift $s\to q^{-n} s$ is invisible after one integrates over $s$ (projects to gauge invariants). 
With $\CD$ b.c., the vortex line does have a mild effect, as it leaves behind a singularity on the boundary.
The shift $s\to q^{-n}s$ could be removed from the monopole sum \eqref{CD-full} by redefining $m\to +n$,
were it not for the additional prefactor $q^{\frac12k_{\rm eff}\text{Tr}(m^2)} s^{k_{\rm eff}m}$, controlled by effective Chern-Simons terms. Due to this extra prefactor, the $\CD$ half-index obeys
\be \II_\CD(q^{-n}s,x) = q^{-\frac12k_{\rm eff}n^2}x^{n}s^{k_{\rm eff}n} \II_\CD(s,x)\,, \label{WV}\ee
where $x$ is the fugacity for the topological $U(1)$ symmetry dual to the bulk $G$ gauge symmetry.
This identity reflects the classic phenomenon that, in the presence of Chern-Simons terms, a vortex line for an abelian gauge symmetry is equivalent to a Wilson line --- here, a Wilson line both for $G$ and the topological $U(1)$.

\subsection{Difference equations}
\label{sec:diff}

The bulk 3d index, $S^3$ partition, and holomorphic blocks of a 3d $\CN=2$ theory all satisfy a common set of difference equations \cite{DGG, DGG-index, BDP-blocks}. On the holomorphic blocks $B(x;q)$, which depend on fugacities $x_i$ for each bulk flavor symmetry,%
\footnote{Here we will assume that the flavor symmetry is abelian. Given a nonabelian symmetry, we work with a maximal torus.} %
the difference equations take the form
\be   f_a(p_i,x_i;q) \cdot B(x,q) = 0\,,  \label{diffB} \ee
where the $f_a$ are a finite set of polynomials in the $q$-commuting operators $p_i,x_i$, which obey
\be p_i x_j = q^{\delta_{ij}}x_j p_i \label{xp} \,.\ee
The $x_i$ operators act on $B(x;q)$ by multiplication, while the $p_i$ operators act by a $q$-shift, sending $x_i\to qx_i$. The 3d index and $S^3$ partition functions obey two sets of equations of the form \eqref{diffB}, involving two mutually commuting copies of the algebra \eqref{xp} and exactly the \emph{same} polynomials $f_a$.

Physically, all these difference equations are a consequence of identities in the algebra of half-BPS line operators of the 3d theory --- in particular, identities among abelian vortex lines and flavor Wilson lines \cite{DGG, DG-Sdual}.%
\footnote{The actual algebra \eqref{xp} has a nice interpretation as arising from 't Hooft and Wilson lines in an Omega-deformed abelian 4d $\CN=2$ theory, \emph{cf.} \cite{GW-surface, GMN-framed}, coupled to a 3d $\CN=2$ theory on its boundary. See \cite{DGG, BDP-blocks} for further discussion.} %
We therefore expect that 3d half-indices would also satisfy \eqref{diffB}. This turns out to be true, up to a small modification controlled by the boundary 't Hooft anomalies.

To understand the modification, we consider a 3d $\CN=2$ theory with $U(1)_x$ flavor symmetry, and a boundary condition that preserves this flavor symmetry. Suppose that the half-index obeys some difference equation
\be f(p,x;q)\cdot \II(x;q) = 0\,,\ee
where $x$ acts by multiplication and $p$ acts by shifting $x\to qx$ (\emph{i.e.} $p$, $x$ represent the insertions of flavor vortex and Wilson lines, respectively). Consider what happens if we were to change the boundary 't Hooft anomaly for $U(1)_x$ by $+1$ units. The shift can be achieved by adding a boundary Fermi multiplet of $U(1)_x$ charge $+1$ and $U(1)_R$ charge zero, whose 2d index is $\text{F}(-q^{\frac12}x;q)=(-q^{\frac12}x;q)_\infty(-q^{\frac12}/x;q)_\infty$. The new half-index is $\II'(x;q)=\II(x;q)\text{F}(-q^{\frac12}x;q)$, and is annihilated by the operator
\be f'(p,x;q) = \text{F}(-q^{\frac12}x;q) f(p,x;q) \text{F}(-q^{\frac12}x;q)^{-1} = f(q^{\frac12}xp,x;q)\,. \ee
Thus, shifting the boundary 't Hooft anomaly modifies the difference operator by replacing $p\to q^{\frac12}xp$. More symmetrically, we might write the modification as a normal-ordered product
\be p\to \;\;:\!xp\!:\;\;= q^{\frac12}xp = q^{-\frac12}px  \,. \label{pxCS} \ee
The redefinition \eqref{pxCS} is an automorphism of the Weyl algebra \eqref{xp}.

This argument is easily generalized to determine how any shift in the boundary 't Hooft anomaly modifies operators acting on the half-index. To express the result concretely, suppose that the difference in boundary 't~Hooft anomalies is encoded in a quadratic polynomial $\delta\CI_2(\mb x_i,\mb r)$. Let $q=e^\beta$, so that we may represent $p_i = \exp\big(\beta x_i\frac{\pd}{\pd x_i}\big)$. Then the shift in anomalies modifies difference operators by conjugation,
\be f(p_i,x_i;q) \to \exp\Big[\!-\!\frac1{2\beta}\delta\CI_2(\log(x_i),i\pi+\tfrac{\beta}{2})\Big] f(p_i,x_i;q)  \exp\Big[\frac1{2\beta}\delta\CI_2(\log(x_i),i\pi+\tfrac{\beta}{2})\Big]  \ee
or simply
\be p_j \to  \exp\Big[\!-\!\frac1{2\beta}\delta\CI_2(\log(x_i),i\pi+\tfrac{\beta}{2})\Big] p_j  \exp\Big[\frac1{2\beta}\delta\CI_2(\log(x_i),i\pi+\tfrac{\beta}{2})\Big]\,. \label{modp}  \ee

Despite the complicated-looking expression, \eqref{modp} is simply a redefinition of $p_i$ by a monomial in the $x$'s. For example, if there is a single $U(1)$ flavor symmetry and we shift the anomaly by $\delta\CI_2(\mb x,\mb r)=\mb x^2$ as above, then
\be p \to e^{-\frac1{2\beta}(\log x)^2}p e^{\frac1{2\beta}(\log x)^2} = e^{-\frac1{2\beta}(\log x)^2} e^{\frac1{2\beta}(\log x+\beta)^2}p = q^{\frac12}xp\,, \ee
just as in \eqref{pxCS}. The appearance of `$i\pi+\tfrac{\beta}{2}$' in \eqref{modp} reflects the fact that the $U(1)_R$ fugacity in the half-index is $\exp\big(i\pi+\tfrac{\beta}{2}\big)=-q^{\frac12}$.

We can now understand which operators will actually annihilate the half-index of a UV boundary condition in a 3d $\CN=2$ theory. Suppose that no flavor symmetry is broken by the boundary condition: so the chirals all have N or D b.c. (not $\text{D}_c$); and in the case of $\CN$ b.c., there is no mixed gauge-flavor anomaly.
If the boundary 't Hooft anomaly were exactly zero, we would expect the very same difference operators \eqref{diffB} to annihilate the half-index
\be f_a(p_i,x_i;q)\cdot \II(x;q) = 0\,. \label{p0} \ee
This is consistent with the fact that the holomorphic blocks of \cite{BDP-blocks} were carefully engineered to have zero boundary anomaly. In addition, given $\CD$ b.c. for the gauge multiplets, there are additional equations expressing independence under $q$-shifts of the $G_\pd$ fugacities `$s$', namely
\be  (p_s-1)\cdot \II(x,s;q) = 0\,. \label{s0} \ee
When the boundary anomaly is \emph{not} zero, we use \eqref{modp} to modify the difference operators accordingly. Explicitly, given a boundary anomaly $\CI_2(\mb x_i,\mb f,\mb r)$ (where $\mb f$ is a $G_\pd$ field strength for $\CD$ b.c.), we conjugate the $p$'s and $p_s$ in \eqref{p0}-\eqref{s0} by
\be \exp\Big[\frac1{2\beta}\delta\CI_2(\log(x_i),\log(s),i\pi+\tfrac{\beta}{2})\Big]\,.\ee
For example, after conjugation, we recover from \eqref{s0} the simple difference equation \eqref{WV} relating $G_\pd$ vortices and Wilson lines.

If a symmetry is broken by the boundary condition, its fugacity (say `$y$') should be subsequently removed from the difference operators by first eliminating the dual operator $p_y$ from (the conjugated versions of) \eqref{p0}, \eqref{s0}, and then setting $y\to 1$.

We will give a few examples of these difference operators below. The fact that the \emph{same} $f_a$ (up to monomial redefinitions) annihilate UV half-indices and holomorphic blocks (and full indices and $S^3$ partition functions) can actually be seen explicitly from the formulaic definitions of all these objects. In each case, simple chiral-matter partition functions are multiplied together, then gauge fugacities are integrated over (and/or monopole sectors summed over). The difference equations can correspondingly be constructed step by step, starting from elementary equations satisfied by free chirals.%
\footnote{In 3d theories associated to 3-manifolds via the 3d-3d correspondence, the difference equations are versions of the ``quantum A-polynomial'' on the 3-manifold side \cite{Gukov-A, Gar-A}, whose analogous step-by-step construction was given by \cite{Dimofte-QRS}.} %
The steps are identical, up to simple monomial redefinitions, no matter which object is being considered.


\section{Particle-vortex triality}
\label{sec:tet}

The simplest dual pair of 3d $\CN=2$ theories is actually part of a mirror ``triality,'' involving theories
\be \begin{array}{c} \CT \\ \text{free chiral} \end{array}\quad\leftrightarrow\quad
\begin{array}{c} \CT' \\ \text{$U(1)_{\frac12}$ + a chiral} \end{array}\quad\leftrightarrow\quad
\begin{array}{c} \CT'' \\ \text{$U(1)_{-\frac12}$ + a chiral} \end{array}
\ee
This is the simplest example of 3d $\CN=2$ ``mirror symmetry'' \cite{dBHOY-N2}, and also follows from deforming the SQED\,$\leftrightarrow$\,XYZ duality of \cite{AHISS} by a large real mass \cite{DGG}. The triality played a fundamental role in the 3d-3d correspondence, where it encoded a $\Z_3$ rotation symmetry of a tetrahredron \cite{DGG}. This triality is a supersymmetric version of classic particle-vortex duality, and can be used to derive both bosonic and fermionic versions of non-supersymmetric particle-vortex duality \cite{KMTW-bos, KMTW-bos2, KarchTong}. 

\subsection{Free chiral}

Let's first focus on theory $\CT$. We define $\CT$ to contain a free 3d $\CN=2$ chiral multiplet $(\Phi,\Psi)$ of charge $\rho=0$ for $U(1)_R$ R-symmetry and charge $+1$ for a $U(1)_x$ flavor symmetry. As in \cite{DGG}, we add $-\frac12$ units of background Chern-Simons coupling for $U(1)_x$ to cancel a parity anomaly. Specifically, the anomaly polynomial encoding the bulk UV (background) Chern-Simons couplings is
\be \mb -\frac12 (\mb f_x-\mb r)^2\,.\ee
The two basic $\CN=(0,2)$ boundary conditions for the free chiral are Neumann (N, $\Psi|_\pd=0$) and Dirichlet (D, $\Phi|_\pd=0$), as in Section \ref{sec:chiral}. The corresponding boundary 't Hooft anomalies (Section \ref{sec:anom}) and half-indices (Section \ref{sec:chiralindex}) are
\be \label{T-anom} \begin{array}{lll}
 &\hspace{.5cm}\text{anomaly} &\hspace{1cm} \text{half-index} \\ 
 \text{N}\,:\quad & \CI_{\rm N}=-(\mb f_x-\mb r)^2 \qquad & \II_{\rm N}(x;q) = (x;q)_\infty^{-1}\,, \\[.2cm]
 \text{D}\,:\quad & \CI_{\rm D}=0 \qquad & \II_{\rm D}(x;q) = (qx^{-1};q)_\infty\,.
\end{array} \ee

Following Section \ref{sec:diff}, we expect that these half-indices satisfy certain difference equations. The difference operator that annihilates the full 3d index of a free chiral is $p+x^{-1}-1$. Since the boundary anomaly for D b.c. is zero, this should also annihilate $\II_{\rm D}(x;q)$, and indeed
\be p \cdot \II_{\rm D}(x;q) = \II_{\rm D}(qx;q) = (x^{-1};q)_\infty = (1-x^{-1})(qx^{-1};q)_\infty = (1-x^{-1})\II_{\rm D}(x;q)\,, \ee
so $(p+x^{-1}-1)\cdot \II_{\rm D}(x;q)=0$. For N b.c., we must conjugate the difference operator by $\exp\big[-\frac1{2\beta}(\log(x)-i\pi-\tfrac\beta2)^2\big]$, corresponding to the boundary anomaly. This modifies
\be p \;\to\; e^{\frac1{2\beta}\big(\log(x)-i\pi-\frac\beta2\big)^2}e^{-\frac1{2\beta}\big(\log(x)-i\pi+\frac\beta2\big)^2}p = -x^{-1}p\,, \ee
and changes the difference operator to $-x^{-1}p+x^{-1}-1$. Thus we expect that $(p-1+x)\cdot \II_{\rm N}(x;q)=0$, and we check that this is true:
\be p\cdot \II_{\rm N}(x;q) = \II_{\rm N}(qx;q) = (qx;q)_\infty^{-1} = (1-x)(x;q)_\infty^{-1} = (1-x)\II_{\rm N}(x;q)\,.\ee

We may also modify the Dirichlet b.c. to D$_c$ ($\Phi|_\pd = c$). This breaks $U(1)_x$ flavor symmetry but preserves $U(1)_R$. As discussed in Section \ref{sec:singind}, the effect is to set $x\to 1$ in the Dirichlet index. Alternatively, we may first add a flavor vortex line of charge $m\in \Z$ as in Section \ref{sec:line}, sending $x\to q^{-m} x$, and subsequently impose D$_c$ b.c.; overall, this sets $x=q^{-m}$ in the Dirichlet index. We have:
\be \begin{array}{lcl}
 &\hspace{-.5cm}\text{anomaly} &\hspace{1cm} \text{half-index} \\ 
\text{D}_c\,:\quad & 0 \qquad & \II_{\text{D}_c}(q) = (q)_\infty\,, \\[.2cm]
\text{D}_{c}+\text{Vortex}_{m}\,:\quad & 0 \qquad & \II_{\text{D}_{c,m}}(q) = (q^m;q)_\infty 
\,.
\end{array} \ee
Note that the vortex line effectively sets $\phi \sim z^m$ at the boundary, which breaks SUSY unless $m\geq 0$. Correspondingly, the half-index $\II_{\text{D}_{c,m}}(q)$ vanishes unless $m\geq 0$.

\subsection{$U(1)_{\frac12}$ + a chiral}

Next, consider the theory $\CT'$. It is a $U(1)$ gauge theory, with a chiral multiplet ($\Phi',\Psi'$) of charge $+1$ under the gauge group and R-charge $\rho=0$. There is also a topological flavor symmetry $U(1)_x$.
The bulk UV Chern-Simons levels required for this theory to be dual to $\CT$ are encoded in the anomaly polynomial
\be \CI'_{\rm bulk} = \frac12(\mb f-\mb r)^2+2\,\mb f\,\mb f_x - \frac12\mb r^2\,. \label{T'bulk}\ee
In particular, the $U(1)$ gauge symmetry (with field strength $\mb f$) has Chern-Simons level $+\frac12$. Recall that under the duality $\CT\,\leftrightarrow\CT'$, the fundamental chiral of $\CT$ maps to a monopole operator of $\CT'$, and (correspondingly) the ordinary flavor symmetry $U(1)_x$ of $\CT$ maps to the topological flavor symmetry of $\CT'$.

We would like to find boundary conditions for $\CT'$ that are dual to N, D, D$_c$, and D$_{c,m}$ in $\CT$. It is most enlightening to begin with N. Notice that the N b.c. leaves the chiral $\Phi$ of $\CT$ free at the boundary. Thus we would expect that a dual boundary condition in $\CT'$ would leave the vev of a \emph{monopole operator} unconstrained at the boundary. There is only one choice of boundary condition for the gauge fields that has this property, namely Dirichlet ($\CD$) as in Sections \ref{sec:bdy-gauge}, \ref{sec:II-D}.
Recall that $\CD$ b.c. has an additional boundary $U(1)_\pd$ flavor symmetry.
We try choosing D b.c. for the chiral of $\CT'$ as well, and examine the half-index. To use the prescription for the half-index from Section \ref{sec:II-D}, we must compute the boundary 't Hooft anomaly:
\be \CI'_{\CD,\text{D}}= \underbrace{\tfrac12(\mb f-\mb r)^2+2\,\mb f\,\mb f_x - \tfrac12\mb r^2}_{\text{bulk CS}}  \underbrace{-\tfrac12\mb r^2}_{\text{$\CD$ for gauge}} +\underbrace{\tfrac12(\mb f-\mb r)^2}_{\text{D for chiral}} =   \mb f^2+2(\mb f_x-\mb r)\mb f  \label{T'-anom}\ee
where now $\mb f$ is the field strength for $U(1)_\pd$.
We read off from this a matrix of effective Chern-Simons levels. The half-index then becomes a monopole sum
\be \II'_{\CD,\text{D}}(x,y;q) = \frac{1}{(q)_\infty}\sum_{m\in \Z} q^{\frac{m^2}{2}}(-q^{-\frac12} x)^m y^m \underbrace{\II_{\rm D}(q^my;q)}_{(q^{1-m}y^{-1};q)_\infty}\,, \label{T'DD} \ee
where we have used `$x$' to denote the topological $U(1)_x$ fugacity and `$y$' to denote the boundary $U(1)_\pd$ fugacity, and the monomial prefactor $q^{\frac{m^2}{2}}(-q^{-\frac12} x)^n y^n$ corresponds to the effective Chern-Simons levels. Due to the $q^{\frac{m^2}{2}}$, this series converges as an element of $\Z[\![q^{\frac12}]\!][x^{\pm 1},y^{\pm 1}]$, and rather beautifully sums up to%
\footnote{For a quick proof that \eqref{T'DD} and \eqref{T'DD-id} are equivalent, it suffices that the two expressions obey the same first-order $q$-difference equations in $x$ and in $y$, and that they agree at a particular value of $(x,y)$. In fact, it suffices to use the single difference equation in $y$, $\II'_{\CD,\text{D}}(x,qy) = -\frac{1}{xy}\II'_{\CD,\text{D}}(x,y;q)$, which follows from $\text{F}(qxy)=-\frac{1}{xy}\text{F}(xy)$ in \eqref{T'DD-id} and follows from simple manipulations of the sum in \eqref{T'DD}. 
Then we check agreement at $y=1$ (for any $x$), which is the well-known identity in \eqref{y1id}.}
\be \II'_{\CD,\text{D}}(x,y;q) = \frac{(xy;q)_\infty(qx^{-1}y^{-1};q)_\infty}{(x;q)_\infty} = \II_{\rm N}(x;q)\, \text{F}(xy;q)\,. \label{T'DD-id} \ee

This calculation strongly suggests that the $(\CD,\text{D})$ b.c. for $\CT'$ is dual to N b.c. for the free chiral of $\CT$ \emph{together with} a free Fermi multiplet of charge $(-1,-1,1)$ for $U(1)_x,U(1)_\pd,U(1)_R$. The appearance of this extra Fermi multiplet is not a surprise. First, since the ($\CD$,D) b.c. for $\CT'$ had a $U(1)_\pd$ symmetry that acted on the boundary but not in the bulk, we would expect a dual b.c. for $\CT$ to have \emph{some} purely two-dimensional degrees of freedom charged under this symmetry.
Second, we may compare anomaly polynomials \eqref{T'-anom}, \eqref{T-anom}, whose difference
\be \CI'_{\CD,\text{D}} - \CI_{\rm N} = (\mb f+\mb f_x-\mb r)^2 \ee
is precisely the contribution of an extra boundary Fermi multiplet $\Gamma$. Thus anomaly matching alone requires $\Gamma$. The map of operators across the duality identifies the boundary value $\Psi'|_\pd$ in theory $\CT'$ (which is gauge-invariant at the boundary) with the composite operator $\Phi|_\pd\Gamma$ in~$\CT$.

Several more dualities may be inferred from \eqref{T'DD-id}. 
The $(\CD,\text{D}_c)$ b.c. for $\CT'$, which sets $\Phi'|_\pd=c$ and breaks the $U(1)_\pd$ boundary symmetry, corresponds to setting $y\to 1$ in the index. Since $(q^{1-m}y^{-1};q)_\infty$ vanishes at $y=1$ unless $m\leq 0$, we find
\begin{align} \II'_{\CD,\text{D}_c}(x;q) &= \frac{1}{(q)_\infty}\sum_{m\leq 0}q^{\frac{m^2}{2}}(-q^{-\frac12}x)^n (q^{1-m};q)_\infty \notag \\
&=  \sum_{m\geq 0}\frac{q^{\frac{m^2}{2}}(-q^{\frac12}x^{-1})^m }{(q)_m} \notag \\
&= \II_{\rm N}(x;q)\, \text{F}(x;q) = \II_{\rm D}(x;q)\,. \label{y1id}
\end{align}
This is actually a well-known q-series identity. It suggests that the $(\CD,\text{D}_c)$ b.c. for $\CT'$ is dual to ordinary D b.c. for $\CT$. Similarly, adding an additional vortex line of charge $n$ and then using $\text{D}_c$ b.c. sets $y\to q^{-n}$
\footnote{Here we used $\text{F}(q^{-m}x) = (-1)^m q^{-\frac{m^2+m}{2}}x^m\text{F}(x)$.}
\be  \II'_{\CD,\text{D}_{c,n}}(x;q) = \frac{1}{(q)_\infty}\sum_{m\leq n}q^{\frac{m^2}{2}}(-q^{-\frac12}x)^m (q^{1+n-m};q)_\infty
= (-1)^n q^{-\frac{n^2+n}{2}}x^n \II_{\rm D}(x;q)\,, \ee
which suggests that the $(\CD,\text{D}_{c,n})$ b.c. is dual to D b.c. for $\CT$ together with a flavor Wilson line of charge $n$ (corresponding to $x^n$) and an extra shift of the background spin and R-charge.

More interestingly, we may compute the half-index of $(\CD,\text{N})$ b.c. for $\CT'$, which give N b.c. to the chiral. Now the boundary 't Hooft anomaly is just $\CI'_{\CD,\text{N}}=\CI'_{\rm bulk} -\tfrac12\mb r^2-\tfrac12(\mb f-\mb r)^2=
2\mb f\,\mb f_x$, so the monopole sum takes the form
\be \II'_{\CD,\text{N}}(x,y;q) = \frac{1}{(q)_\infty} \sum_{n\in \Z} x^n \II_{\rm N}(q^n y;q)\,. \ee
It is convenient to rewrite $\II_{\rm N}(q^n y;q) = \II_{\rm D}(q^n y;q) / \text{F}(q^n y;q) = (-1)^n q^{\frac{n^2-n}{2}}y^n  \II_{\rm D}(q^n y;q) / \text{F}(y;q) = (-1)^n q^{\frac{n^2-n}{2}}y^n  \II_{\rm D}(q^n y;q)\, \text{C}(y;q) $, whence
\begin{align} \II'_{\CD,\text{N}}(x,y;q) &= \text{C}(y;q)  \frac{1}{(q)_\infty} \sum_{n\in \Z}  (-1)^n q^{\frac{n^2-n}{2}}y^n x^n \II_{\rm D}(q^n y;q) \\ & \overset{\eqref{T'DD}}{=} \text{C}(y;q) \II'_{\CD,\text{D}}(x,y;q) \\ & \overset{\eqref{T'DD-id}}{=}  \II_{\rm N}(x;q)\, \text{F}(xy;q)\text{C}(y;q)\,.\end{align}
This suggests that the $(\CD,\text{N})$ b.c. for $\CT'$ is dual to the N b.c. for $\CT$, coupled to both a boundary Fermi multiplet $\Gamma$ and a boundary chiral multiplet $C$. The symmetries and R-charges are just right for a boundary superpotential coupling
\be \int d\theta^+\, \Phi\big|_\pd \Gamma C\,. \label{W-T'} \ee
Notice how the chiral $C$ may be understood as \emph{flipping} D to N b.c. for the chiral $(\Phi',\Psi')$ of theory $\CT'$, by a boundary superpotential coupling $\Psi'\big|_\pd C$; the dual of this coupling in $\CT$ is precisely \eqref{W-T'}.

There are also a family of boundary conditions for $\CT'$ that use Neumann ($\CN$) for the gauge multiplet, preserving dynamical $U(1)$ gauge symmetry at the boundary. Some of these were first discussed in \cite{GGP-walls}. Suppose we combine $\CN$ with N b.c. for the charged chiral. The boundary anomaly polynomial is $2\,\mb f \,\mb f_x$, where $\mb f$ is now the curvature of the dynamical gauge symmetry at the boundary. There is no gauge anomaly, so no additional boundary matter is required, but we see that the topological $U(1)_x$ symmetry will be broken by a mixed anomaly. Following Section \ref{sec:II-N}, the index is simply computed as
\be \II'_{\CN,\text{N}}(q) = (q)_\infty \oint\frac{ds}{2\pi is} \II_{\rm N}(s;q) \label{T'NN} \ee
where $s$ is the gauge fugacity. This can be evaluated by residues to give
\be \II'_{\CN,\text{N}}(q) = \sum_{a=0}^\infty \frac{1}{(q^{-1};q^{-1})_a} =(q)_\infty = \II_{\text{D}_c}(q)\,,\ee
suggesting that $(\CN,\text{N})$ b.c. is dual to D$_c$ b.c. for $\CT$.
This is actually a physically sensible answer: the Neumann b.c. for an abelian gauge multiplet sets the scalar $\sigma$ field equal to a boundary FI parameter, and more generally (as explained in Section \ref{sec:photon}) sets the chiral dual-photon field equal to a complexified FI parameter $t_{2d}$. In the quantum gauge theory, this means that the monopole operator of $\CT'$ should be set to a nonzero constant $\sim e^{t_{2d}}$ at the boundary. The dual statement in theory $\CT$ is that the chiral field should be set to a nonzero constant, which is exactly what D$_c$ b.c. does.

A generalization of \eqref{T'NN} is to add a Wilson line of charge $n$ ending on the boundary. This leads to a projection onto boundary operators of gauge charge $-n$. The half-index becomes
\be \II'_{\CN_m,\text{N}}(q) = (q)_\infty \oint\frac{ds}{2\pi is} s^n \II_{\rm N}(s;q) = \sum_{a=0}^\infty \frac{q^{-an}}{(q^{-1};q^{-1})_a} = (q^{1-n};q)_\infty = \II_{D_{c,-n}}(q)\,.  \ee
In other words, we find D b.c. for $\CT$ deformed by a flavor vortex of charge $-n$. This reflects the standard fact that a Wilson line in abelian gauge theory is equivalent to a vortex line for the topological flavor symmetry --- \emph{i.e.} the ordinary flavor symmetry in $\CT$.

It is worth observing that the $(\CN,\text{N})$ should be obtainable from $(\CD,\text{N})$ by gauging the $U(1)_\pd$ boundary symmetry.
This breaks $U(1)_x$, but it is useful to keep $x$ in the calculation until the very end. Then the 2d gauging prescription 
is to multiply $\II'_{\CD,\text{N}}(x,y;q)$ by $(q)^2_\infty$ and take the residue at $y=1$, leading to $\II_{\rm N}(x;q)\, \text{F}(x;q)$.
Setting $x=1$ we recover  $\II_{\text{D}_c}(q)$. 

The physical interpretation of this calculation is somewhat interesting: on the mirror side we have an N b.c. coupled to a boundary theory 
consisting of an a $U(1)$ gauge theory coupled to a single 2d chiral and a single 2d Fermi multiplet. It appears that the 2d theory 
flows to a single Fermi multiplet in the IR, the ``meson'' $\mu = \Gamma C$, with a dynamically generated 
fermionic superpotential $c \mu$, which then flips the boundary condition to $\text{D}_c$. 

We could also combine $\CN$ b.c. for the gauge fields with D b.c. for the chiral, but this boundary condition does have a gauge anomaly, since its anomaly polynomial is \eqref{T'-anom}. 
In principle, we could also attempt to modify the $\CN$ b.c. with additional boundary matter in order to cancel the mixed anomaly for the topological $U(1)_x$ symmetry. 
Because of the sign of the anomaly, though, we cannot do so by boundary Fermi multiplets: we need boundary chiral multiplets. 
For example, $(\CN,\text{D})$ b.c. together with a boundary chiral multiplet of charges $(1,1,0)$ for $U(1),U(1)_x,U(1)_R$ has a total boundary anomaly polynomial $(\mb f-\mb r)^2+2\,\mb f\,\mb f_x -(\mb f+\mb f_x-\mb r)^2=-(\mb f_x-\mb r)^2+\mb r^2$\,, so only an 't Hooft anomaly remains.

If we couple the boundary chiral multiplet to the bulk chiral multiplet by a bi-linear fermionic superpotential we simply convert $D$ b.c. back to $N$ b.c. and 
we do not get anything new. If we do not add such a coupling, the naive half-index takes the form
\begin{align} \label{badN} \II'_{\CN,\text{D}\text{+chiral}}(x;q) &= (q)_\infty\oint \frac{ds}{2\pi is} \II_{\rm D}(s;q)\,\text{C}(sx;q)
\\ & =  (q)_\infty\oint \frac{ds}{2\pi is} \frac{(qs^{-1};q)_\infty}{(sx;q)_\infty(qs^{-1}x^{-1};q)_\infty}\,,
\notag\end{align}
but there is no obvious prescription to deal with the infinite line of poles of $C(sx;q)$ at $s=q^nx^{-1}$ ($n\in \Z$). 

Instead, we can go back to the $(\CD,\text{D})$ boundary condition, add the 2d chiral field to cancel the $U(1)_\pd$ 't Hooft anomalies and then gauge $U(1)_\pd$
as a 2d gauge symmetry. The resulting index would have an overall factor of $\II_{\rm N}(x;q)$ but would vanish, as 
on the mirror side one has a boundary theory consisting of a 2d $U(1)_{2d}$ gauge theory coupled to a Fermi multiplet of charge $(-1,1)$ for $U(1)_{2d},U(1)_R$ and a chiral multiplet of charges $(1,0)$ for $U(1)_{2d},U(1)_R$, which cancel each other in the index
and leave no poles to be picked.  

We can interpret this tentatively as a manifestation of spontaneous SUSY breaking. This agrees with the physical picture we 
discussed for $(\CN,\text{N})$: in the absence of the coupling $\Phi|_{\pd} \mu$, the dynamically generated 
$c \mu$ fermionic superpotential for the $U(1)_{2d}$ gauge theory will break SUSY. 

We summarize the plethora of dual boundary conditions we have found so far:
\be \label{BB'} \begin{array}{c|c} \text{$\CB$ (theory $\CT$)} & \text{$\CB'$ (theory $\CT'$)} \\\hline
 \text{N + 2d fermi} & (\CD,\text{D}) \\
  \text{N + 2d fermi/chiral} & (\CD,\text{N}) \\
 \text{D} & (\CD,\text{D}_c) \\
 \text{D + Wilson$_n$} & (\CD, \text{D}_{c})+\text{Vortex}_n \\
 \text{D$_c$} & (\CN,\text{N}) \\
 \text{D$_{c}$}+\text{Vortex}_n & \quad (\CN,\text{N})+\text{Wilson}_{-n} 
\end{array}
\ee

\subsection{$U(1)_{-\frac12}$ + a chiral}

The theory $\CT''$ is similar to $\CT$. However, its Neumann boundary conditions turn out to be better behaved, while some of its Dirichlet boundary conditions exhibit bad behavior.

We define theory $\CT''$ as a $U(1)$ gauge theory with a chiral of gauge charge $+1$ and R-charge zero, with bulk UV Chern-Simons levels encoded by the anomaly polynomial
\be \CI''_{\rm bulk}= -\tfrac12(\mb f-\mb r)^2-2(\mb f-\mb r)\mb f_x-\mb f_x^2-\tfrac12\mb r^2\,.\ee
As before, $\mb f$ is the field strength of the dynamical gauge field, while $-\mb f_x$ is the topological $U(1)_x$ flavor symmetry.

We start with the Neumann ($\CN$) boundary conditions for the gauge fields, which all behave nicely. Giving N b.c. to the chiral as well, we find a boundary anomaly
\be \CI''_{\CN,\text{N}}= \CI''_{\rm bulk} -\tfrac12(\mb f-\mb r)^2+\tfrac12\mb r^2= -(\mb f+\mb f_x-\mb r)^2\,, \ee
which can be neatly cancelled by a boundary Fermi multiplet of charges $(1,1,-1)$ for $U(1)\times U(1)_x\times U(1)_R$. Note that the extra boundary Fermi cancels \emph{both} the gauge and the mixed gauge-$U(1)_x$ anomalies, so this boundary condition preserves the $U(1)_x$ topological symmetry.
Just as in $\CT'$, we expect that $(\CN,\text{N})$ is dual to a Dirichlet-like b.c. for $\CT$; in fact, since $U(1)_x$ is preserved, we expect pure D b.c. Indeed, the half-index shows
\begin{align} \II''_{\CN,\text{N}\text{+fermi}}(x;q) &= (q)_\infty \oint \frac{ds}{2\pi is} {\text{F}(sx;q)} \II_{\rm N}(s;q) \\[.2cm]
 &=(q)_\infty \oint \frac{ds}{2\pi is} \frac{\text{F}(sx;q)}{(s;q)_\infty} \notag \\[.2cm]
 &\overset{\text{residues}}{=} \sum_{a\geq 0}\frac{\text{F}(q^{-a}x;q)}{(q^{-1})_a} = \text{F}(x)\sum_{a\geq 0} \frac{x^a}{(q)_a} \notag \\[.2cm]
 &= \frac{\text{F}(x)}{(x;q)_\infty} = (qx^{-1};q)_\infty = \II_{\rm D}(x;q)\,. \notag
\end{align}

The $(\CN,\text{N})$ b.c. can be modified by a bulk Wilson line of charge $n$ that ends on the boundary, which keeps boundary operators of gauge charge $-n$. We expect this to be dual to a bulk flavor-vortex in theory $\CT$, which shifts the spins of operators charged under $U(1)_x$. We check:
\be \II''_{\CN_n,\text{N+fermi}}(x;q) =  (q)_\infty \oint \frac{ds}{2\pi is} s^n {\text{F}(sx;q)} \II_{\rm N}(s;q) 
 = \II_{\rm D}(q^{-n}x;q)\,, \ee
as expected.

The $(\CN,\text{D})$ b.c. is even simpler: the boundary anomaly is $\CI''_{\CN,\text{D}}= \CI''_{\rm bulk} +\tfrac12(\mb f-\mb r)^2+\tfrac12\mb r^2=-2(\mb f-\mb r)\mb f_x-\mb f_x^2$, so $U(1)_x$ is anomalous, but there is no gauge anomaly to cancel. We would expect this to be dual to D$_c$ b.c. for the free chiral, and indeed
\be \II''_{\CN,\text{D}}(q) = (q)_\infty \oint \frac{ds}{2\pi is} \II_{\rm D}(s;q) = (q)_\infty \oint \frac{ds}{2\pi is} (qs^{-1};q)_\infty = (q)_\infty = \II_{\text{D}_c}(q)\,.\ee
Additionally, bulk Wilson lines in $\CT''$ are dual to vortex lines in $\CT$:
\be \II''_{\CN_n,\text{D}}(q) = (q)_\infty \oint \frac{ds}{2\pi is} s^n \II_{\rm D}(s;q) 
= (-1)^n q^{\frac{n(n+1)}{2}}(q^n;q)_\infty = (-1)^n q^{\frac{n(n+1)}{2}}\II_{D_{c,n}}(q)\,.\ee

In contrast to $\CN$ b.c., the half-indices for Dirichlet ($\CD$) b.c. on the gauge multiplet behave very badly due to the negative bulk Chern-Simons coupling. For example, $(\CD,\text{N})$ and $(\CD,\text{D)}$ b.c. have boundary 't Hooft anomalies $\CI''_{\CD,\text{N}}= \CI''_{\rm bulk} -\tfrac12(\mb f-\mb r)^2-\tfrac12\mb r^2=-(\mb f+\mb f_x-\mb r)^2-\mb r^2$ and $\CI''_{\CD,\text{D}}= \CI''_{\rm bulk} +\tfrac12(\mb f-\mb r)^2-\tfrac12\mb r^2=-2(\mb f-\mb r)\mb f_x-\mb f_x^2-\mb r^2$, respectively. Letting $y$ denote the fugacity for the boundary $U(1)_\pd$ symmetry, we find putative half-indices
\begin{align} \label{tet-diverge}
 \II''_{\CD,\text{N}}(x,y;q) &= \frac{1}{(q)_\infty} \sum_{m\in \Z} q^{-\frac{m^2}{2}}y^{-m}(-q^{\frac12}x^{-1})^m \frac{1}{(q^mx;q)_\infty}\,, \\[.2cm]
 \II''_{\CD,\text{D}}(x,y;q) &= \frac{1}{(q)_\infty} \sum_{m\in \Z} x^{-m}(q^{1-m}y^{-1};q)_\infty\,. 
 \end{align}
Both sums diverge badly as series in $q$: for $(\CD,\text{N})$, the $n$-th term in the sum begins with the large negative power $q^{-\frac{n(n-1)}{2}}$ for both positive and negative $n$; while for $(\CD,\text{D})$, the $n$-th term begins with $q^{-\frac{n(n-1)}{2}}$ for negative $n$. In the analysis of bulk indices, such behaviour is usually indicative of 
a ``bad'' setup where some operators, such as these boundary monopoles, hit the unitarity bound along the RG flow and the $U(1)_R$ in the IR 
contains emergent symmetries. Our tools are thus insufficient to study the problem. 

There is one choice of $\CD$ b.c. for which the half-index does make sense. If we give D$_{c}$ b.c. to the chiral (possibly with a vortex of charge $n$) and break $U(1)_\pd$ flavor symmetry, we find
\begin{align} \II_{\CD,D_{c,n}}(x;q) &= \frac{1}{(q)_\infty} \sum_{m\in \Z} x^{-m}(q^{1+n-m};q)_\infty
 =  \frac{1}{(q)_\infty} \sum_{m\geq 0} x^{m-n}(q^{m+1};q)_\infty \\[.2cm]
 &= x^{-n} \sum_{m\geq 0} \frac{x^m}{(q)_m} = x^{-n} \II_{\rm N}(x;q) \notag
\end{align}
Thus, we seem to recover N b.c. for a free chiral, with a flavor Wilson line of charge~$-n$.

Summarizing:
\be \label{BB''} \begin{array}{c|c} \text{$\CB$ (theory $\CT$)} & \text{$\CB''$ (theory $\CT''$)} \\\hline
 \text{N} & (\CD,\text{D}_c) \\
  \text{N + Wilson$_{-n}$} \quad & (\CD,\text{D}_{c})+\text{Vortex}_n \\
 \text{D} & (\CN,\text{N})\text{ + fermi} \\
  \text{D + Vortex$_{-n}$} \quad & \quad (\CN,\text{N})\text{ + fermi, Wilson$_n$} \\
   \text{D$_{c}$} & \quad (\CN,\text{D})  \\
 \text{D$_{c}$}+\text{Vortex}_n &\quad (\CN,\text{D})\text{ + Wilson$_n$}
\end{array}
\ee

\subsection{Left vs. right boundary conditions, and CS levels}
\label{sec:signs}

It may appear that there is a curious asymmetry between the available boundary conditions for theories $\CT'$ and $\CT''$, and their respective duals for the free chiral $\CT$.  One may well wonder how the sign of a Chern-Simons level, $+\frac12$ for $\CT'$ and $-\frac12$ for $\CT''$ can make such a difference. The distinction is indeed irrelevant in the 3d $\CN=2$ bulk. However, when constructing boundary conditions, we implicitly chose
\begin{itemize}
\item a ``right'' (vs. left) boundary condition, for a 3d theory on $x^\perp\leq 0$ (rather than $x^\perp\geq 0$)
\item a 2d $\CN=(0,2)$ (vs. $\CN=(2,0)$) SUSY algebra to preserve.
\end{itemize}
The first choice (right vs. left) controls the sign of anomalies coming from bulk UV Chern-Simons levels. The second choice ($\CN=(0,2)$ vs. $\CN=(2,0)$) controls the sign of anomalies from bulk matter and gauginos, since it determines how left-handed vs. right-handed fermions pair up in SUSY multiplets with the bosonic fields. \emph{Both} choices break the symmetry between positive and negative Chern-Simons levels in the bulk. If we were to reverse both choices simultaneously, then the behavior of boundary conditions for $\CT'$ and $\CT''$ would be exchanged.


\section{Duality interfaces}
\label{sec:int}

We would next like to construct a duality interface between the free chiral $\CT$ and the abelian gauge theories $\CT'$ or $\CT''$ from section \ref{sec:tet}. In preparation, we discuss some of the general structure of a duality interface.

\subsection{Factorizing the identity}
\label{sec:id}

\begin{figure}[htb]
\centering
\includegraphics[width=4.5in]{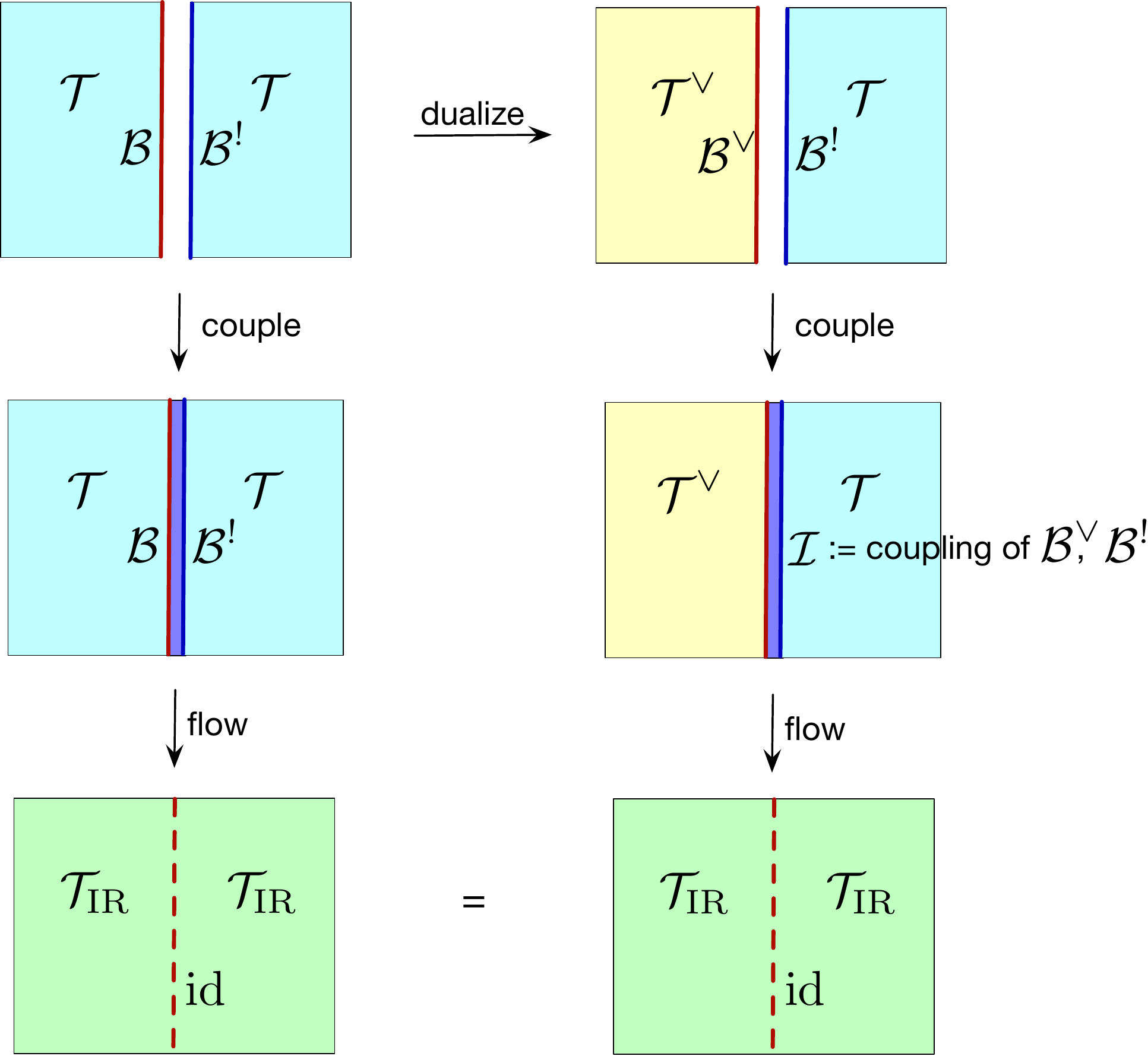}
\caption{Defining a duality interface $\CI$ between dual 3d theories $\CT^\vee$ and $\CT$ by starting with a factorization of the identity interface in $\CT$ (on the LHS) and dualizing half the space. The interface has the necessary property (that it flows to the identity in the IR) provided that the ``diagram commutes''; in particular the coupling on the RHS must involve the appropriate duals of the operators on the LHS.}
\label{fig:identity}
\end{figure}

A systematic way to construct a duality interface for a pair of dual 3d theories $\CT$ and $\CT^\vee$ (these could be any theories) involves starting with a pair of UV boundary conditions $\CB$ and $\CB^!$ for $\CT$ alone that can be coupled together in such a way that they flow to the trivial/identity interface in the IR. This ``factorization of the identity'' is illustrated on the LHS of Figure \ref{fig:identity}, with $\CB$ as a right b.c. and $\CB^!$ as a left b.c. It is not always guaranteed that a suitable pair $\CB,\CB^!$ exists. If it does, however, the duality interface can be obtained by dualizing $(\CT,\CB)$ to a dual boundary condition $(\CT^\vee,\CB^\vee)$ for the dual theory --- \emph{i.e.} dualizing $\CT$ on the left half-space --- and then coupling $(\CT^\vee,\CB^\vee)$ back to the boundary condition $\CB^!$ on the right half-space, as on the RHS of Figure \ref{fig:identity}.

In 3d $\CN=2$ UV gauge theories, we can always find suitable $(\CB,\CB^!)$, often in many different ways. The pair is built up from certain complementary choices of boundary conditions for the matter multiplets and for the gauge multiplets.

Consider a free 3d chiral multiplet. It is not hard to see that taking $\CB=\text{N}$ and $\CB^!=\text{D}$ (or vice versa) factorizes the identity as desired, if the two are coupled together by a quadratic superpotential. To be explicit, let us denote the 2d $\CN=(0,2)$ multiplets on the left of the interface ($x^\perp\leq 0$) as $(\Phi,\Psi)$, and on the right $(x^\perp\geq 0$) as $(\Phi',\Psi')$. We may deform the product of N b.c. for $(\Phi,\Psi)$ and D b.c. for $(\Phi',\Psi')$ by the superpotential
\be \int d^2x\,d\theta^+ \Phi\big|_\pd  \Psi'\big|_\pd\,. \label{W-glue} \ee
In the IR, this has the effect of simultaneously setting $\Psi\big|_\pd = \Psi'\big|_\pd$ and $\Phi\big|_\pd = \Phi'\big|_\pd$, thereby gluing the theory back together.

We may also check anomalies. Suppose there is a background Chern-Simons level $k$ for the chiral multiplet's $U(1)$ flavor symmetry. Then the flavor symmetry has boundary 't Hooft anomalies that depend both on the boundary condition and whether it is on the left or right:
\be \begin{array}{c|cc}
\text{b.c.}\backslash\raisebox{.2cm}{\text{location}} & \;x^\perp\leq 0 \;&\; x^\perp\geq 0 \\\hline
\text{N} &  k-\frac12 & -k-\frac12 \\
\text{D} & k+\frac12 & -k+\frac12
\end{array} \ee
Thus $\CB=\text{N}$ b.c. on $x^\perp\leq 0$ can be paired with $\CB^!=\text{D}$ b.c. on $x^\perp\geq 0$ to leave behind zero anomaly, as must be the case if the system flows to a trivial interface in the IR. (Alternatively, $\CB=\text{D}$ can be paired with $\CB^!=\text{N}$.)

In a similar way, the $\CN$ and $\CD$ boundary conditions for a 3d gauge multiplet constitute a suitable $(\CB,\CB^!)$ pair.
Indeed, there is a canonical coupling between $\CN$ and $\CD$, given by using the dynamical $G$ gauge symmetry on $\CN$ to gauge the $G_\pd$ boundary flavor symmetry on $\CD$. This obviously glues the gauge theory back together, and extends supersymmetrically to identify the entire 3d gauge multiplet across the interface. Moreover, anomalies cancel in essentially the same way as above. For example, for simple $G$ at Chern-Simons level $k$, the boundary gauge or 't Hooft anomalies are
\be \begin{array}{c|c@{\quad}c}
\text{b.c.}\backslash\raisebox{.2cm}{\text{location}} & \;x^\perp\leq 0 \;&\; x^\perp\geq 0 \\\hline
\CN &  (k+h)\text{Tr}(\mb f^2) & (-k+h)\text{Tr}(\mb f^2) \\
\CD & (k-h)\text{Tr}(\mb f^2) & (-k-h)\text{Tr}(\mb f^2)
\end{array} \ee
and cancel between $\CN$ and $\CD$.

In a full 3d gauge theory, we may now construct a suitable pair $(\CB,\CB^!)$ by simply combining a choice of (N,D) or (D,N) for each matter multiplet and a choice of $(\CN,\CD)$ or $(\CD,\CN)$ for the gauge group. The procedure should work even in the presence of a bulk superpotential~$W$, as long as sufficiently many D b.c. can be chosen on each side to ensure the vanishing of~$W$ at the interface. 

\subsection{Particle-vortex duality interface}
\label{sec:int-tet}

Let's illustrate the above construction for theory $\CT$ (a free chiral) and its duals $\CT'$ ($U(1)_{\frac12}$+a chiral)  and $\CT''$ ($U(1)_{-\frac12}$+a chiral) from Section \ref{sec:tet}.

The simplest way to begin is by ``factorizing the identity'' in theory $\CT'$, using the complementary boundary conditions $\CB'=(\CD,\text{D})$ (on $x^\perp\leq 0$) and $\CB'{}^! = (\CN,\text{N})$ (on $x^\perp\geq 0$).
The two half-spaces are coupled by using the gauge symmetry on the right to gauge the $U(1)_\pd$ flavor symmetry on the left, and by a superpotential of the form \eqref{W-glue}.
Then we dualize the $x^\perp\leq 0$ half-space. Recall from \eqref{BB'} that the dual of $\CB'=(\CD,\text{D})$ is $\CB^\vee = (\text{N}$ + 2d fermi). Explicitly, if we denote the bulk 3d chiral multiplet of $\CT'$ as $(\Phi',\Psi')$, the bulk 3d chiral of $\CT$ as $(\Phi,\Psi)$, and the 2d Fermi as $\Gamma$, the map of boundary operators relates
\be \text{Theory $\CT'$:}\quad \Psi'\qquad\leftrightarrow\qquad \text{Theory $\CT$:}\quad \Phi\Gamma\,. \ee

After re-coupling the $\CB^\vee = (\text{N}$ + 2d fermi) b.c. for $\CT$ to $\CB'{}^! = (\CN,\text{N})$ b.c. for $\CT'$, we find that the interface has a fairly symmetric description. It is built by:
\begin{itemize}
\item Placing $\CT$ on a half-space $x^\perp\leq 0$ with N b.c. (leaving the chiral $\Phi$ unconstrained at the boundary)
\item Placing $\CT'$ on a half-space $x^\perp\geq 0$ with $\CB'{}^!=(\CN$,N) b.c. (leaving the chiral $\Phi'$ unconstrained at the boundary)
\item Adding a 2d Fermi multiplet $\Gamma$ at the interface $x^\perp=0$
\item Coupling the two halves via a cubic superpotential $\int d\theta^+\,\Phi\Gamma \Phi'$, and identifying (gauging) the $U(1)_\pd$ flavor symmetry on the left with the $U(1)$ gauge symmetry on the right.
\end{itemize}
We schematically represent this interface as
\be \hspace{.5in} \overset{\CT}{\text{N}}\big|\Gamma\big|\overset{\CT'}{(\CN,\text{N})} \qquad (J_\Gamma=\Phi\Phi')\label{int-T'T} \ee
Notice that the charges of the various multiplets
\be \begin{array}{c|ccc}
& \Phi & \Gamma & \Phi' \\\hline
U(1)_\pd = U(1)_{\rm gauge} & 0 & -1 & 1 \\
U(1)_x & 1 & -1 & 0 \\
U(1)_R & 0 & 1 & 0
\end{array} \ee
are just right to ensure that the interface superpotential $\Phi\Gamma \Phi'$ preserves gauge, flavor, and R-symmetry. More so, by construction, anomalies for all symmetries cancel perfectly: from N b.c. on the left we have $-(\mb f_x-\mb r)^2$; from ($\CN$,N) b.c. on the right we have $-\big[\frac12(\mb f-\mb r)^2+2\,\mb f\,\mb f_x - \frac12\mb r^2\big] +\tfrac12\mb r^2 -\frac12(\mb f-\mb r)^2$ (note the opposite sign from bulk CS levels, due to the orientation), and from $\Gamma$ we have $(\mb f+\mb f_x-\mb r)^2$, which add up to zero. Interestingly, anomaly cancellation forces the interface Fermi multiplet $\Gamma$ to be charged under the topological $U(1)_x$ symmetry of $\CT'$. The cubic superpotential then identifies $U(1)_x$ with the usual flavor symmetry of $\CT$.

Notice how the Fermi multiplet $\Gamma$ mediates the map of operators under the duality. The superpotential $\int d\theta^+\,\Phi\Gamma \Phi'$ modifies the N b.c. for the chirals on either side, from the usual $\Psi\big|_\pd=0$, $\Psi'\big|_\pd=0$ to
\be \Psi = \Gamma\Phi'\,,\qquad  \Phi\Gamma = \Psi'\,. \ee
In addition, if we move onto the (bulk) moduli space of $\CT$ by giving $\Phi$ a vev, the Fermi multiplet $\Gamma$ will become massive. Integrating it out at one loop generates a ``field-dependent FI term'' on the interface, encoded in the superpotential%
\footnote{Superpotentials like this were described and generalized in many ways in, \emph{e.g.}, \cite{AdamsBasuSethi, McOristMelnikov, Tong-heterotic, BGR-02, QuigleySethi, GGP-triality}. At its heart, the superpotential is an $\CN=(0,2)$ version of the ``$\Sigma\log\Sigma$'' effective twisted superpotential that appears in $\CN=(2,2)$ theories and played a fundamental role in 2d mirror symmetry \cite{Witten-phases, HoriVafa}.}
\be \int d\theta^+ \Upsilon' \log \Phi\,, \label{t-eff} \ee
where $\Upsilon'$ is the field strength multiplet in theory $\CT'$, restricted to the boundary. The superpotential \eqref{t-eff} now carries the contribution to the mixed gauge-topological anomaly from $\Gamma$; in particular, by itself, \eqref{t-eff} breaks the $U(1)_x$ symmetry under which $\Phi$ is charged.
We now recall from Section \ref{sec:photon} that $\CN$ b.c. on the gauge multiplet of Theory $\CT'$ is a D b.c. on its dual-photon multiplet $S'{}^\vee$. In turn, the effective superpotential \eqref{t-eff} modifies the D b.c. on the dual photon to $S'{}^\vee\big|_\pd = J_{\Upsilon'} = \log\Phi$. Since the monopole operator of $\CT'$ may be defined in the UV as $V'\sim\exp(S'{}^\vee)$, this induces a relation
\be  V' \sim \Phi  \ee
at the interface, which identifies the vev of the monopole operator of $\CT'$ with the vev of the free chiral of $\CT$.

We may similarly define a duality interface between $\CT$ and $\CT''$ ($U(1)_{-\frac12}$+a chiral). Now the simplest procedure is to ``factorize the identity'' for $\CT$ by coupling D b.c. on $x^\perp\leq 0$ to N b.c. on $x^\perp\geq 0$, and then dualize the left half to $\big[(\CN,\text{N})$+fermi$\big]$ for $\CT''$. We obtain:
\begin{itemize}
\item $\CT''$ on $x^\perp\leq 0$ with $(\CN,\text{N})$ b.c.
\item $\CT$ on $x^\perp \geq 0$ with N b.c.
\item a 2d Fermi multiplet $\Gamma$ of charge $(-1,-1,1)$ for $U(1)_\pd=U(1)_{\rm gauge}\times U(1)_x\times U(1)_R$, coupled to the bulk via a cubic superpotential $\int d\theta^+\, \Phi'' \Gamma \Phi$.
\end{itemize}
It is again straightforward to check that all anomalies cancel. The map of operators also proceeds the same way, mediated by $\Gamma$.

The collision between the $\CT''\big|\CT$ interface and the $\CT\big|\CT'$ interface now suggests a UV definition for a duality interface between $\CT''$ and $\CT'$. Since sandwiching $\CT$ between two N b.c. leaves behind an ordinary 2d chiral multiplet $\Phi$, we expect the $\CT''\big|\CT'$ duality interface to contain two 2d Fermi multiplets $\Gamma,\Gamma'$ and the 2d chiral $\Phi$, coupled to the bulk chirals $\Phi''$, $\Phi'$ on either side by a superpotential
\be \int d\theta^+\big( \Phi''\Gamma\Phi + \Phi\tilde\Gamma \Phi')\,. \ee

We emphasize that the duality interfaces constructed here are all \emph{uni-directional}. For example, the interface above between $\CT$ and $\CT'$ requires $\CT$ to sit on the half-space $x^\perp\leq 0$ and $\CT'$ to sit on $x^\perp\geq 0$, and not the other way around. 
The asymmetry comes from our choice in preserving 2d $\CN=(0,2)$ SUSY (rather than $(2,0)$), as well as the relative signs of the bulk Chern-Simons couplings --- much as in Section \ref{sec:signs}.
The general methods of Section \ref{sec:id} can be used, in principle, to construct duality interfaces going in the opposite directions. However, the construction seems to result in non-perturbative superpotential couplings at the interface, involving boundary monopole operators.%
\footnote{This is reminiscent of the 3d $\CN=4$ abelian duality interfaces defined in \cite{BDGH}. We indeed expect that the 3d $\CN=4$ interfaces can be reduced to 3d $\CN=2$ interfaces, which may occasionally involve nonperturbative couplings.}
We hope to pursue this further elsewhere.


\section{SQED and XYZ}
\label{sec:XYZ}

We may extend the techniques used to understand particle-vortex duality to produce dual boundary conditions and a duality interface for the pair of dual 3d $\CN=2$ theories
\be \overset{\displaystyle\text{SQED}}
{\text{$U(1)$ w/ 2 chirals of charge $\pm1$}}\qquad\leftrightarrow\qquad \overset{\displaystyle\text{XYZ model}}{\text{3 chirals, $W=X_{3d}Y_{3d}Z_{3d}$}}\,. \ee
This basic duality was introduced in \cite{AHISS}, and may be derived from abelian 3d $\CN=4$ mirror symmetry. One new ingredient is the bulk superpotential $W$ in the XYZ model. As discussed in Section \ref{sec:mx}, boundary conditions for the XYZ model may require extra boundary degrees of freedom to factorize $W$. We shall mainly consider boundary conditions are ``sufficiently Dirichlet,'' as in Section \ref{sec:XYZmf}, so that $W$ is automatically set to zero on the boundary. However, the duality interface will indeed involve a nontrivial factorization.

We use the following conventions. In SQED, the chirals, decomposed into $\CN=(0,2)$ multiplets, are denoted $(\Phi,\Psi)$, $(\tilde \Phi,\tilde \Psi)$. In XYZ the chirals are $(X,\Psi_X)$, $(Y,\Psi_Y)$, $(Z,\Psi_Z)$, and duality maps the meson $\Phi\tilde\Phi$ of SQED to $X$, and the monopole operators of SQED to $Y$ and $Z$.
The global symmetry group of SQED comprises a $U(1)_a$ axial symmetry, $U(1)_y$ topological symmetry, and the usual $U(1)_R$ symmetry, under which the various chiral fields have charges
\be \begin{array}{c|cc|ccc}
  & \Phi & \tilde\Phi & X & Y & Z \\\hline
 U(1)_{\rm gauge} & 1 & -1 & 0 & 0 & 0 \\
 U(1)_a & 1 & 1 & 2 & -1 & -1 \\
 U(1)_y & 0 & 0 & 0 & 1 & -1 \\
 U(1)_R & 0 & 0 & 0 & 1 & 1
\end{array} \ee
The bulk UV Chern-Simons levels for SQED just contain the coupling to the topological $U(1)_y$ symmetry, and may be encoded in the anomaly polynomial $2\,\mb\, \mb f \ \mb y$ (where $\mb y$ is the $U(1)_y$ field strength). This matches identically zero bulk CS terms for XYZ. This identification follows from a careful comparison of sphere partition functions, which are sensitive to background Chern-Simons terms (contact terms), see \emph{e.g.} \cite{DGG}.

\subsection{A fundamental duality and the interface}
\label{sec:XYZ-DDD}

The most interesting dual pair of boundary conditions --- and the one that we use to build the duality interface --- involves Dirichlet $(\CD)$ for the gauge multiplet in SQED and D for both chirals. We'll call this ($\CD$,D,D). We can derive its mirror in the XYZ model from a simple physical analysis!

The ($\CD$,D,D) b.c. in SQED should leave the two monopole operators unconstrained at the boundary, while setting the meson $\Phi\tilde\Phi$ to zero. Therefore, we expect that the dual b.c. in the XYZ model is Neumann for $Y$ and $Z$, and Dirichlet for $X$.%
\footnote{The same conclusion was reached in \cite{OkazakiYamaguchi}.} %
 We call this putative dual (D,N,N). Note that, a priori, it requires no further boundary d.o.f., since it sets $W|_\pd=0$. However, just as in particle-vortex duality:

- ($\CD$,D,D) b.c. for SQED has a boundary $U(1)_\pd$ flavor symmetry that's missing in XYZ;

- there is a mismatch in boundary 't Hooft anomalies between ($\CD$,D,D) and (D,N,N).
There is an easy way to fix both of these problems by adding an extra boundary Fermi multiplet on the XYZ side. Indeed, the difference in boundary 't Hooft anomalies (with $\mb f,\mb a,\mb y$ denoting the field strenghts for $U(1)_\pd,U(1)_a,U(1)_y$)
\be \begin{array}{l}
\CI_{\text{$\CD$,D,D}}=2\,\mb f\,\mb y - \frac12\mb r^2+ \frac12(\mb f+\mb a-\mb r)^2+\frac12(-\mb f+\mb a-\mb r)^2 \\[.2cm]
\CI_{\text{D,N,N}}=\frac12(2\mb a-\mb r)^2-\frac12(\mb y-\mb a)^2-\frac12(\mb y+\mb a)^2
\end{array}  \label{DDD-anom} \ee
is $\CI_{\text{$\CD$,D,D}}-\CI_{\text{D,N,N}}=(\mb f+\mb y)^2$, and is precisely made up by a 2d Fermi multiplet $\Gamma$ with charges $(-1,0,-1,0)$ for $U(1)_\pd\times U(1)_a\times U(1)_y\times U(1)_R$. This multiplet ``carries'' the $U(1)_\pd$ symmetry on the XYZ side. We surmise that
\be (\CD,\text{D,D})\qquad\leftrightarrow\qquad \text{(D,N,N) + Fermi $\Gamma$}\,. \label{DDD} \ee

This putative duality of boundary conditions immediately leads to a construction of the duality interface, following the logic of Section \ref{sec:id}. We may start with a factorization of the identity interface in SQED that couples the ($\CD$,D,D) b.c. on $x^\perp\leq 0$ to ($\CN$,N,N) on $x^\perp \geq 0$. Dualizing the left half-space leads to XYZ on $x^\perp\leq 0$ with (D,N,N)+$\Gamma$ b.c., coupled to SQED with ($\CN$,N,N) on $x^\perp \geq 0$. We might schematically, and more symmetrically, denote this as
\be \overset{\text{XYZ}}{\text{(D,N,N)}} \big| \Gamma \big| \overset{\text{SQED}}{(\CN,\text{N,N})}\,. \label{XYZ-int1} \ee
Here the $U(1)$ gauge symmetry on the right is used to gauge the $U(1)_\pd$ flavor symmetry of~$\Gamma$. In addition, there must be some boundary superpotential couplings! There is essentially a single possibility compatible with the global symmetries and with factorization of $W$; we propose
\be \label{W-XYZ} \int d\theta^+\big[ \Psi_X \Phi\tilde \Phi + Y \Gamma \Phi \big]\,,\qquad E_\Gamma = Z\tilde \Phi\,,\ee
where $\Gamma$ is given both J and E terms $J_\Gamma=Y\Phi$, $E_\Gamma=Z\tilde \Phi$. This could actually be deduced from the original coupling in the factorization of the identity interface and the map of operators across the duality. The coupling \eqref{W-XYZ} actually treats $\Phi,\tilde\Phi$ and $Y,Z$ symmetrically, though it does not look so, since we may always replace $\Gamma$ with its ``T-dual'' $\Gamma^\dagger$ whose E and J terms are swapped (\emph{cf.} Appendix \ref{app:T}). To check that \eqref{W-XYZ} provides a matrix factorization of the bulk superpotential on the XYZ side, we first observe that the $\Psi_X \Phi\tilde\Phi$ coupling sets
\be X = \Phi \tilde\Phi \label{Xphi} \ee
at the interface; then
\be J_\Gamma E_\Gamma = (Y\Phi)(Z\tilde\Phi) \overset{\eqref{Xphi}}= XYZ\, \ee
as required.

The proposed interface gives the expected map of operators. The relation between $X$ and the meson $\Phi\tilde\Phi$ is explicit in \eqref{Xphi}. In addition, the E and J terms for $\Gamma$ modify the N b.c. for Y, Z, $\Phi$, $\tilde \Phi$ as in Section \ref{sec:2dmatter}:
\be \Psi_Y = \Gamma\Phi\,,\quad \Psi_Z = \Gamma \tilde \Phi\,;\qquad
   Y\Gamma = \Psi\,,\quad Z\Gamma=\tilde \Psi\,. \ee
Finally, the relations involving monopole operators may be deduced from quantum corrections to the interface superpotential (just like in \eqref{t-eff} for particle-vortex duality). We may give a vev to either $Y$ or $Z$ (but not both, due to the bulk $X_{3d}Y_{3d}Z_{3d}$ superpotential) and make $\Gamma$ massive at the boundary. Integrating $\Gamma$ out at one-loop generates interface superpotentials
\be  \text{$Y\neq 0$}\;\Rightarrow\; \int d\theta^+\Upsilon \log Y\,,\qquad   \text{$Z\neq 0$}\;\Rightarrow\; -\int d\theta^+\Upsilon \log Z\,,\ee
where $\Upsilon$ is the field strength in SQED, restricted to the interface. These
 modify the D b.c. for the monopoles $V_\pm$ of SQED to
\be V_+ \sim Y\,,\qquad V_-\sim Z\,. \label{XYZ-UZ}\ee

Unlike the case of particle-vortex dualities, the duality interface between SQED and XYZ is actually symmetric. That is, we can sensibly define an interface in the opposite direction
\be \overset{\text{SQED}}{\text{($\CN$,N,N)}} \big| \Gamma \big| \overset{\text{XYZ}}{(\text{D,N,N})}\, \label{XYZ-int2} \ee
in essentially the same way as above, but now with a 2d Fermi multiplet $\Gamma$ of charges $(-1,0,1,0)$ under $U(1)_{\rm gauge}\times U(1)_a\times U(1)_y\times U(1)_R$.%
\footnote{This ``reverse'' interface may be derived systematically by starting with a factorization of the identity interface in XYZ, of the form $\text{(N,D,D)}|\text{(D,N,N)}$, then using a duality from Section \ref{sec:XYZ-other} to dualize the left side to SQED with $(\CN,\text{N,N})$ b.c. coupled to a boundary Fermi $\Gamma$.} %
The reader may verify that all anomalies cancel. The important difference between the SQED-XYZ duality and the particle-vortex duality of Section \ref{sec:tet} is that the former has vanishing bulk Chern-Simons levels for dynamical gauge symmetry. This makes the SQED-XYZ duality more symmetric with respect to left vs. right boundary conditions.

Finally, we may check the duality \eqref{DDD} with a half-index computation. For SQED, we compute the ``effective'' boundary CS levels from the anomaly \eqref{DDD-anom}, which simplifies to $\mb f^2+2\,\mb f\,\mb y+(\mb a-\mb r)^2-\frac12\mb r^2$. Following the rules of Section \ref{sec:II-D}, this leads to a half-index
\begin{align} \II^{\,\rm SQED}_{(\CD,\text{D,D})}(s,a,y;q) &= \frac{1}{(q)_\infty}\sum_{n\in \Z} q^{\frac{n^2}{2}}s^ny^n \underbrace{\II_{\rm D}(q^nsa;q)}_{(q^{1-n}s^{-1}a^{-1};q)_\infty} \underbrace{\II_{\rm D}(q^{-n}s^{-1}a;q)}_{(q^{1+n}sa^{-1};q)_\infty}\,, \label{IIDDD}
\end{align}
where $s,a,y$ are the $U(1)_\pd,U(1)_a,U(1)_y$ fugacities. This turns out to equal
\begin{align} \II^{\,\rm XYZ}_{(\text{D,N,N})+\Gamma}(s,a,y;q) &= (-q^{\frac12}sy)_\infty(-q^{\frac12}s^{-1}y^{-1})_\infty \frac{(qa^{-2};q)_\infty}{(-q^{\frac12}a^{-1}y;q)_\infty(-q^{\frac12}a^{-1}y^{-1};q)_\infty} \label{IIDNN} \\
&= \text{F}\Big(\frac{-q^{\frac12}}{sy}\Big) \II_{\rm D}(a^2;q) \II_{\rm N}\Big(\frac{-q^{\frac12}y}{a};q\Big)\II_{\rm N}\Big(\frac{-q^{\frac12}}{ay};q\Big)\,\notag
\end{align}
as desired.

Just as in \eqref{T'DD}, the equivalence of \eqref{IIDDD} and \eqref{IIDNN} may be established by showing that both expressions obey the same first-order difference equations in $s$, $a$, and $y$, and that they are equal at a particular point. The difference equations, which may all be interpreted as identities for line operators (Section \ref{sec:diff}), are
\be \begin{array}{c} \II(q^{-1}s,a,y;q) = q^{-\frac12}sy \II(q^{-1}s,a,y;q)\,,\quad
\II(q^{-\frac12}s,q^{\frac12}a,q^{\frac12}y;q) = \frac{1-a}{1+q^{-\frac12}a^{-1}y^{-1}}\II(s,a,y;q)\,,\\
  \II(q^{-1}s,a,qy;q) = \frac{1+q^{\frac12}a^{-1}y}{1+q^{-\frac12}a^{-1}y^{-1}}\II(s,a,y;q)
\end{array}  \label{diffXYZ}
\ee
and follow either directly from \eqref{IIDNN} or by simple manipulations inside the sum of \eqref{IIDDD}. A convenient evaluation point is $s=a=y=1$, where both  \eqref{IIDDD} and \eqref{IIDNN} reduce fairly trivially to $(q)_\infty$.
Alternatively, by sending $s/a\to 0$ while holding $sa$ and $sy$ fixed, we may reduce to the identity in \eqref{T'DD}--\eqref{T'DD-id}. This corresponds physically to turning on a large real mass to integrate out $\tilde \Phi$ in SQED and $X,Z$ in XYZ, recovering the basic particle-vortex duality.

\subsection{Other dual boundary conditions}
\label{sec:XYZ-other}

There are many other dual pairs of boundary conditions for SQED and XYZ. As we have already mentioned, this pair of 3d theories is even better behaved than the particle-vortex duality of Section \ref{sec:tet}, due to vanishing bulk Chern-Simons levels. Thus both Dirichlet and Neumann b.c. are available for the gauge multiplet of SQED, with interesting duals on the XYZ side. We summarize the proposed dualities in Table \ref{tab:XYZ}.%
\footnote{As mentioned in the introduction, some of the dualities involving Neumann b.c. for the gauge multiplet of SQED have appeared previously in \cite{GGP-walls}.} %
For each of these pairs of boundary conditions, the boundary anomalies match up perfectly, and there is an associated half-index identity. We discuss a few of the more interesting cases below. Various vortex/Wilson lines can also be incorporated in a fairly straightforward way, following the models of Section \ref{sec:tet}.

\begin{table}[htb]
\centering
$\begin{array}{c@{\quad}|@{\quad}c}
\text{SQED} & \text{XYZ} \\\hline
 (\CD,\text{D,D}) & \text{(D,N,N)+$\Gamma_{1/(sy)}$} \\
 (\CD,\text{D,D$_c$}) & \text{(D,N,D)} \\
 (\CD,\text{D$_c$,D}) & \text{(D,D,N)} \\
 (\CD,\text{D$_c$,D$_c$}) & \text{(D$_c$,N,D)} \simeq \text{(D$_c$,D,N)} \\
 (\CD,\text{N,D}) & \text{(D,N,N)}+\Gamma_{1/(sy)}+ C_{as};\;J_\Gamma=CY \\
 (\CD,\text{D,N}) & \text{(D,N,N)}+\Gamma_{1/(sy)}+\tilde C_{a/s};\; E_\Gamma=\tilde CZ \\
 (\CD,\text{N,N}) & \begin{array}{l} \text{(D,N,N)}+\Gamma_{1/(sy)}+ C_{as}+\tilde C_{a/s}; \\
   \hspace{.3in} J_{\Psi_X}=C\tilde C,J_\Gamma=CY,E_\Gamma=\tilde CZ \end{array} \\
 (\CN,\text{N,N})+\Gamma_{s/y} & \text{(N,D,D)} \\
 (\CN,\text{N,N})+\Gamma_{s/y}+\Gamma'_{-q^{1/2}/a^2};\; J_{\Gamma'}=\Phi\tilde\Phi & \text{(D,D,D)} \\
 (\CN,\text{N,D}) & \text{(N,D$_c$,D)}\simeq \text{(D,D$_c$,N)} \\
 (\CN,\text{D,N}) & \text{(N,D,D$_c$)}\simeq \text{(D,N,D$_c$)}
\end{array}$
\caption{Elementary pairs of dual boundary conditions for SQED and XYZ. Additional boundary Fermi and chiral multiplets are denoted $\Gamma$ and $C,\tilde C$, respectively, with their charge under $U(1)_{\rm gauge}$ (or $U(1)_\pd$), $U(1)_a$ and $U(1)_y$ encoded by fugacities $s$, $a$, and $y$ in the subscripts.}
\label{tab:XYZ}
\end{table}

\subsubsection*{$(\CD,\text{D$_c$,D$_c$})\;\leftrightarrow\; \text{(D$_c$,N,D) or (D$_c$,D,N)}$}

The new feature here is that there seem to be two distinct UV boundary conditions for the XYZ model that are IR equivalent. This becomes less surprising upon closer inspection. The (D$_c$,N,D) b.c. sets $X|_\pd=c\neq 0$ and $Z|_\pd=0$ while seemingly leaving $Z$ unconstrained. However, the bulk F-term $\pd W/\pd Z= XY$ (identified as the $\CN=(0,2)$ E-term for $\Psi_Z$) must also vanish in a supersymmetric vacuum. Since at the boundary the F-term is $XY|_\pd = c Y|_\pd$, we see that $Y|_\pd$ is also set to zero in the IR. Thus, the (D$_c$,N,D) b.c. seems to look identical to (D$_c$,D,N).

We can verify that boundary 't Hooft anomalies match. We compute:
\be \begin{array}{c@{\,:\quad}c}
\text{(D$_c$,N,D)} & \frac12(2\mb a-\mb r)^2-\frac12(\mb y-\mb a)^2+\frac12(\mb y+\mb a)^2\big|_{2\mb a=0} = \frac12\mb r^2\,, \\
\text{(D$_c$,D,N)} & \frac12(2\mb a-\mb r)^2+\frac12(\mb y-\mb a)^2-\frac12(\mb y+\mb a)^2\big|_{2\mb a=0} = \frac12\mb r^2\,,
\end{array}
\ee
which agree.
Moreover, these match the boundary anomaly of ($\CD$,D$_c$,D$_c$) b.c. for SQED:
\be\textstyle \text{($\CD$,D$_c$,D$_c$)}\,:\quad 2\,\mb f\,\mb y-\frac12\mb r^2+\frac12(\mb f+\mb a-\mb r)^2+\frac12(-\mb f+\mb a-\mb r)^2\big|_{\mb f=\mb a=0} = \tfrac12\mb r^2\,. \ee
Curiously, the anomaly for (D$_c$,D,D), which might also have been expected to be IR equivalent to these b.c., equals $\frac12(2\mb a-\mb r)^2+\frac12(\mb y-\mb a)^2+\frac12(\mb y+\mb a)^2\big|_{2\mb a=0}=\mb y^2+\frac12\mb r^2$, and does not agree.

The half-index identity corresponding to the putative duality of boundary conditions is obtained by taking the identity \eqref{IIDDD}-\eqref{IIDNN} of Section \ref{sec:XYZ-DDD} and setting to one the fugacities for both of the chirals in SQED, which are given D$_c$ b.c., namely $as=1$ and $a/s=1$, or simply $a=s=1$. This results fairly trivially in a half-index on either side equal to $(q)_\infty$, independent of $y$. 

A stronger argument for the duality can be made by deforming the $(\CD,\text{D,D})\leftrightarrow\text{(D,N,N)+$\Gamma$}$ duality of Section \ref{sec:XYZ-DDD} by boundary superpotentials. To reach $(\CD,\text{D$_c$,D$_c$})$ in SQED, we add a superpotential
\be \int d\theta^+(c\Psi+c'\tilde\Psi)\,. \ee
If we turn these terms on one at a time, we may view the $c\Psi$ coupling as modifying the b.c. to $\Phi|_\pd =c$, and then the $c'\tilde\Psi$ coupling as subsequently modifying the boundary vev of the \emph{meson} operator to $\Phi\tilde \Phi|_\pd = cc'$. Correspondingly, in SQED we add superpotentials for the dual operators: first we add $c\Gamma Y$ (since $\Gamma Y$ is dual to $\Psi|_\pd$), and then we add $cc'\Psi_X$ (since $X$ is dual to the meson). The superpotential
\be \int d\theta^+(c\Gamma Y+cc'\Psi_X) \ee
has the effect of ``flipping'' the b.c. on $Y$ from N to D, and deforming the b.c. on $X$ to $X|_\pd=cc'$, which is precisely (D$_c$,D,N). Alternatively, had we turned on the superpotential terms in the opposite order, we would have obtained (D$_c$,N,D). 

\subsubsection*{($\CD$,N,N) $\leftrightarrow$ (D,N,N) with a Fermi and two chirals}

The boundary condition here for XYZ is a true matrix factorization, very similar to the one that entered the duality interface. Indeed, this duality can neatly be obtained by colliding the interface \eqref{XYZ-int1} with ($\CD$,N,N) b.c. for SQED. The ``sandwich'' between ($\CN$,N,N) (on the right side of the interface) and ($\CD$,N,N) kills the $U(1)$ gauge symmetry, and leaves behind two 2d chirals $C,\tilde C$, coming from the reductions of $\Phi,\tilde \Phi$ on the segment. The interface itself contributes the 2d Fermi $\Gamma$, and we are left with (D,N,N) b.c. for XYZ coupled to $\Gamma,C,\tilde C$ via the superpotential
\be \int d\theta^+\big[ \Psi_X C\tilde C+\Gamma CY\big]\,,\qquad E_\Gamma=\tilde CZ\,.\ee
As in the interface, this sets $X|_\pd = C\tilde C$, so that $E_\Gamma J_\Gamma = C\tilde CYZ = XYZ =W$.

The relevant index identity is
\begin{align} &\frac{1}{(q)_\infty} \sum_{n\in \Z} q^{-\frac{n^2}{2}}s^{-n}y^n \II_{\rm N}(q^nsa;q)\II_{\rm N}(q^{-n}s^{-1}a;q) \\
& \hspace{.5in}=  \text{F}\Big(\frac{-q^{\frac12}}{sy}\Big) \text{C}(as)\text{C}(a/s) \II_{\rm D}(a^2;q) \II_{\rm N}\Big(\frac{-q^{\frac12}y}{a};q\Big)\II_{\rm N}\Big(\frac{-q^{\frac12}}{ay};q\Big)\,, \notag
\end{align}
and follows easily from \eqref{IIDDD}-\eqref{IIDNN} by rewriting  $\II_{\rm N}(q^nsa;q)\II_{\rm N}(q^{-n}s^{-1}a;q)$ \\ as $\text{C}(as)\text{C}(a/s) q^{n^2}s^{2n}\II_{\rm D}(q^nsa;q)\II_{\rm D}(q^{-n}s^{-1}a;q)$ on the LHS.

\subsubsection*{($\CN$,N,N)+Fermi $\leftrightarrow$ (N,D,D)}

The ($\CN$,N,N) boundary condition in SQED kills the monopole operators while leaving the meson unconstrained, so we would naively expect it to be dual to (N,D,D) in XYZ. In fact, cancellation of a gauge anomaly on the SQED side requires the introduction of an additional boundary Fermi multiplet $\Gamma$. Together with this modification, the duality seems to hold as expected.

Let us consider the anomalies explicitly. On the SQED side, we have
\be \tfrac12\mb r^2 +2\,\mb f\,\mb y-\tfrac12(\mb f+\mb a-\mb r)^2-\tfrac12(-\mb f+\mb a-\mb r)^2
= -\mb f^2+2\mb f\,\mb y-(\mb a-\mb r)^2+\tfrac12\mb r^2\,.
\ee
Adding a 2d Fermi multiplet $\Gamma$ of charge $(-1,0,1,0)$ for $U(1)_{\rm gauge}\times U(1)_a\times U(1)_y\times U(1)_R$ contributes $(\mb f-\mb y)^2$, modifying the anomaly to $\mb y^2 -(\mb a-\mb r)^2+\tfrac12\mb r^2$, and in particular canceling the terms involving the dynamical field strength $\mb f$. We may compare this with (N,D,D) b.c. for SQED, which has anomaly
\be -\tfrac12(2\mb a-\mb r)^2+\tfrac12(\mb y-\mb a)^2+\tfrac12(\mb y+\mb a)^2 = \mb y^2 -(\mb a-\mb r)^2+\tfrac12\mb r^2\,, \ee
matching perfectly.

The relevant half-index identity is
\begin{align} & (q)_\infty \oint \frac{ds}{2\pi is} \II_{\rm N}(as;q)\II_{\rm N}(a/s;q)\text{F}(-q^{\frac12}y/s;q) 
  = (q)_\infty \oint \frac{ds}{2\pi is} \frac{\text{F}(-q^{\frac12} y/s;q)}{(as;q)_\infty(a/s;q)_\infty} \notag \\
 &\hspace{.5in}\overset{\text{residues}}{=} \sum_{n\geq 0} (-1)^nq^{\frac12n(n+1)}\frac{\text{F}(-q^{\frac12-n}y/a;q)}{(q)_n(q^na^2;q)_\infty} \notag \\
 &\hspace{.5in} = \frac{\text{F}(-q^{\frac12}y/a;q)}{(a^2;q)_\infty} \sum_{n\geq 0} (-q^{\frac12}y/a)^n\frac{(a^2;q)_n}{(q)_n} \notag\\ 
 &\hspace{.5in}\overset{\text{q-binomial}}{=} \frac{\text{F}(-q^{\frac12}y/a;q)}{(a^2;q)_\infty} \frac{(-q^{\frac12}ay;q)_\infty}{(-q^{\frac12} y/a;q)_\infty} \notag\\
 &\hspace{.5in} = \frac{(-q^{\frac12}a/y;q)_\infty (-q^{\frac12}ay;q)_\infty}{(a^2;q)_\infty} = \II_{\rm N}(a^2;q)\II_{\rm D}(-q^{\frac12}y/a;q)\II_{\rm D}(-q^{\frac12}/(ay);q)\,.
\end{align}
The same identity appeared when studying holomorphic block dualities \cite{BDP-blocks}, later interpreted via half-indices in \cite{GGP-walls}.

\subsubsection*{($\CN$,N,N)+2 fermis $\leftrightarrow$ (D,D,D)}

This is a simple flip of the preceding boundary condition: an additional 2d Fermi multiplet $\Gamma'$ is used to flip the meson in SQED, and to flip N to D b.c. for $X$ in XYZ.

\subsubsection*{($\CN$,D,N) $\leftrightarrow$ (N,D,D$_c$) or (D,N,D$_c$)}

Finally, we take a look at another duality where the XYZ boundary condition has multiple UV descriptions. In SQED, the ($\CN$,D,N) b.c. does not have a gauge anomaly, but it does have a mixed anomaly: we compute a boundary polynomial
\be \tfrac12\mb r^2+2\,\mb f\,\mb y+\tfrac12(\mb f+\mb a-\mb r)^2-\tfrac12(-\mb f+\mb a-\mb r)^2
 = \tfrac12\mb r^2+2\,\mb f(\mb y+\mb a-\mb r)\,.
\ee
This breaks the symmetry with field strength $\mb y+\mb a-\mb r$, which is precisely the symmetry under which the chiral $Z$ in the XYZ model is charged. Thus, in XYZ, we expect a D$_c$ b.c. for $Z$. Moreover, ($\CN$,D,N) sets the meson to zero so we expect its dual to give D b.c. to $X$. Comparing 't Hooft anomalies then suggests that the dual b.c. should be (D,N,D$_c$).

As before, the BPS equations on the (D,N,D$_c$) b.c. actually set $Y|_\pd=0$ as well in a supersymmetric vacuum. This fits the proposed duality, since $\CN$ b.c. in SQED certainly kills both monopole operators, including the one dual to $Y$. We suspect that the (D,N,D$_c$) b.c. can equivalently be described as (N,D,D$_c$).

The half-index identity is fairly simple:
\begin{align} &(q)_\infty \oint\frac{ds}{2\pi is} \II_{\rm D}(as;q)\II_{\rm N}(a/s;q) = (q)_\infty \oint \frac{ds}{2\pi is} \frac{(q/(as);q)_\infty}{(a/s;q)_\infty} \notag\\
&\hspace{.5in}\overset{\text{residues}}{=} \sum_{n\geq 0} q^{\frac{n^2}{2}}(-q^{\frac12})^n   \frac{(q^{1-n}a^{-2};q)_\infty}{(q)_n} \notag \\
&\hspace{.5in}= (qa^{-2};q)_\infty \sum_{n\geq 0} (qa^{-2})^n\frac{(a^2;q)_n}{(q)_n} \notag \\
&\hspace{.5in}\overset{\text{q-binomial}}= (q)_\infty  = \II_{\rm N}(a^2;q)\II_{\rm D}(-q^{\frac12}y/a;q)\II_{\rm D}(-q^{\frac12}/(ay);q)\big|_{y=-q^{\frac12}/a}\,.
\end{align}

\subsubsection{Gauging $\CD$ and 2d dualities}
It is again instructive to take $\CD$-type boundary conditions and gauge the 2d $U(1)_\pd$ symmetry, comparing the 
result with proposed mirrors of $\CN$-type boundary conditions.

The richest example is to gauge $(\CD,\text{N,N})+\Gamma_{s/y}$ in order to obtain $(\CN,\text{N,N})+\Gamma_{s/y}$.
On the mirror side, we get $(D,\text{N,N})$ coupled to an intricate boundary theory: a 2d $U(1)_{s}$ gauge theory with two Fermi multiplets $\Gamma_{s/y}$ and $\Gamma_{1/(sy)}$ and two chiral multiplets $C_{as}$ and $\tilde C_{a/s}$. 
From Table 1, we expected this mirror to be equivalent to (N,D,D). 
This is only possible if the 2d boundary gauge theory flows in the IR to a combination of a chiral multiplet and two Fermi multiplets.

The picture is supported by a 2d index calculation, which involves picking a single pole from the positively charged $C_{as}$.
Indeed, if we add an extra Fermi multiplet to analyze $(\CN,\text{N,N})+\Gamma_{s/y}+\Gamma'_{-q^{1/2}/a^2}$,
the RG flow of the 2d theory to the two Fermi multiplets that flip (D,N,N) to (D,D,D) is the very simplest example of 
$(0,2)$ triality \cite[Fig. 2]{GGP-triality}, with $N_1=N_2=1$ and $N_3=0$. 

\subsection{Generalization: $U(N)$ SQCD with $N_f=N$}
\label{sec:detYZ}

The basic SQED\,$\leftrightarrow$\,XYZ duality has a natural generalization, in which SQED is replaced by $U(N)$ SQCD with $N_f=N$ fundamental and antifundamental chirals (quarks and antiquarks), and the dual theory is another ``Landau-Ginzburg model'' with a cubic superpotential capturing the low-energy dynamics of SQCD \cite{AHISS, Aharony}. Specifically, the dual theory contains chirals $Y_{3d},Z_{3d}$ that match the monopole operators of SQCD,  an $N\times N$ matrix $M_{3d}$ matching the mesons of SQCD, and a superpotential
\be W = \det(M_{3d})Y_{3d}Z_{3d}\,. \ee
We'll call this the ``detYZ'' model.
Just as in SQED\,$\leftrightarrow$\,XYZ, the bulk UV Chern-Simons levels are identically zero on both sides. 
The dual boundary conditions for SQCD and detYZ end up following an identical pattern as those of SQED\,$\leftrightarrow$\,XYZ, and we describe a few of them here.

Let us denote the quarks of SQED, decomposed into $\CN=(0,2)$ multiplets, as $(Q^i_a,\Psi_i^a)$, $(\tilde Q_i^{\bar a},\tilde\Psi^i_{\bar a})$, where $i,a,\bar a$ are indices for $U(N)_{\rm gauge}$ and $SU(N),\wt{SU(N)}$ flavor symmetries, respectively. Similarly, we decompose the chirals of detYZ into $(M_a{}^{\bar a},\mu^a{}_{\bar a})$, $(Y,\Psi_Y)$, and $(Z,\Psi_Z)$. The charges of the $(0,2)$ chiral halves are
\be \begin{array}{c|c|ccccc}
 & U(N)_{\rm gauge} & SU(N) & \wt{SU(N)} & U(1)_a & U(1)_y & U(1)_R \\\hline
Q & \mb N & \mb N & \mb 1 & 1 & 0 & 0 \\
\tilde Q & \ol{\mb N} & \mb 1 & \ol{\mb N} & 1 & 0 & 0 \\\hline
M & 0 & \mb N & \ol{\mb N} & 2 & 0 & 0 \\
Y & 0 & 0 &0 & -N & 1 & 1 \\
Z & 0 & 0 & 0 & -N & -1 & 1 \end{array}
\ee

As before, there are two basic pairs of dual boundary conditions that can be used to construct duality interfaces going in both directions. In compact notation:
\be \begin{array}{c@{\quad\leftrightarrow\quad}c}
 \text{($\CD$,D,D)}  & \text{(D,N,N)}+\Gamma \\[.2cm]
 \text{($\CN$,N,N)}+\Gamma' & \text{(N,D,D)}\,,
\end{array} \label{detBB}\ee
where $\Gamma$ and $\Gamma'$ are 2d Fermi multiplets in representations
\be \begin{array}{c|c|ccccc}
 & \text{$U(N)$ or $U(N)_\pd$} & SU(N) & \wt{SU(N)} & U(1)_a & U(1)_y & U(1)_R \\\hline
\Gamma & \text{det}^{-1} & \mb 1 & \mb 1 & 0 & -1 & 0 \\
\Gamma' & \text{det} & \mb 1 & \mb 1 & 0 & -1 & 0 
 \end{array}
\ee

Let us check that the boundary anomalies match for \eqref{detBB}. First consider ($\CD$,D,D), \emph{i.e.} Dirichlet for the $U(N)$ gauge multiplet as well as the quarks and anti-quarks of SQED. Following \eqref{Un-anom} for the computation of $U(N)$ anomalies, we find
\begin{align} \CI_\text{($\CD$,D,D)} &\;=\; -N\text{Tr}(\mb s^2)+(\text{Tr}\,\mb s)^2-\tfrac12N^2\mb r^2 + 2(\text{Tr}\,\mb s)\mb y \qquad \text{(gauginos, FI)} \\
 &\quad +\tfrac12(N\text{Tr}(\mb x^2)+N\text{Tr}(\mb s^2) + 2N(\text{Tr}\,\mb s)(\mb a-\mb r)+N^2(\mb a-\mb r)^2) \qquad (Q) \notag \\
 &\quad+\tfrac12(N\text{Tr}(\tilde{\mb x}^2)+N\text{Tr}(\mb s^2) - 2N(\text{Tr}\,\mb s)(\mb a-\mb r)+N^2(\mb a-\mb r)^2)\qquad (\tilde Q) \notag \\[.2cm]
 &\;=\; (\text{Tr}\,\mb s)^2+2(\text{Tr}\,\mb s)\mb y +\tfrac12N\big[\text{Tr}(\mb x^2)+\text{Tr}(\tilde{\mb x}^2)\big]+N^2(\mb a-\mb r)^2-\tfrac12N^2\mb r^2 \notag
\end{align}
where, in order to match fugacities in the index, we have used $\mb s$ to denote the field strength of $U(N)_\pd$, and $\mb x,\tilde{\mb x}$ to denote the field strengths of the $SU(N),\wt{SU(N)}$ flavor symmetry. On the dual side, the (D,N,N) boundary condition (D b.c. for $M$, N b.c. for $Y,Z$) has an anomaly
\begin{align} \CI_\text{(D,N,N)} \;=\;& \tfrac12(N\text{Tr}\,\mb x+N\text{Tr}\,\tilde{\mb x}+N^2(2\mb a-\mb r)^2) \qquad (M) \\
 & -\tfrac12(-N\mb a+\mb y)^2-\tfrac12(-N\mb a-\mb y)^2 \qquad (Y,Z)
 \notag
\end{align}
The difference in these two quantities,
\be  \CI_\text{($\CD$,D,D)} = \CI_\text{(D,N,N)} + (\text{Tr}\,\mb s+\mb y)^2\,,\ee
is just right to match the additional contribution of the Fermi multiplet $\Gamma$ on the detYZ side. Similarly, the anomalies for the other pair of boundary conditions are $\CI_\text{($\CN$,N,N)}=- \CI_\text{($\CD$,D,D)}+4(\text{Tr}\,\mb s)\mb y$ and $\CI_\text{(N,D,D)}=-\CI_\text{(D,N,N)}$, so
\be \CI_\text{($\CN$,N,N)} + (\text{Tr}\,\mb s-\mb y)^2  = \CI_\text{(N,D,D)}\,, \ee
with the difference made up for by the $\Gamma'$ Fermi multiplet.
In particular, the anomaly of $\Gamma'$ is just right to cancel the gauge and mixed gauge-topological anomaly for $\CN$ b.c. in SQCD.

We may also check these proposed dual boundary conditions by a half-index computation.
Following the rules of Section \ref{sec:index}, we compute
\begin{align} \label{ind-SQCD}
\hspace{-.2in} \II_{(\text{$\CD$,D,D})} = \frac{1}{(q)_\infty^N}\sum_{m\in \Z^N} \frac{q^{\frac12 m\cdot m} s^m y^{\sum_i m_i}}
   {\prod_{i\neq j}(q^{1+m_i-m_j}s_i/s_j;q)_\infty} \prod_{i,\alpha=1}^{N} \II_{\rm D}\Big(q^{m_i}ax_\alpha s_i;q\Big)\II_{\rm D}\Big(\frac{a}{q^{m_i}\tilde x_\alpha s_i};q\Big)\,,
\end{align}
\begin{align} \label{ind-detYZ}
\hspace{-.4in} \II_{(\text{D,N,N})+\Gamma} = \text{F}\Big(\frac{-q^{\frac12}y}{\textstyle \prod_i s_i};q\Big) \II_{\rm N}(-q^{\frac12}a^{-N}y;q) \II_{\rm N}(-q^{\frac12}a^{-N}y^{-1};q) \prod_{\alpha,\beta=1}^N \II_{\rm D}\Big(\frac{a^2x_\alpha}{\tilde x_\beta};q\Big)\,,
\end{align}
where $s=(s_1,...,s_N)$ are the $U(N)_\pd$ fugacities, $m=(m_1,...,m_N)\in \Z^N\simeq \text{cochar}(U(N))$ are the monopole charges, $a,y$ are the axial and topological fugacities, and $x=(x_1,...,x_N)$, $\tilde x=(\tilde x_1,...,\tilde x_N)$ are the $SU(N),\wt{SU(N)}$ fugacities, constrained to satisfy $\prod_\alpha x_\alpha=\prod_\beta \tilde x_\beta=1$.
We have checked these formulas for a variety of $N$ in \texttt{Mathematica} up to order $q^{10}$.
\footnote{It seems feasible that the equality $\II_{(\text{$\CD$,D,D})}=\II_{(\text{D,N,N})+\Gamma}$ could be established in general by comparing difference equations, much like \eqref{diffXYZ}. This approach is promising because the index \eqref{ind-detYZ} is simply a product of $q$-Pochhammer symbols, which obey simple first-order equations.}

By following the general prescription of Section \ref{sec:id}, we may now use \eqref{detBB} to construct duality interfaces. They have the same form as in SQED$\;\leftrightarrow\;$XYZ. In one direction,
\be  \overset{\text{detYZ}}{\text{(D,N,N)}} \big| \Gamma \big| \overset{\text{SQCD}}{(\CN,\text{N,N})}\,, \label{SQCD-int1} \ee
with an interface superpotential $\int d\theta^+\big[Y \Gamma \text{det}(Q)+ Q\mu\tilde Q\big]
$ and $E_\Gamma = Z\text{det}(\tilde Q)$. Note that at the interface this sets $M=Q\tilde Q$ and provides a matrix factorization of the bulk superpotential $E_\Gamma J_\Gamma = YZ\det(Q)\det(\tilde Q) = \det(M)YZ$.
In the other direction,
\be \overset{\text{SQCD}}{(\CN,\text{N,N})}   \big| \Gamma' \big|   \overset{\text{detYZ}}{\text{(D,N,N)}}\,, \label{SQCD-int2} \ee
with superpotential $\int d\theta^+\big[Y \Gamma' \text{det}(\tilde Q)+ Q\mu\tilde Q\big]
$ and $E_{\Gamma'} = Z\text{det}(Q)$.
The Fermi muliplets $\Gamma$ and $\Gamma'$ mediate the map of bulk operators across the duality interface, essentially the same way as in SQED$\;\leftrightarrow\;$XYZ. In particular, giving a vev to $Y$ or $Z$ generates a superpotential $\int d\theta^+ (\Tr\Upsilon)\log Y$ or $-\int d\theta^+ (\Tr\Upsilon)\log Z$, generalizing \eqref{XYZ-UZ}, which sets the monopole operators of SQCD equal to $Y$ and $Z$.

Simple modifications of \eqref{detBB} and/or collisions with the interfaces may be used to construct many, many other dual pairs of boundary conditions, analogous to all of Table~\ref{tab:XYZ}.

It is also instructive to take $\CD$-type boundary conditions and gauge the 2d $U(1)_\pd$, comparing the 
result with proposed mirrors of $\CN$-type boundary conditions. Again, we can start from $(\CD,\text{D,D})$,
add fundamental and anti-fundamental chiral fields to flip to $(\CD,\text{N,N})$ and Fermi multiplet $\Gamma'$.

On the mirror side we get $(D,N,N)$ coupled to a 2d $U(N)_\pd$ gauge theory with fundamental and anti-fundamental chiral fields
and two Fermi multiplets in the determinant representation. Consistency of our proposal requires the 2d theory to flow 
to a bi-fundamental chiral and two Fermi multiplets to flip $(D,N,N)$ to $(N,D,D)$. 

This is reasonable. Indeed, if we add an extra bi-fundamental Fermi multiplet,
the RG flow of the 2d theory to the two Fermi multiplets which flip $(D,N,N)$ to $(D,D,D)$ is a simple example of 
$(0,2)$ triality \cite{GGP-triality}, with $N_1=N_2=N$ and $N_3=0$.


\section{Level-rank dualities}
\label{sec:LR}

So far we have discussed pairs of 3d $\CN=2$ dual theories with nontrivial gauge degrees of freedom on one side only, but potentially interesting matter content on both sides. We now turn to the opposite scenario: gauge groups on both sides, and trivial matter. We investigate boundary conditions in 3d $\CN=2$ level-rank duality.

Level-rank dualities comprise a familiar set of tractable examples that have their origins in equivalences of two-dimensional chiral algebras. Viewing the associated conformal field theories as the edge modes of three dimensional topological theories, the level-rank duality ascends to a duality of Chern-Simons theories. The classic non-supersymmetric dualities~\cite{NS-LR1, NS-LR2, NT-LR, CLZ-LR, NS-LR3} (and more modern extensions, \emph{e.g.}  \cite{Aharony, HsinSeiberg-LR}) readily admit $\mathcal{N}=2$ supersymmetric completions \cite{GiveonKutasov, BCC-Seiberg}.
For instance, one basic equivalence in 3d $\mathcal{N}=2$ Chern-Simons theory is
\be U(N)_{k+N} \;\leftrightarrow\; U(k)_{-k-N}\,. \ee
Another is $SU(N)_{k+N}\;\leftrightarrow\; U(k)_{-k-N,-N}$, where the notation on the left indicates a different level for the $U(1)$ factor, discussed at the end of Section \ref{sec:LR-D}.

These level-rank dualities follow from several dualities we will meet later with additional matter, such as Aharony, Giveon-Kutasov, or other Seiberg-like dualities, either by setting $N_f=0$ or by adding masses to the matter multiplets and integrating them out until one obtains a pure $\CN=2$ gauge theory.

The supersymmetric dualities are related to non-supersymmetric ones in a straightforward way: by integrating out (or in!) massive vectormultiplet scalars and gauginos, and shifting the Chern-Simons level due to the gauginos. The shift is always controlled by the adjoint anomaly.
If $G$ is a simple group, an $\CN=2$ $G_\kappa$ Chern-Simons theory flows to $G_{\kappa-h\,\text{sign}(\kappa)}$ non-supersymmetric Chern-Simons in the IR. If $G$ is not simple, the level shift may be more complicated, as in Section \ref{sec:anom}.

In the deep infrared, both the supersymmetric and non-supersymmetric Chern-Simons theories (which we think of as defined by a path integral over gauge connections, with the usual Chern-Simons action) flow to a massive, topological quantum field theory. We will refer to the topological theory corresponding to $G_k$ non-SUSY Chern-Simons as $TFT[G_k]$. It has no local degrees of freedom, but does have a category of line operators.

The level-rank duals of boundary conditions may be understood to follow from two more fundamental UV-IR relations.
Let $G$ be a simple group and $k>0$.
We propose that
\begin{itemize}
\item[1)] 3d $\CN=2$ $G_{k+h}$ Yang Mills-Chern-Simons theory with a Dirichlet $(\CD)$ boundary condition flows to a left-moving (chiral) $G_k$ WZW model coupled to $TFT[G_k]$. 
\item[2)] 3d $\CN=2$ $G_{-k-h}$ Yang-Mills-Chern-Simons theory with a Neumann $(\CN)$ boundary condition must be coupled to a left-moving chiral algebra $\CT_{2d}$ (which may be part of an $\CN=(0,2)$ boundary theory) in order to cancel the gauge anomaly. In the IR, this boundary condition flows to the coset $\CT_{2d}/G_{-k}$ coupled to $TFT[G_{-k}]$.
\end{itemize}
Both relations are fully compliant with 't Hooft anomaly matching. 
Analogous statements should hold for non-simple $G$ with modified level shifts.

The idea that Neumann b.c. for Yang-Mills-Chern-Simons theory leads to coset models is familiar in the literature, \emph{cf.}  \cite{DHSV, GGP-fivebranes, ArmoniNiarchos}, though we believe the statement about Dirichlet boundary conditions is new. We will give evidence of both statements for $G=U(N)$ and $SU(N)$ (and, in the case of $\CD$ b.c., general simple $G$) by computing half-indices.
With $\CD$ b.c., the half-index constitutes an ``abelianized'' formula for WZW characters.
We will also check the identification of bulk (UV) Wilson lines with modules for the boundary chiral algebras.

We note that the sign of the Chern-Simons level is important in the above statements. As discussed in Section \ref{sec:signs}, the choice of left vs. right boundary condition breaks the symmetry between positive and negative bulk levels. Recall that 3d $\CN=2$ $G_\kappa$ YM-CS theory breaks SUSY in the bulk if $|\kappa|<h$  \cite{Witten-3dindex, BHKK-break, Ohta-break, IS-new}. 

With our conventions for boundary conditions, the half-indices of $G_{k+h}$ theory with $\CD$ b.c. are badly behaved. Presumably, boundary monopole operators 
hit some unitarity bound along the RG flow. Notice that there are no known ${\cal N}=2$ supersymmetric anti-chiral WZW models, which would have been an obvious 
candidate for the IR boundary physics. 

On the other hand, $\CN$ b.c. for $k<0$ require some anti-chiral boundary matter in the UV, such as $(0,2)$ chiral multiplets, making index calculations somewhat trickier. The IR physics 
should involve some $(0,2)$ $G$ gauge theory coupled both to a $G_{-k}$ WZW model and the UV boundary matter. It goes beyond the scope of this paper. 

The supersymmetric statements above have obvious non-supersymmetric versions:
\begin{itemize}
\item[1')] Non-SUSY $G_k$ YM-CS theory with a Dirichlet b.c. flows to a chiral $G_k$ WZW model coupled to $TFT[G_k]$.
\item[2')] Non-SUSY $G_{-k}$ YM-CS theory with Neumann b.c. coupled to chiral algebra $\CT_{2d}$ to cancel the gauge anomaly flows to a chiral $\CT_{2d}/G_{-k}$ coset, coupled to $TFT[G_{-k}]$. 
\end{itemize}
(Now there are no subtleties about simple vs. non-simple $G$.) We became aware of these relations in discussions with Kevin Costello
and in other projects (such as \cite{Gaiotto:2017euk}).

The supersymmetric and non-supersymmetric statements about RG flows
lead to level-rank-dual boundary conditions when appropriate choices for $G$ and $\CT_{2d}$ are made --- so that a particular coset $\CT_{2d}/G_k$ (coming from $\CN$ b.c.) happens to be equivalent to a $\tilde G_{\tilde k}$ WZW model (coming from $\CD$ b.c.).
We will discuss this in Section \ref{sec:LR-N}.

\subsection{Dirichlet boundary conditions}
\label{sec:LR-D}

We present a few examples of statement (1) about RG flows with Dirichlet boundary conditions. In the IR, the bulk theory is purely topological and has no local operators. We should thus find that half-indices compute the characters of appropriate WZW models that appear in the IR on the boundary. We indeed find a match with the Weyl-Kac character formula \cite{Kac, Difrancesco, kass} for general simple $G$.

\subsection*{U(1)}

Let us first consider 3d $\CN=2$ $G=U(1)_k$ Yang-Mills-Chern-Simons theory. The half-index of $\CD$ b.c. is very simply
\be \II_\CD[U(1)_k] = \frac{1}{(q)_\infty} \sum_{m\in \Z} q^{\frac12 km^2}x^{km}y^k\,, \label{U1k} \ee
where $x$ is the fugacity for the $U(1)_\pd$ boundary flavor symmetry, and $y$ is the topological $U(1)_y$ fugacity.
For $k=1$, this is an ordinary theta-function $\II_\CD[U(1)_1] = (-q^{\frac12}xy;q)_\infty(-q^{\frac12}/(xy);q)_\infty = \text{F}(-q^{\frac12}xy;q)$, which is the vacuum character of $U(1)_1$ WZW, or equivalently a left-handed 2d fermion, \emph{a.k.a} an $\CN=(0,2)$ Fermi multiplet. For general $k>0$, \eqref{U1k} is a higher-level theta-function $\II_\CD[U(1)_k]=\frac{(q^k;q^k)_\infty}{(q)_\infty}(-q^{\frac k2}x^ky;q^k)_\infty (-q^{\frac k2}x^{k}y^{-1};q^k)_\infty$, which matches the vacuum character for $U(1)_k$ WZW (intended as a lattice VOA),
\be \II_\CD[U(1)_k] = \chi_0[U(1)_k]\,.\ee

Notice that the sum over boundary monopole sectors is crucial in reproducing the WZW character, rather than the Kac-Moody current algebra character. The current algebra character just counts modes of the chiral $U(1)$ current $J\sim i\pd\phi$, reproducing the $1/(q)_\infty$ prefactor. The WZW character further includes the vertex operators $e^{im\phi}$, which correspond to boundary $U(1)$ monopoles.

When $k<0$, the monopole sum diverges badly, similar to what we saw in particle-vortex duality \eqref{tet-diverge} when the boundary anomaly was negative. We will restrict ourselves to $k>0$.

To obtain characters of other modules, we add Wilson lines. Following Section \ref{sec:line}, we see that a Wilson line of charge $n\in \Z$ in the UV $G=U(1)_k$ Yang-Mills-Chern-Simons theory modifies the half-index as
\be  \II_\CD[U(1)_k,n] = \frac{1}{(q)_\infty} \sum_{m\in \Z} q^{\frac12 km^2}x^{km}y^{km}\times (q^mx)^n
 =  \frac{1}{(q)_\infty} \sum_{m\in \Z} q^{\frac12 km^2+mn}x^{km+n}y^{km} \,. \label{U1kn} \ee
This agrees with the WZW character up to a mild prefactor,
\be  \chi_n[U(1)_k] = q^{-\frac{n^2}{2k}}y^{-\frac nk}\, \II_\CD[U(1)_k,n]\,.\ee

\subsection*{SU(2)}

Next, consider $SU(2)_{k}$, at $k>0$. A general spin $j$ character%
\footnote{Henceforth, we will write all characters without the conventional prefactor, sometimes called the modular anomaly, so that the $q$-series begins at order $q^0$. For a character associated to an integrable representation of some level $k$ affine Kac-Moody algebra with highest weight $\Lambda$ the modular characteristic may be written as $q^{s(\Lambda)}$ where $s(\Lambda) = |\Lambda + \rho|^2/(2(k + h)) - |\rho|^2/2h$ and $\rho$ denotes the Weyl vector. \label{fn:modular}} %
may be written as
\begin{equation} \label{chiSU2}
\chi_j[SU(2)_k] = \frac{1}{(q)_\infty(q x^2;q)_\infty(q x^{-2};q)_\infty} \sum_{m = - \infty}^\infty q^{(k+2)m^2 + (2j+1) m} \chi_{(k+2)m+j}(x)
\end{equation}
where $\chi_{j}(x) = \frac{x^{2j + 1} - x^{-2j-1}}{x-x^{-1}}$. Notice that there are negative terms in the sum, 
as $\chi_{-j}(x) = - \chi_{j-1}(x)$. 
In particular, for the vacuum character we can write
\begin{equation}
\chi_0[SU(2)_k] = \frac{1}{(q)_\infty(q x^2;q)_\infty(q x^{-2};q)_\infty} \left[\sum_{m = 0}^\infty q^{(k+2)m^2 + m} \chi_{(k+2)m}(x)-\sum_{m = 1}^\infty q^{(k+2)m^2 - m} \chi_{(k+2)m-1}(x) \right]
\end{equation}
This shows the sequence of nested Verma modules of spin $0$, $k+1$, $k+2$, $2k+3, \ldots$ which give a resolution of
the vacuum module. 

Another simple manipulation of \eqref{chiSU2} expresses the vacuum character as
\begin{align}
\chi_0[SU(2)_k] &=  \frac{1}{(q)_\infty}\sum_{m = - \infty}^\infty q^{(k+2)m^2}x^{2(k+2)m} \frac{( q^{-m}- x^2 q^m) }{(1-x^2)(q x^2;q)_\infty(q x^{-2};q)_\infty} \notag \\
&= \frac{1}{(q)_\infty} \sum_{m = - \infty}^\infty q^{k m^2}x^{2k m} \frac{1}{(q^{1+2m} x^2;q)_\infty(q^{1-2m} x^{-2};q)_\infty}\notag \\
&= \II_\CD[SU(2)_{k+2}],
\end{align}
where in the middle step we employed the usual identity $\text{F}(x;q)=q^{\frac{n^2}{2}}(-q^{-\frac12}x)^n\text{F}(q^nx;q)$ for $\text{F}(x;q) = (x;q)_\infty(qx^{-1};q)_\infty$. This final expression agrees precisely with the half-index of 3d $\CN=2$ $SU(2)_{k+2}$ CS-YM theory with $\CD$ b.c., as we had hoped.

In a similar way, the spin-$j$ character may be rewritten as
\begin{equation}
\chi_j[SU(2)_k] =  \frac{1}{(q)_\infty} \sum_{m = - \infty}^\infty q^{k m^2}x^{2k m} \frac{1}{(q^{1+2m} x^2;q)_\infty(q^{1-2m} x^{-2};q)_\infty}
\chi_{j}(q^{m} x)
\end{equation}
which agrees with the half-index $\II_\CD[SU(2)_{k+2},\mb j]$ in the presence of a spin-$j$ Wilson line.
Notice also the difference equation
\begin{equation}
\chi_j[SU(2)_k](q^{\mp 1} x) = x^{\pm k} q^k \chi_j[SU(2)_k](x),
\end{equation} which follows from the discussion in Sections \ref{sec:line}, \ref{sec:diff}.

As a concrete example, the half-index of the theory $G=SU(2)_4$ with Dirichlet boundary conditions with a Wilson line of spin $1/2$ is
\be \II_{\rm D}[SU(2)_4,\mb{\tfrac12}]= \chi_{\frac12}(x)+(q+2q^2)\big[\chi_{\frac12}(x)+\chi_{\frac32}(x)\big] + q^3\big[4\chi_{\frac12}(x)+3\chi_{\frac32}(x)+\chi_{\frac52}(x)\big]+\ldots \ee
One can check explicitly that this is exactly the character in $\widehat{\mathfrak{su}}(2)_2$ with highest weight vector $\Lambda = (1, 1)$ (written via its Dynkin labels), as expected. (The vacuum, or basic, representation of $\widehat{\mathfrak{su}}(2)_2$ has highest weight vector $\Lambda = (k, 0) = (2, 0)$ in this notation). One can perform an identical check for the $j=1$ Wilson line against the representation with highest weight vector $\Lambda = (0, 2)$, as well as similar checks for general $k, j$.

\subsection*{SO(3)}

Similarly, we can begin with the WZW vacuum character for $SO(3)_k$. This can be obtained by combining the vacuum character of $SU(2)_{2 k}$ with the spin $k$ character of $SU(2)_{2 k}$, which has dimension $\frac{k(k+1)}{2k+2} = \frac{k}{2}$. It can be compactly written as a sum over a lattice refined by a factor of two:
\begin{equation}
\chi_0[SO(3)_k] = \frac{1}{(q)_\infty(q x;q)_\infty(q x^{-1};q)_\infty} \sum_{m = - \infty}^\infty (-1)^m q^{\frac{k+1}{2}m^2 + \frac{m}{2}} \chi_{(k+1)m}(x^{\frac12})\,,
\end{equation}
and rearranged the same way as before to
\begin{equation}
\chi_0[SO(3)_k] = \frac{1}{(q)_\infty} \sum_{m = - \infty}^\infty q^{\frac{k}{2} m^2}x^{\frac{k}{2} m} \frac{1}{(q^{1+m} x;q)_\infty(q^{1-m} x^{-1};q)_\infty} = \II_\CD[SO(3)_{k+2}] \,,
\end{equation}
which matches the Dirichlet half-index.

\subsection*{Any simple G}

Analogously, we expect the vacuum WZW character for general simple group $G$ at level $k>0$ to be computed by the half-index
\be \II_\CD[G_{k+h}] = \frac{1}{(q)_\infty^{r}} \sum_{m\in \Lambda^\vee} \frac{q^{\frac 12 k (m,m)}x^{km}}{\prod_{\alpha\in\Phi} (q^{1+m\cdot\alpha} x^\alpha;q)_\infty}\,,  \ee
where $r=\text{rank}(G)$, $\Lambda^\vee$ denotes the cocharacter lattice of $G$, and $\Phi$ is the set of roots of $G$.
Just as in the $SU(2)$ example above, this can be rewritten as 
\begin{equation}
\II_\CD[G_{k+h}]  = \frac1{(q)_\infty^r} \sum_{m\in \Lambda^\vee} q^{\frac{1}{2} k(m,m)}x^{k m} \frac{\prod_{\alpha \in \Phi^+} (1-q^{m\cdot\alpha}  x^\alpha)}{\prod_{\alpha \in \Phi^+} \text{F}(q^{m\cdot \alpha} x^\alpha;q)}\,.
\end{equation}
Then using $\text{F}(q^{m\cdot \alpha} x^\alpha;q)=q^{-\frac12(m\cdot\alpha)^2}(-q^{\frac12}x^{-\alpha})^{m\cdot\alpha}\text{F}(x^\alpha;q)$ as well as $\sum_{\alpha\in \Phi^+}(m\cdot\alpha)^2=h(m,m)$ and $\prod_{\alpha\in \Phi^+}x^{\alpha(m\cdot\alpha)}=x^{hm}$  this becomes
\be\label{eq:ourWK} \II_\CD[G_{k+h}]  =  \frac1{(q)_\infty^r\prod_{\alpha\in\Phi^+}(qx^\alpha;q)_\infty(qx^{-\alpha};q)_\infty} \sum_{m \in \Lambda^{\vee}} q^{\frac12(k+h)(m,m)} x^{(k+h)m} \prod_{\alpha\in \Phi^+} (-q^{\frac12})^{-m\cdot \alpha} \frac{1-q^{m\cdot\alpha}x^\alpha}{1-x^\alpha}\,. \ee
In Appendix \ref{app:WK}, we show that this expression is equivalent to the usual expression for the Weyl-Kac character formula (up to the overall factor of the modular anomaly) when $G$ is simply connected.

We can also write down the half-index with a Wilson loop insertion in representation $\mb R$:
\be \II_\CD[G_{k+h},\mb R] = \frac{1}{(q)_\infty^r} \sum_{m\in \Lambda^\vee} \frac{q^{\frac12 k(m,m)}x^{km}}{\prod_{\alpha\in \Phi^+} (q^{1+m\cdot\alpha} x^\alpha;q)_\infty} \text{Tr}_{\mb R}(q^mx)\,, \label{KW-R} \ee
which we expect to coincide with the character of other highest-weight modules.  
Of course, there should be identifications between Wilson lines. 
These are captured by the identity
\begin{equation}
\II_\CD[G_{k+h},\mb R](q^{(\delta,\cdot)}x;q)  = x^{k \delta} q^{-\frac{k}{2} (\delta,\delta)} \II_\CD[G_{k+h},\mb R](x;q)\,, \end{equation}
which follows easily from a shift of summation in \eqref{KW-R}\,.

\subsection*{U(N)}

For level-rank duality, it is useful to consider $U(N)$ groups as well. There are two well-behaved possibilities.

Suppose that we wish to reproduce the $U(N)_k$ WZW characters from a half-index, at $k>0$. Then we need to look for a 3d $\CN=2$ theory whose boundary 't Hooft anomaly is $k\Tr(\mb f^2)$. A naive guess is $\CN=2$ $U(N)_{k+N}$ theory; but the computation in \eqref{Un-anom} shows that this won't quite work, due to an extra $(\Tr\,\mb f)^2$ contribution from the gauginos. What works instead is an $\CN=2$ YM-CS theory with a different level for the $U(1)$ part of $U(N)$. 

We use the following notation (coincident with \cite{Aharony-LR, HsinSeiberg-LR}): We write $U(N)_\kappa=U(N)_{\kappa,\kappa}$ for a Chern-Simons theory whose (UV) action corresponds to the anomaly polynomial
\be U(N)_\kappa\,:\qquad \kappa\,\Tr(\mb f^2)\,, \ee
where $\mb f$ is the $U(N)$ field strength. We write  $U(N)_{\kappa,\kappa+p}$ for the more general possibility
\be U(N)_{\kappa,\kappa+p}\,:\qquad \kappa\,\Tr(\mb f^2) + \frac{p}{N}(\Tr\,\mb f)^2\,. \label{kkp}\ee
Notice that if we restrict $\mb f$ to be diagonal with all entries equal to $f/\sqrt{N}$ (corresponding to the $U(1)$ field strength inside $U(N)$), the polynomial \eqref{kkp} reduces to $(\kappa+p)f^2$. Thus the effective $U(1)$ level is $\kappa+p$.

In order to reproduce the $U(N)_k$ WZW vacuum character, we should use 3d $\CN=2$ $U(N)_{k+N,k}$ theory. The polynomial encoding its bulk CS levels is $\CI_{\rm bulk} = (k+N)\Tr(\mb f^2) -(\Tr\,\mb f)^2$, so that the boundary anomaly on a $\CD$ b.c. is
\be \CI_\CD = \CI_{\rm bulk} -\big[ N\Tr(\mb f^2)-(\Tr\,\mb f)^2+\tfrac{N^2}{2}\mb r^2\big] = k\Tr(\mb f^2)-\tfrac{N^2}{2}\mb r^2\,,
\ee
which (modulo the $\mb r^2$ term) matches the anomaly of $U(N)_k$ WZW. If we include a topological $U(1)_y$ symmetry as well, which gives an extra term $2(\Tr\, \mb f)\mb y$ in the anomaly polynomial,
 then the half-index is
\be \II_\CD[U(N)_{k+N,k}] = \frac{1}{(q)_\infty^N} \sum_{m\in \Z} \frac{q^{\frac k2 m\cdot m} x^{km}y^{\sum_im_i} }{\prod_{i\neq j}(q^{1+m_i-m_j}x_i/x_j;q)_\infty} \overset{?}= \chi_0[U(N)_k]\,.
\ee

Alternatively, we may consider 3d $\CN=2$ $U(N)_{k+N}$ YM-CS theory. Its bulk CS levels are encoded by $(k+N)\Tr(\mb f^2)$, so the boundary anomaly on $\CD$ b.c. is now
\be \CI_\CD = k\Tr(\mb f^2) +(\Tr\,\mb f)^2 +2(\Tr\,\mb f)\mb y-\tfrac{N^2}{2}\mb r^2\,\,. \label{UNkN-anom} \ee
The half-index
\be \label{IIDUNkN} \II_\CD[U(N)_{k+N}]  =  \frac{1}{(q)_\infty^N} \sum_{m\in \Z} \frac{q^{\frac k2 m\cdot m+\frac12(\sum_i m_i)^2} x^{km}(xy)^{\sum_im_i} }{\prod_{i\neq j}(q^{1+m_i-m_j}x_i/x_j;q)_\infty} \overset{?}= \chi_0[U(N)_{k,k+N}]
\ee
is now expected to compute the vacuum character of $U(N)_{k,k+N}$ WZW, which matches the anomaly \eqref{UNkN-anom}.

\subsection{Neumann boundary conditions and dualities}
\label{sec:LR-N}

We now consider statement (2): that 3d $\CN=2$ YM-CS theory with $\CN$ b.c., and appropriate degrees of freedom to cancel the gauge anomaly, flows to a coset model. Since this has already been studied in the literature, we only give a few examples, and focus on explaining how the statement may be used productively to generate dual boundary conditions. 

\subsection*{U(1)$_{-k}\leftrightarrow$SU(k)$_{k+1}$}

Consider 3d $\CN=2$ $U(1)_{-k}$ theory, now with $\CN$ b.c.  The boundary anomaly is simply
\be \CI_{\CN} = -k\,\mb f^2 \,.\ee
If $k>0$, a simple way to cancel the anomaly is by introducing $k$ boundary Fermi multiplets $\Gamma_i$ of gauge charge $+1$ (and R-charge zero). The corresponding half-index is
\be \label{INU1} \II_{\CN+\Gamma}[U(1)_{-k}] = (q)_\infty\oint \frac{ds}{2\pi is} \prod_{i=1}^k \text{F}(-q^{\frac12}x_is;q)\,, \ee
where we have introduced additional fugacities $x=(x_1,...,x_k)$ with $\prod_i x_i=1$ for the new boundary $SU(k)$ flavor symmetry that rotates the Fermi multiplets. We now observe that the boundary anomaly has been shifted to a 't Hooft anomaly
\be \CI_\CN+\CI_\Gamma =  \Tr(\mb x^2) \ee
for the $SU(k)$ flavor symmetry. Such an anomaly could be matched by an $SU(k)_1$ WZW model in the IR. Indeed the $SU(k)_1$ WZW is exactly what we would expect from the coset construction
\be \frac{\text{$k$ free fermions}}{U(1)_{-k}} = \frac{U(k)_1}{U(1)_{-k}} = SU(k)_1\,. \ee

The index \eqref{INU1} is actually easy to evaluate directly, since $\text{F}(-q^{\frac12}z) = \frac{1}{(q)_\infty} \sum_{n\in \Z}q^{\frac{n^2}{2}}z^n$ and taking the coefficient of $s^0$ in the integrand gives
\be  \II_{\CN+\Gamma}[U(1)_{-k}] = \frac{1}{(q)_\infty^{k-1}} \sum_{n\in \Z^k \atop \sum_i n_i=0} q^{\frac12 n\cdot n}x^n = \chi_0[SU(k)_1]\,. \label{SUk1} \ee
This is the lattice sum for the vacuum character of $SU(k)_1$ WZW. For example, at $k=2$ we may use $SU(2)$ fugacities $(x_1,x_2)=(x,x^{-1})$ and obtain
\be \chi_0[SU(2)_1]= 1+q\chi_1(x)+q^2\big[\chi_1(x)+\chi_0(x)\big] + q^3\big[2\chi_1(x)+\chi_0(x)]+q^4\big[\chi_2(x)+2\chi_1(x)+2\chi_0(x)\big]+...\ee
where $\chi_j(x)$ is the spin-$j$ character of $SU(2)$, just as in \eqref{chiSU2}.

In Section \ref{sec:LR-D} we encountered another boundary condition whose half-index was an $SU(k)_1$ character, namely $\CD$ b.c. for 3d $\CN=2$ $SU(k)_{k+1}$ theory. The Dirichlet half-index of this latter theory matches \eqref{SUk1}. We deduce a duality of SUSY boundary conditions
\be U(1)_{-k}\quad \CN+\text{($k$ fund. fermis)} \quad\leftrightarrow \quad
 SU(k)_{k+1}\quad \CD\,. \ee

We could in principle have considered $k<0$ as well.
Then the $\CN$ b.c. gauge anomaly for $U(1)_{-k}$ would have to be cancelled by chiral rather than Fermi multiplets. 
However, this leads to the same sort of difficulties in the index computation that we encountered in particle-vortex duality, \emph{cf.} \eqref{badN}.
 To avoid potential problems with free chirals on the boundary, we restrict ourselves to $k>0$.

\subsection*{U(N)$_{-k-N,-k}\leftrightarrow$SU(k)$_{k+N}$}

To generalize the $U(1)_{-k}$ example above, let us consider 3d $\CN=2$ $U(N)_{-k-N,-k}$ with $k,N>0$.The level convention \eqref{kkp} means that the bulk CS levels are
\be \CI_{\rm bulk} = -(k+N)\Tr(\mb f^2) + (\Tr\,\mb f)^2\,. \ee
The boundary anomaly on $\CN$ b.c. due to gauginos then becomes
\be \CI_\CN = \CI_{\rm bulk} + N \Tr(\mb f^2)-(\Tr\,\mb f)^2+\tfrac{N^2}{2}\mb r^2 = -k \Tr(\mb f^2) +\tfrac{N^2}{2}\mb r^2 \,.\ee
We may cancel this by introducing $k$ fundamental Fermi multiplets $\Gamma_i$ (of R-charge zero), transforming under an $SU(k)$ boundary flavor symmetry. They shift the anomaly by $\CI_\Gamma = k\,\Tr(\mb f^2)+N\Tr(\mb x^2)$, where $\mb x$ is the $SU(k)$ field strength, to
\be \CI_\CN + \CI_\Gamma = N\Tr(\mb x^2)+\tfrac{N^2}{2}\mb r^2 \,.\ee
This matches perfectly with the 't Hooft anomaly of a Dirichlet b.c. on 3d $\CN=2$ $SU(k)_{k+N}$ YM-CS theory (up to a shift of the R-R anomaly, which can be put in by hand). Thus we may expect a duality of boundary conditions 
\be   U(N)_{-k-N,-k}\quad \text{$\CN$ + $k$ fund. fermis} \quad\leftrightarrow \quad SU(k)_{k+N} \quad \CD\,. \label{LR-USU} \ee

We are claiming slightly more than this, namely that both sides flow to an $SU(k)_N$ WZW model on the boundary. On the right side this follows from Section \ref{sec:LR-D}. On the left side, we expect a coset model, which agrees by standard level-rank duality of chiral algebras:
\be \frac{\text{$Nk$ free fermions}}{U(N)_{-k}} = \frac{U(Nk)_1}{U(N)_{-k}} \simeq SU(k)_N \ee
In addition, the Neumann b.c. for $U(N)_{-k-N,-k}$ theory has a half-index
\be \II_{\CN+\Gamma}[U(N)_{-k-N,-k}] = (q)_\infty^N \oint \prod_{i=1}^N \frac{ds_i}{2\pi is_i} \prod_{i\neq j}\Big(q\frac{s_i}{s_j};q\Big)_\infty\;{\prod_{i=1}^N\prod_{a=1}^k \text{F}(-q^{\frac12}x_a s_i;q)} \ee
whose agreement with the vacuum character $\chi_0[SU(k)_N]$ we have checked computationally for various values of $N$ and $k$.

\subsection*{SU(N)$_{-k-N}\leftrightarrow$U(k)$_{k+N,N}$}

Conversely, we might start with Neumann b.c. for 3d $\CN=2$ $SU(N)_{-k-N}$. The boundary gauge anomaly is again cancelled by $k$ fundamental Fermi multiplets (\emph{i.e.} coupling to $Nk$ free fermions), which now introduce a $U(k)$ boundary flavor symmetry. We expect the Neumann b.c. to flow to the coset model
\be \frac{\text{$Nk$ free fermions}}{SU(N)_{-k}} \simeq U(k)_N\,. \ee
The same result is obtained on a Dirichlet b.c. for 3d $\CN=2$ $U(k)_{k+N,N}$, whence
\be \label{LR-SUU}   SU(N)_{-k-N}\quad \text{$\CN$ + $k$ fund. fermis} \quad\leftrightarrow \quad U(k)_{k+N,N} \quad \CD\,. \ee

\subsection*{U(N)$_{-k-N}\leftrightarrow$U(k)$_{k+N}$}

Finally, let us apply the same analysis starting with a Neumann b.c. for 3d $\CN=2$ $U(N)_{-k-N}$ theory, with $k,N>0$. It is useful to keep track of the $U(1)_y$ topological symmetry, for which we introduce a field strength $\mb y$ as usual. Then the bulk CS levels are
\be \CI_{\rm bulk} = -(k+N)\Tr(\mb f^2) +2(\Tr\,\mb f)\mb y\,, \ee
leading to a boundary anomaly
\be \CI_\CN = \CI_{\rm bulk}+ N \Tr(\mb f^2)-(\Tr\,\mb f)^2+\tfrac{N^2}{2}\mb r^2 = -k \Tr(\mb f^2)-(\Tr\,\mb f)^2+2(\Tr\mb f)\mb y+\tfrac{N^2}{2}\mb r^2\,.\ee

Adding $k$ fundamental Fermi multiplets $\Gamma$ on the boundary is not sufficient to cancel the $-(\Tr\,\mb f)^2$ part of the gauge anomaly. However, we can deal with this by introducing one more ``baryonic'' Fermi multiplet $\eta$, which transforms under the $U(1)$ part of $U(N)$. We can even cancel the mixed gauge-topological anomaly by giving $\eta$ nontrivial $U(1)_y$ charge.

The multiplets $\Gamma$ and $\eta$ introduce an extra $U(k)$ flavor symmetry on the boundary. This is the commutant $U(N)\times U(1)_y$ in $U(Nk+1)$. Under both gauge and flavor symmetry, these Fermi multiplets may be taken to have charges
\be \label{LR-Ge}  \begin{array}{c|cc|cc}
 & U(N) & U(k) & U(1)_y & U(1)_R \\\hline
 \Gamma & \mb N & \mb k & 0 & 0 \\
 \eta & \text{det} & \text{det}^{-1} & -1 & 0
\end{array} \ee
Then their joint contribution to the anomaly is
\begin{align} \CI_\Gamma + \CI_\eta  &= \big[k\Tr(\mb f^2)+N\Tr(\mb x^2)+2(\Tr\,\mb f)(\Tr\,\mb x)\big] + (\Tr\,\mb f-\Tr\,\mb x-\mb y)^2  \\
 &= k\Tr(\mb f^2)+N\Tr(\mb x^2) + (\Tr\,\mb f)^2 + (\Tr\,\mb x)^2+2 \mb y(\Tr\,\mb x-\Tr\,\mb f)-\mb y^2\,, \notag
\end{align}
which cancels the gauge part of the $\CN$ b.c. anomaly and leaves behind
\be \CI_\CN + \CI_\Gamma+\CI_\eta = N\Tr(\mb x^2)+(\Tr\,\mb x)^2+2 \mb y(\Tr\,\mb x) - \mb y^2+\tfrac{N^2}{2}\mb r^2\,. \label{UU-anom}\ee

The 't Hooft anomaly \eqref{UU-anom} matches that of Dirichlet b.c. in $U(N)_{k+N}$ theory \eqref{UNkN-anom}, up to a shift of the $\mb r^2$ and $\mb y^2$ terms. Thus, we may surmise that
\be \label{LR-UU}  U(k)_{-k-N}\quad \CN+\Gamma,\eta \quad\leftrightarrow \quad U(k)_{k+N}\quad \CD\,. \ee
In the IR, we expect the right side to flow to a $U(k)_{k+N,N}$ WZW model, and the left side to flow to a coset, essentially
\be \frac{\text{$Nk+1$ free fermions}}{U(N)_{-k-N,-k}}\simeq U(k)_{k+N,N}\,. \ee
This is the duality of chiral algebras responsible for level-rank duality of non-superymmetric $U(N)_{-k-N,-k}$ and $U(k)_{k+N,N}$ Chern-Simons theories, \emph{cf.} \cite{Aharony-LR}.

We may also compare half-indices. The Neumann b.c. has
\be \II_{\CN+\Gamma+\eta}[U(N)_{-k-N}] =  (q)_\infty^N \oint \prod_{i=1}^N \frac{ds_i}{2\pi is_i} \prod_{i\neq j}\Big(q\frac{s_i}{s_j};q\Big)_\infty\;
\text{F}\Big(\frac{-q^{\frac12}\prod_i s_i}{y\prod_a x_a};q\Big)
{\prod_{i=1}^N\prod_{a=1}^k \text{F}(-q^{\frac12}x_a s_i;q)}\,, \ee
and we have checked computationally that this indeed matches \eqref{IIDUNkN} in several cases.

\subsection{Interfaces}
\label{sec:LR-int}

We may use the proposed dual boundary conditions above to construct duality interfaces for level-rank dual theories. The construction works the same way as in Sections \ref{sec:int} and \ref{sec:XYZ}. Just as in the case of particle-vortex duality, the interfaces turn out to be directional: in our conventions, they must necessarily have negative CS levels on the left to positive CS levels on the right.

For example, the duality interfaces for $U(N)_{-k-N,-k}\;\leftrightarrow\; SU(k)_{k+N}$ and $SU(N)_{-k-N}\;\leftrightarrow\; U(k)_{k+N,N}$ have $\CN$ b.c. for the gauge fields on either side, coupled to $Nk$ free Fermi multiplets $\Gamma$ in a bi-fundamental representation of the gauge groups. Schematically, 
\be U(N)_{-k-N,-k}\big|\Gamma|SU(k)_{k+N}\,,\qquad SU(N)_{-k-N}\big|\Gamma\big| U(k)_{k+N,N}\,. \ee
Again we emphasize that anomalies on the interfaces \emph{only} cancel for $k,N>0$.

Similarly, the duality interface for  $U(N)_{-k-N}\;\leftrightarrow\; U(k)_{k+N}$ involves both bi-fundamental and bi-baryonic Fermi multiplets $\Gamma,\eta$ as in \eqref{LR-Ge}:
\be U(N)_{-k-N}\big|\Gamma,\eta\big| U(k)_{k+N}\,. \ee

It is worth pointing out that this is a ``supersymmetrization'' of a non-supersymmetric level-rank duality interface,
defined in terms of bi-fundamental fermions rather than Fermi multiplets. 

\subsection{From U to SU}
\label{sec:UtoSU}

In bulk 3d dualities, one has a great deal of freedom to ``move'' $U(1)$ gauge groups around. This typically proceeds  by gauging topological symmetries, which can have the effect of un-gauging dynamical $U(1)$'s \cite{Witten-SL2, KapustinStrassler, 4d3d-dualities}. The bulk $U(N)$ and $SU(N)$ level-rank dualities above are known to be related by such operations.
The gauging and un-gauging operations may be applied in the presence of dual boundary conditions as well, being careful to account for the boundary degrees of freedom.

We illustrate this by reducing the duality of boundary conditions in $U(N)_{-k-N}\;\leftrightarrow\; U(k)_{k+N}$ theories \eqref{LR-UU} to $U(N)_{-k-N,-k}\;\leftrightarrow\; SU(k)_{k+N}$ as in \eqref{LR-USU}. For a perfect match of boundary 't Hooft anomalies, the bulk CS levels in $U(N)_{-k-N}\;\leftrightarrow\; U(k)_{k+N}$ duality may be defined as
\be\begin{array}{ll} U(N)_{-k-N}\,:\quad & \CI_{\rm bulk}= -(k+N)\Tr(\mb f^2)+2(\Tr\,\mb f)\mb y+ \mb y^2-\tfrac{N^2}{2}\mb r^2\,, \\[.2cm]
U(k)_{k+N}\,:\quad & \CI_{\rm bulk}= (k+N)\Tr(\mb x^2)+2(\Tr\,\mb x)\mb y +\tfrac{k^2}{2}\mb r^2\,. \end{array}\ee

Now we simultaneously gauge the $U(1)_y$ topological symmetry in both theories. In the $U(k)_{k+N}$ theory, this has the effect of un-gauging the $U(1)$ part of $U(N)$, to which the topological symmetry has a mixed CS coupling. Thus we are left with an $SU(k)_{k+N}$ bulk theory. The boundary condition is still $\CD$.

On the other hand, in the $U(N)_{-k-N}$ theory, due to the extra $\mb y^2$ CS term,  gauging the $U(1)_y$ symmetry has virtually no effect in the bulk. Indeed, once $U(1)_y$ is gauged, we can redefine abelian gauge charges in a way that corresponds to sending $\mb y\to \mb y-\Tr\,\mb f$. The bulk CS levels become
\be \CI_{\rm bulk}' = -(k+N)\Tr(\mb f^2) + (\Tr\,\mb f)^2+ \mb y^2-\tfrac{N^2}{2}\mb r^2\,, \ee
and we see that the bulk theory has become a decoupled $U(N)_{-k-N,-k}\times U(1)_1$. On the boundary, the extra 2d Fermi multiplets now have charges 
\be \begin{array}{c|cc|cc}
 & U(N) & SU(k) & U(1)_y & U(1)_R \\\hline
 \Gamma & \mb N & \mb k & 0 & 0 \\
 \eta & \mb 1 &\mb 1 & -1 & 0
\end{array} \ee
Thus $\Gamma$ is only charged under $U(N)$ and $\eta$ only under the new $U(1)$, giving decoupled boundary conditions
\be \big(\text{$U(N)_{-k-N,-k}$ w/ $\CN+\Gamma$}\big)\otimes \big(\text{$U(1)_1$ w/ $\CN+\eta$}\big)\,. \ee
However, $U(1)_1$ with a boundary Fermi multiplet $\eta$ is a trivial theory in the IR, both on the bulk and on the boundary: the coset chiral algebra is $\text{(1 free fermion)}/U(1)_1 = U(1)_1/U(1)_1$, and we can easily verify that the half-index is
\be \II_{\CN+\eta}[U(1)_1]= (q)_\infty \oint \frac{dy}{2\pi iy} \text{F}(-q^{\frac12}/y) = \oint \frac{dy}{2\pi iy}\sum_{n\in \Z}q^{\frac{n^2}{2}}y^n = 1\,. \ee
Altogether, the effect of gauging the $U(1)_y$ symmetry is to reduce the $U(N)_{-k-N}$ side of the duality to $U(N)_{-k-N,-k}$ with $\CN$ b.c. coupled to $\Gamma$. This is precisely what we want for \eqref{LR-USU}.

\section{Aharony duality}
\label{sec:Aharony}

In this section we identify several dual pairs of boundary conditions for Aharony-dual theories with $U(N)$ gauge groups \cite{Aharony}.
Aharony dualities generalize the SQED$\,\leftrightarrow\,$XYZ and SQCD$\,\leftrightarrow\,$detYZ dualities from Section \ref{sec:XYZ}, and flow from 4d Seiberg duality \cite{Seiberg-duality} (\emph{cf.} \cite{4d3d-dualities}).
Just as in SQED and SQCD in Section \ref{sec:XYZ}, all the bulk UV Chern-Simons levels in Aharony duality are identically zero, and the matter fields are ``non-chiral'' --- in that equally numbers of fundamental and anti-fundamental 3d chiral multiplets occur. This leads to a structure of dual boundary conditions that is very similar to that found in Section \ref{sec:XYZ}, and duality interfaces that are effectively bi-directional.

\subsection{Fundamental duality and interface}

Aharony duality relates the 3d $\CN=2$ theories
\be \begin{array}{l}\text{Theory A\,:}\quad  \left\{\begin{array}{l} \text{$U(N_c)$ gauge theory} \\  \text{with $N_f$ fundamental ($Q_{\rm 3d}$) and $N_f$ anti-fundamental ($\tilde Q_{\rm 3d}$) chirals} \end{array} \right. \\[.5cm]
 \text{Theory B\,:}\quad \left\{\begin{array}{l} \text{$U(N_c')$ gauge theory, $\boxed{N_c':=N_f-N_c}$} \\
 \text{with $N_f$ fundamental ($Q_{\rm 3d}'$) and $N_f$ anti-fundamental ($\tilde Q_{\rm 3d}'$) chirals,} \\
 \text{a $N_f\times N_f$ matrix of singlets $(M_{\rm 3d})$, and two more singlets ($V_{\rm 3d}^\pm$),} \\
 \text{\qquad and $W = M_{\rm 3d}Q_{\rm 3d}'\tilde Q_{\rm 3d}' + U_{\rm 3d}^-V^+_{\rm 3d}+U_{\rm 3d}^+V^-_{\rm 3d}$}
\end{array}\right.\end{array}
\ee
We assume that $N_f>N_c$ (so $N_c'>0$); otherwise, at $N_f=N_c$ one finds instead the SQED\,$\leftrightarrow$\,XYZ and SQCD\,$\leftrightarrow$\,detYZ dualities of Section \ref{sec:XYZ}.

Recall that the bulk duality maps the mesons $Q_{\rm 3d}\tilde Q_{\rm 3d}$ of Theory A to the singlets $M_{\rm 3d}$ of Theory B, and it maps the two scalar monopole operators of Theory A to the singlets $V^\pm_{\rm 3d}$ of Theory B.
Theory B has its own mesons $Q_{\rm 3d}'\tilde Q_{\rm 3d}'$ and scalar monopole operators $U^\pm$; however, these operators are killed by the bulk superpotential.

The global symmetry is $SU(N_f)\times \widetilde{SU(N_f)} \times U(1)_a\times U(1)_y\times U(1)_R$, where as usual the latter three $U(1)$'s are axial, topological, and R-symmetry.
We split the bulk chirals into $\CN=(0,2)$ chiral and Fermi multiplets as usual, namely
\be (Q,\Psi)\,,\quad (\tilde Q,\tilde \Psi)\,;\qquad (Q',\Psi')\,,\quad (\tilde Q',\tilde \Psi')\,,\quad (M,\mu)\,,\quad (V^\pm,\upsilon^\pm)\,.
\ee
The $(0,2)$ chirals can be taken to have charges
\be
\begin{array}{c|cc|ccccc}
& U(N_c) & U(N_c') & SU(N_f) & \widetilde{SU(N_f)} & U(1)_a& U(1)_y & U(1)_R \\\hline
Q & \mb N_c & \mb 1 & \mb N_f & \mb 1 & 1&0&0 \\
\tilde Q  & \ol{\mb N_c} & \mb 1 & \mb 1 & \ol{\mb N_f} & 1 &0&0 \\\hline
Q' & \mb 1 & \mb N_c & \ol{\mb N_f} & \mb 1 & -1&0&1 \\
\tilde Q' & \mb 1 & \ol{\mb N_c} & \mb 1 & \mb N_f & -1 &0&1 \\
M & \mb 1 & \mb 1 & \mb N_f & \ol{\mb N_f} & 2 &0&0 \\
V^\pm &\mb 1 & \mb 1 & \mb 1 & \mb 1 & -N_f & \pm 1 & N_c'+1
\end{array}
\ee
and as usual the (0,2) fermis have the conjugate flavor charges and $(1-\rho_{\rm chiral})$ R-charge. We again emphasize that the bulk UV Chern-Simons levels (both for gauge and global symmetries) are identically zero.

We propose the following two pairs of dual boundary conditions:
\be \begin{array}{l@{\quad\leftrightarrow\quad}l}
  (\CD,\overset{Q}{\text{D}},\overset{\tilde Q}{\text{D}})  & (\CN,\overset{Q'}{\text{N}},\overset{\tilde Q'}{\text{N}},\overset{M}{\text{D}},\overset{V^\pm}{\text{N}}) + \Gamma,\eta \\[.2cm]
   (\CN,\overset{Q}{\text{N}},\overset{\tilde Q}{\text{N}}) + \Gamma',\eta'  & (\CD,\overset{Q'}{\text{D}},\overset{\tilde Q'}{\text{D}},\overset{M}{\text{N}},\overset{V^\pm}{\text{D}})\,,
\end{array} \label{Abc}\ee
where $\Gamma,\eta,\Gamma',\eta'$ are 2d boundary Fermi multiplets with charges
\be \label{AGe}
\begin{array}{c|cc|ccccc}
& U(N_c) & U(N_c') & SU(N_f) & \widetilde{SU(N_f)} & U(1)_a& U(1)_y & U(1)_R \\\hline
\Gamma & \ol{\mb N_c} & \ol{\mb N_c'} &\mb 1&\mb 1&0&0&0 \\
\eta & \text{det} & \text{det}^{-1} &\mb1 & \mb 1&0&1&0 \\
\Gamma' & \ol{\mb N_c} & \ol{\mb N_c'} &\mb 1&\mb 1&0&0&0 \\
\eta' & \text{det} & \text{det}^{-1} &\mb1 & \mb 1&0&-1&0
\end{array}\ee
For example, $\Gamma,\eta$ are charged under the dynamical $U(N_c')$ of Theory B, but also carry charge under a $U(N_c)_\partial$ boundary flavor symmetry, consistent with $\CD$ b.c. in Theory A. The opposite is true of $\Gamma',\eta'$.
The presence of these 2d fermis and their charges can be inferred from matching boundary anomalies, as we show momentarily. There are also two general consistency checks showing that the pairs of boundary conditions in \eqref{Abc} are \emph{reasonable}:
\begin{itemize}
\item[1)] Neither requires a matrix factorization of the superpotential in Theory B. In particular, both $(\overset{Q'}{\text{N}},\overset{\tilde Q'}{\text{N}},\overset{M}{\text{D}})$ and $(\overset{Q'}{\text{N}},\overset{\tilde Q'}{\text{N}},\overset{M}{\text{D}})$ set the cubic term in the superpotential to zero, satisfying the requirements of Section \ref{sec:mx}. In addition, $(\CD,\overset{V^\pm}{\text{D}})$ explicitly sets the monopole term in the superpotential to zero, while $(\CN,\overset{V^\pm}{\text{N}})$ does so by killing the $U^\pm$ monopole operators with $\CN$ b.c. on the gauge multiplet.
\item[2)] The choice of D vs. N b.c. for the various matter fields either fully aligned or fully anti-aligned with the signs of the fields' axial charges. This turns out to be important in recovering pure level-rank duality by turning on a large axial mass and integrating out all the bulk matter --- in particular, it ensures that after the bulk matter is integrated out, no edge modes are left behind. We will say more Section \ref{sec:Aflow}.
\end{itemize}

Two natural duality interfaces can be produced from \eqref{Abc} by applying the procedure of Section \ref{sec:id}:%
\be
\overset{\text{Theory B}}{ (\CN,\overset{Q'}{\text{N}},\overset{\tilde Q'}{\text{N}},\overset{M}{\text{D}},\overset{V^\pm}{\text{N}})} \big|\Gamma,\eta\big|   \overset{\text{Theory A}}{ (\CN,\overset{Q}{\text{N}},\overset{\tilde Q}{\text{N}}) } \qquad\text{or}\qquad
\overset{\text{Theory A}}{(\CN,\overset{Q}{\text{N}},\overset{\tilde Q}{\text{N}})}\big|\Gamma',\eta'\big|    \overset{\text{Theory B}} {(\CN,\overset{Q'}{\text{N}},\overset{\tilde Q'}{\text{N}},\overset{M}{\text{D}},\overset{V^\pm}{\text{N}})}\,.
\label{Aint}
\ee
Each interface carries a superpotential,
\be \int d\theta^+\big[ \mu Q\tilde Q+ Q' \Gamma Q]\quad \big(E_\Gamma = \tilde Q'\tilde Q)\qquad\text{or}\qquad
  \int d\theta^+\big[  Q\tilde Q\mu+ Q \Gamma' Q']\quad \big(E_{\Gamma'} = \tilde Q\tilde Q') \ee
that factorizes the \emph{cubic} part of the bulk superpotential from Theory B, by a now familiar mechanism.
Namely, the $\mu Q\tilde Q$ coupling sets $M=Q\tilde Q$ at the interface, and then $E_\Gamma \cdot J_\Gamma = Q\tilde Q Q'\tilde Q'$ (with all gauge and flavor indices contracted), which equals the bulk $W = MQ'\tilde Q'$. 

We would expect the Fermi multiplet $\eta$ to play a role in 1) factorizing the $U^-V^++U^+V^-$ part of the bulk superpotential at the interface; and 2) mediating the relation between monopole operators of Theory A and the $V^\pm$ singlets of Theory B at the interface.
The latter function was accomplished by $\Gamma$ in SQED $\leftrightarrow$ XYZ and SQCD $\leftrightarrow$ deyYZ in Section \ref{sec:XYZ}, \emph{cf.}~\eqref{XYZ-UZ}.
The precise couplings of $\eta$ to the bulk and the mechanism by which it accomplishes (1) and (2)  are not yet clear, and would be interesting to understand.

Other dual pairs of boundary conditions can be generated from \eqref{Abc}, either by deforming both sides, or by colliding with the interface \eqref{Aint}. The pattern is a generalization of Table \ref{tab:XYZ} for SQED\,$\leftrightarrow$\,XYZ.

\subsection{Anomalies}

Now, let us check more closely that boundary anomalies match for the proposed dual pairs \eqref{Abc}.
We denote the field strengths for various gauge and global symmetries as
\be
\begin{array}{cc|ccccc}
 U(N_c) & U(N_c') & SU(N_f) & \widetilde{SU(N_f)} & U(1)_a& U(1)_y & U(1)_R \\\hline
\mb s & \mb s' & \mb x & \tilde{\mb x} & \mb a & \mb y & \mb r
\end{array}
\ee
As usual, `$\mb s$' may refer to the dynamical field strength on $\CN$ b.c., or the $U(N_c)_\pd$ flavor field strength on $\CD$ b.c., and similarly for $\mb s'$.
Then for $ (\CD,\overset{Q}{\text{D}},\overset{\tilde Q}{\text{D}})$, the boundary anomaly is
\begin{align} \CI_{\text{($\CD$,D,D)}}&=  -N_c\Tr(\mb s^2) +(\Tr\,\mb s)^2-\tfrac12 N_c^2\mb r^2 + 2(\Tr\,\mb s)\mb y \qquad (\text{gauginos, FI}) \\
&\quad+\tfrac12 \big[ N_c\Tr(\mb x^2)+N_f\Tr(\mb s^2)+2N_f(\Tr\,\mb s)(\mb a-\mb r)+N_cN_f(\mb a-\mb r)^2\big] \qquad (Q) \notag\\
&\quad+\tfrac12 \big[ N_c\Tr(\tilde{\mb x}^2)+N_f\Tr(\mb s^2)-2N_f(\Tr\,\mb s)(\mb a-\mb r)+N_cN_f(\mb a-\mb r)^2\big] \qquad (\tilde Q)  \notag\\
&= N_c'\Tr(\mb s^2)+(\Tr\,\mb s)^2+2(\Tr\,\mb s)\mb y+\tfrac12N_c\big[\Tr(\mb x^2)+\Tr(\tilde{\mb x}^2)\big]+N_cN_f(\mb a-\mb r)^2-\tfrac12 N_c^2\mb r^2\,, \notag
\end{align}
and $\CI_{\text{($\CN$,N,N)}}=-\CI_{\text{($\CD$,D,D)}}+4(\Tr\,\mb s)\mb y $. Similarly, the $ (\CN,\overset{Q'}{\text{N}},\overset{\tilde Q'}{\text{N}},\overset{M}{\text{D}},\overset{V^\pm}{\text{N}})$ anomaly is
\begin{align} \CI_{\text{($\CN$,N,N,D,N)}} &= +N_c'\Tr(\mb s'{}^2) -(\Tr\,\mb s')^2+\tfrac12 N_c'{}^2\mb r^2 + 2(\Tr\,\mb s')\mb y \qquad (\text{gauginos, FI}) \\
&\quad -\tfrac12 \big[ N_c'\Tr(\mb x^2)+N_f\Tr(\mb s'{}^2)+2N_f(\Tr\,\mb s)(-\mb a)+N_c'N_f(-\mb a)^2\big] \qquad (Q') \notag\\
&\quad -\tfrac12 \big[ N_c'\Tr(\tilde{\mb x}^2)+N_f\Tr(\mb s'{}^2)-2N_f(\Tr\,\mb s)(-\mb a)+N_c'N_f(-\mb a)^2\big] \qquad (\tilde Q')  \notag\\
&\quad + \tfrac12 \big[ N_f(\Tr(\mb x^2)+\Tr(\tilde{\mb x}^2)+N_f^2(2\mb a-\mb r)^2\big] \qquad (M) \notag \\
&\quad -\tfrac12\big[ (-N_f\mb a+\mb y+N_c'\mb r)^2+(-N_f\mb a-\mb y+N_c'\mb r)^2\big] \qquad (V^\pm)\,, \notag 
\end{align}
and $\CI_{\text{($\CD$,D,D,N,D)}} = -\CI_{\text{($\CN$,N,N,D,N)}}+4(\Tr\,\mb s')\mb y$. Quite beautifully, we find
\be \label{A-match} \begin{array}{c} \CI_{\text{($\CD$,D,D)}} = \CI_{\text{($\CN$,N,N,D,N)}} + \CI_\Gamma+\CI_\eta\,, \\[.2cm]
\CI_{\text{($\CN$,N,N)}} + \CI_{\Gamma'}+\CI_{\eta'} = \CI_{\text{($\CD$,D,D,N,D)}}\,,
\end{array}
\ee
where $\CI_\Gamma=\CI_{\Gamma'}=N_c'\Tr(\mb s^2)+2(\Tr\,\mb s)(\Tr\,\mb s')+N_c\Tr(\mb s'{}^2)$ and $\CI_\eta=(\Tr\,\mb s-\Tr\,\mb s'+\mb y)^2$, $\CI_{\eta'}=(\Tr\,\mb s-\Tr\,\mb s'-\mb y)^2$ are precisely the anomalies of the additional 2d Fermi multiplets we proposed.

In particular, the ($\CN$,N,N,D,N) boundary condition in Theory B has gauge and mixed gauge-topological anomalies that are precisely cancelled by coupling to the boundary fermis $\Gamma,\eta$; and the ($\CN$,N,N) b.c. in theory A has similar anomalies that are cancelled by coupling to $\Gamma',\eta'$.

\subsection{Half-indices}

To compute half-indices of the boundary conditions in \eqref{Abc}, we introduce fugacities
\be
\begin{array}{cc|ccccc}
 U(N_c) & U(N_c') & SU(N_f) & \widetilde{SU(N_f)} & U(1)_a& U(1)_y & U(1)_R \\\hline
 s,m &  z,n &  x & \tilde{ x} &  a &  y & -q^{\frac12}
\end{array}
\ee
constrained so that $\prod_{\alpha=1}^{N_f} x_\alpha = \prod_{\alpha=1}^{N_f} \tilde x_\alpha=1$.
Then we find
\be \label{II-AD1} \II_{\text{($\CD$,D,D)}} = \frac{1}{(q)_\infty^{N_c}}\sum_{m\in \Z^N} \frac{q^{\frac12\big[N_c'm\cdot m+(\sum_i m_i)^2\big]}s^{N_c'm}\big(y\,{\textstyle\prod_i s_i}\big)^{\sum_i m_i}}
{\prod_{i\neq j} (q^{1+{m_i-m_j}}s_i/s_j;q)_\infty } \prod_{1\leq i\leq N\atop 1\leq \alpha\leq N_f} \II_{\rm D}(q^{m_i}as_ix_\alpha;q) \II_{\rm D}\Big(\frac{a}{q^{m_i}s_i\tilde x_\alpha;q}\Big)\,,
\ee
\begin{align} \label{II-AN1} \II_{\text{($\CN$,N,N,D,N)}+\Gamma,\eta}  &=
\II_{\rm N}\Big(\frac{(-q^{\frac12})^{N_c'+1}y}{a^{N_f}}\Big)\II_{\rm N}\Big(\frac{(-q^{\frac12})^{N_c'+1}}{a^{N_f}y}\Big)
 \prod_{\alpha,\beta=1}^{N_f} \II_{\rm D}\Big(\frac{a^2x_\alpha}{\tilde x_\beta}\Big) \\
 &\quad \times
 \frac{(q)_\infty^{N_c'}}{N_c'!} \oint \prod_{i=1}^{N_c'}\frac{dz_i}{2\pi i z_i}
 \prod_{i\neq j} (qz_i/z_j;q)_\infty \prod_{1\leq i\leq N_c'\atop 1\leq \alpha\leq N_f} \II_{\rm N}\Big(\frac{-q^{\frac12}z_i}{ax_\alpha};q\Big)\II_{\rm N}\Big(\frac{-q^{\frac12}\tilde x_\alpha}{az_i};q\Big) \notag \\
  &\hspace{1.3in}\times  \text{F}\Big(\frac{-q^{\frac12}y\prod_i  s_i}{\prod_j z_j};q\Big) \prod_{1\leq i \leq N_c\atop 1 \leq j \leq N_c'}\text{F}(-q^{\frac12}s_iz_j;q)\,,
\notag \end{align}
as well as
\begin{align} \II_{\text{($\CN$,N,N)}+\Gamma',\eta'} &= \frac{(q)_\infty^{N_c}}{N_c!}
\oint \prod_{i=1}^{N_c}\frac{ds_i}{2\pi i s_i}  \prod_{i\neq j} (qs_i/s_j;q)_\infty
\prod_{1\leq i\leq N\atop 1\leq \alpha\leq N_f} \II_{\rm N}(as_ix_\alpha;q) \II_{\rm N}\Big(\frac{a}{s_i\tilde x_\alpha};q\Big)  \\
  &\hspace{1.3in}\times  \text{F}\Big(\frac{-q^{\frac12}\prod_i  s_i}{y\prod_j z_j};q\Big) \prod_{1\leq i \leq N_c\atop 1 \leq j \leq N_c'}\text{F}(-q^{\frac12}s_iz_j;q)\,, \notag
\end{align}
\begin{align} \II_{\text{($\CD$,D,D,N,D)}}  &=
\II_{\rm D}\Big(\frac{(-q^{\frac12})^{N_c'+1}y}{a^{N_f}}\Big)\II_{\rm D}\Big(\frac{(-q^{\frac12})^{N_c'+1}}{a^{N_f}y}\Big)
 \prod_{\alpha,\beta=1}^{N_f} \II_{\rm N}\Big(\frac{a^2x_\alpha}{\tilde x_\beta}\Big) \\
 &\hspace{-.5in} \times
 \frac{1}{(q)_\infty^{N_c'}} \sum_{n\in \Z^{N_c'}} 
 \frac{ q^{\frac12\big[N_cn\cdot n+(\sum_i n_i)^2\big]}z^{N_cn}\big(y\,{\textstyle\prod_i z_i}\big)^{\sum_i n_i} } {\prod_{i\neq j} (q^{n_i-n_j}z_i/z_j;q)_\infty}
  \prod_{1\leq i\leq N_c'\atop 1\leq \alpha\leq N_f} \II_{\rm D}\Big(\frac{-q^{\frac12+n_i}z_i}{ax_\alpha};q\Big)\II_{\rm D}\Big(\frac{-q^{\frac12-n_i}\tilde x_\alpha}{az_i};q\Big)\,. \notag 
\end{align}
We expect that
\be \begin{array}{c} \II_{\text{($\CD$,D,D)}}(s,x,\tilde x,q,y;q) = \II_{\text{($\CN$,N,N,D,N)}+\Gamma,\eta}(s,x,\tilde x,q,y;q) \\[.2cm]
 \II_{\text{($\CN$,N,N)}+\Gamma',\eta'}(z,x,\tilde x,q,y;q) = \II_{\text{($\CD$,D,D,N,D)}} (z,x,\tilde x,q,y;q)\,,
\end{array} \ee
and have checked these and have checked these in a variety of examples by comparing (at least) the first ten terms in the $q$-series expansion.%
\footnote{For comparison, the full superconformal index for many Aharony dual pairs was computed in \cite{Bashkirov-Aharony}.}

For illustrative purposes, let us examine one nontrivial example,  the duality $\II_{\text{($\CD$,D,D)}}(s,x,\tilde x,q,y;q) = \II_{\text{($\CN$,N,N,D,N)}+\Gamma,\eta}(s,x,\tilde x,q,y;q)$ with $N_c = N_c'=2, N_f = 4$. For those who wish to reproduce some of our computations in \texttt{Mathematica}, we note that in most examples it is convenient to make the substitution $a \rightarrow a/q$, which has the nice feature of improving convergence in the monopole sum. The expressions quickly become unwieldy, so we only display a few terms:
\begin{align*}
&\II(s,x,\tilde x,q,y;q) = \\
&1+\left(\frac{s_1}{s_2}+\frac{s_2}{s_1}+2\right) q+\frac{q^{3/2} \left(s_2^4 s_1^6 y^2+s_2^5 s_1^5 y^2+s_2^6 s_1^4 y^2+s_1^2+s_2 s_1+s_2^2\right)}{s_1^3 s_2^3 y} \\
&+q^2
   (6-\frac{s_1 \tilde{ x}_1}{a}-\frac{s_1 \tilde{ x}_2}{a \tilde{ x}_1}-\frac{s_1 \tilde{ x}_3}{a \tilde{ x}_2}-\frac{s_1}{a \tilde{ x}_3}-\frac{s_2 \tilde{ x}_1}{a}-\frac{s_2 \tilde{ x}_2}{a \tilde{ x}_1}-\frac{s_2 \tilde{ x}_3}{a \tilde{ x}_2}-\frac{s_2}{a
   \tilde{ x}_3}-\frac{x_3}{a s_2}-\frac{1}{a s_2 x_1}-\frac{x_1}{a s_2 x_2}-\frac{x_2}{a s_2 x_3} \\ &-\frac{x_3}{a s_1}-\frac{1}{a s_1 x_1}-\frac{x_1}{a s_1 x_2}-\frac{x_2}{a s_1
   x_3}+2 \left(\frac{s_1}{s_2}+\frac{s_2}{s_1}\right)+\frac{s_1^2+s_2 s_1}{s_2^2}+\frac{s_2^2+s_1 s_2}{s_1^2}) + \ldots
\end{align*}

\section{Flows from Aharony duality}
\label{sec:Aflow}

It is well known (\emph{cf.} \cite{BCC-Seiberg, 4d3d-dualities}) that Aharony dualities can be deformed by real masses in such a way that they flow to interesting ``chiral'' dualities with non-zero Chern-Simons levels, including Giveon-Kutasov duality \cite{GiveonKutasov} and level-rank dualities in pure gauge theory from Section \ref{sec:LR}. 
With some care, the flows can be performed in the presence of boundary conditions. Then the basic dual boundary conditions \eqref{Abc} in Aharony duality can be used to derive (or re-derive) dual boundary conditions for other pairs of bulk theories. We discuss the basic technique, and illustrate it in a few examples.

\subsection{$U(N)_{k+N}\leftrightarrow U(k)_{-k-N}$ level-rank duality}
\label{sec:A-UU}

Consider the Aharony duality of Section \ref{sec:Aharony} with $N_c=N$ and $N_f=k+N$, where $k,N>0$. Turning on a large real mass for the axial $U(1)_a$ symmetry allows all matter fields to be integrated out in the \emph{bulk} of both Theory A and Theory B. Integrating out fermions shifts the Chern-Simons level in a manner that depends on the sign of the mass:
\begin{itemize}
\item With a large positive axial mass, the bulk Theory A flows to pure $\CN=2$ $U(N)_{k+N}$ Yang-Mills-Chern-Simons theory, and Theory B flows to a pure $\CN=2$ $U(k)_{-k-N}$ YM-CS theory. The global symmetry is reduced to $U(1)_y\times U(1)_R$ (the remainder of the flavor group acts trivially).
There are some non-trivial background Chern-Simons terms; after an equal shift of the R-R term on both sides (by $-(\tfrac12 N^2+kN)\mb r^2$) we find that the bulk CS terms correspond to the anomaly polynomials
\be \label{apos}  \hspace{-.1in}\begin{array}{ll}
\text{Theory A:\quad $U(N)_{k+N}$} \qquad &\CI^A_{\rm bulk} = (k+N) \Tr(\mb s^2)+2\mb y  \Tr\,\mb s+\tfrac12 N^2\mb r^2\,,\\[.2cm]
\text{Theory B:\quad $U(k)_{-k-N}$} \qquad  & \CI^B_{\rm bulk} = -(k+N)\Tr(\mb s'{}^2) +2\mb y \Tr\,\mb s' - \mb y^2 +\tfrac12 N^2\mb r^2\,.
\end{array}\ee
\item With a large negative axial mass, we instead find 
\be \label{aneg}  \hspace{-.1in}\begin{array}{ll}
\text{Theory A:\quad $U(N)_{-k-N}$} \qquad &\CI^A_{\rm bulk} = -(k+N) \Tr(\mb s^2)+2\mb y  \Tr\,\mb s-\tfrac12 N^2\mb r^2\,,\\[.2cm]
\text{Theory B:\quad $U(k)_{k+N}$} \qquad   &\CI^B_{\rm bulk} = (k+N)\Tr(\mb s'{}^2) +2\mb y \Tr\,\mb s' + \mb y^2 -\tfrac12 N^2\mb r^2\,.
\end{array}\ee
\end{itemize}
These coincide with the level-rank duality of Section \ref{sec:LR}.

We can derive the dual boundary conditions in level-rank duality from dual boundary conditions in Aharony duality, but in doing so we must be a little careful. Turning on a real mass $m_\Phi$ for a 3d $\CN=2$ chiral multiplet allows it to be integrated out in the bulk. However, in the presence of a boundary condition, it may acquire a massless edge mode, in the form of a purely 2d $\CN=(0,2)$ chiral or Fermi multiplet.
We recall from Section \ref{sec:anom} that
\begin{itemize}
\item a 3d $\CN=2$ chiral with N b.c. has an $\CN=(0,2)$ chiral edge mode if $m_\Phi>0$
\item a 3d $\CN=2$ chiral with D b.c. has an $\CN=(0,2)$ chiral edge mode if $m_\Phi<0$
\end{itemize}
The presence of these edge modes is necessary to ensure that the boundary anomaly remains constant during a flow.

The simplest way to flow from Aharony to level-rank duality on a half-space is to choose boundary conditions that ensure that no edge modes survive. Recall the two dual pairs of b.c. from \eqref{Abc}:
\be \begin{array}{ccc}
\text{Theory A} && \text{Theory B} \\
  (\CD,\overset{Q}{\text{D}},\overset{\tilde Q}{\text{D}})  &\quad\leftrightarrow\quad& (\CN,\overset{Q'}{\text{N}},\overset{\tilde Q'}{\text{N}},\overset{M}{\text{D}},\overset{V^\pm}{\text{N}}) + \Gamma,\eta \\[.2cm]
   (\CN,\overset{Q}{\text{N}},\overset{\tilde Q}{\text{N}}) + \Gamma',\eta'  &\quad\leftrightarrow\quad& (\CD,\overset{Q'}{\text{D}},\overset{\tilde Q'}{\text{D}},\overset{M}{\text{N}},\overset{V^\pm}{\text{D}})\,,
\end{array} \label{Abc-flow}\ee
We use the top pair of boundary conditions when turning on a large, \emph{positive} axial mass $m_a$. Since the effective real mass of each 3d chiral is proportional to its axial charge, this means that $Q,\tilde Q,M$ will have positive mass while $Q',\tilde Q',V^\pm$ have negative mass; therefore, no edge modes will arise during the flow. We deduce that
\be \text{$U(N)_{k+N}$ YM-CS with $\CD$ b.c.} \quad\leftrightarrow\quad
 \text{$U(k)_{-k-N}$ YM-CS with $\CN$ b.c. + $\Gamma,\eta$}\,,  \label{A-LR1}\ee
where $\Gamma,\eta$ are bifundamental and bi-det Fermi multiplets, just as in level-rank duality, Section \ref{sec:LR-N}. Alternatively, we can use the bottom pair of b.c. when turning on a large, \emph{negative} axial mass, resulting in
\be \text{$U(N)_{-k-N}$ YM-CS with $\CN$ b.c. + $\Gamma',\eta'$} \quad\leftrightarrow\quad
 \text{$U(k)_{k+N}$ YM-CS with $\CD$ b.c.}\,. \label{A-LR2} \ee
This is of course the same as \eqref{A-LR1}.

In a similar way, we can flow from the interfaces \eqref{Aint} in Aharony duality to the interface \eqref{LR-UU} in level-rank duality. No matter whether we turn on a large positive mass in the first interface of \eqref{Aint} or a large negative mass in the second interface, we end up with the same level-rank interface
\be (U(N)_{-k-N},\CN)\big|\Gamma,\eta\big|(U(k)_{k+N},\CD)\,, \ee
which necessarily has a negative level on the left and a positive level on the right.

In principal, one could consider turning on an axial mass of the ``wrong'' sign. This does generate edge modes, which need to be kept track of carefully. In particular, it may introduce additional 2d $\CN=(0,2)$ chiral multiplets on the boundary. Charged 2d $\CN=(0,2)$ chirals can be problematic, as we first found in particle-vortex duality, \emph{cf.} \eqref{badN}.

For example, suppose that we introduce a large negative axial mass in the top pair of boundary conditions in \eqref{Abc-flow}. This will lead to chiral edge modes of $Q_{3d}',\tilde Q_{3d}'$, which are potentially problematic. A possibly way to remedy the problem is to first flip the D b.c. on the $Q_{3d},\tilde Q_{3d}$ chirals in Theory A, by introducing boundary (0,2) chiral multiplets $q,\tilde q$ of the same charges. Then, in the presence of negative axial mass, Theory A flows to $U(N)_{-k-N}$ CS-YM theory with $\CD$ b.c., while Theory B flows to $U(N)_{k+N}$ YM-CS theory with $\CN$ b.c. coupled to a large collection of 2d multiplets:

- The fermis $\Gamma,\eta$

- A Fermi $\mu$ that's a surviving edge mode of $M_{3d}$

- Chirals $Q',\tilde Q',V^\pm$ that are surviving edge modes of the corresponding bulk fields

- Chirals $q,\tilde q$ from flipping $Q,\tilde Q$ in Theory A \\
These various 2d multiplets are coupled together by a boundary superpotential $\int d\theta^+\big[\mu q\tilde q+q \Gamma Q'\big]$, as well as $E_\mu = -Q'\tilde Q'$ and $E_\Gamma=\tilde q \tilde Q'$. The $V^\pm$ are free 2d chirals, but not necessarily problematic because they are uncharged under the gauge symmetry.

\subsection{$U(N)_{k+N-\frac{N_f}2}\leftrightarrow U(k)_{-k-N+\frac{N_f}{2}}$ with $N_f$ fundamentals}
\label{sec:A-F}

It is not necessary to integrate out all the matter when flowing from Aharony duality. Here and in Section \ref{sec:A-AF} we discuss two particularly well-behaved scenarios in which some 3d chiral matter remains.

Consider the basic bulk Aharony duality of Section \ref{sec:Aharony}, and write
\be N_c=N\,,\qquad N_f=k+N\,,\qquad N_c'=k \ee
as before.
Suppose we want integrate out all the anti-fundamental chirals $\tilde Q_{3d}$ in Theory A, leaving behind fundamentals $Q_{3d}$. This can be done by first redefining axial charge in Theory~A by adding $-1$ times the charge under the $U(1)$ part of the gauge group, and by $N_f$ times the topological charge. Then introducing a large positive (say) axial mass gives $\tilde Q_{3d}$ a positive mass, keeps $Q_{2d}$ massless, and (thanks to the mixing with  topological charge) maintains a vanishing effective FI parameter. The bulk Theory A flows to
\be \text{Theory A: \quad $U(N)_{\frac{N_f}{2}} = U(N)_{k+N-\frac{N_f}{2}}$ with $N_f$ fundamental chirals $Q_{3d}$}\,, \ee
and the flow generates bulk Chern-Simons terms for gauge and global symmetries encoded in the polynomial
\be \label{I-AF1} \CI_{\rm bulk}^A = \tfrac{N_f}{2}\Tr(\mb s^2)+(\Tr\,\mb s)\big[2\mb y+N_f\mb r\big]+\tfrac{N N_f}2 \mb r^2\,. \ee
By inspecting effective Chern-Simons levels on the Coulomb branch of the theory, one also finds that there is a single scalar, gauge-invariant monopole remaining.

We may track the effect of the same deformation in Theory B.
We similarly start by redefining axial charge by adding $N_f$ times topological charge and (now) $+1$ times $U(1)'$ gauge charge. Then introducing a large positive axial mass gives the $\tilde Q_{3d}'$ anti-fundamental chirals and the $V^-_{3d}$ singlet a negative real mass, gives all the singlets $M_{3d}$ a positive mass, and keeps $Q_{3d}'$ and $V^+_{3d}$ massless. The bulk theory flows to
\be  \text{Theory B: \quad $U(k)_{-\frac{N_f}{2}} = U(k)_{-k-N+\frac{N_f}{2}}$ with $N_f$ fund's $Q_{3d}'$ and a singlet $V^+_{3d}$}\,, \ee
with bulk Chern-Simons terms
\be \label{I-AF2} \CI_{\rm bulk}^B  = -\tfrac{N_f}{2}\Tr(\mb s'{}^2)+2(\Tr\,\mb s')\mb y+k\,\mb r\,\mb y-\tfrac{N^2}{2}\mb r^2-\tfrac12\mb y^2\,. \ee
We summarize the charges of the remaining matter fields in Theories A and~B:
\be 
\begin{array}{c|cc|ccc}
& U(N) & U(k) & SU(N_f) & U(1)_y & U(1)_R \\\hline
Q & \mb N & \mb 1 & \mb N_f &0&0 \\\hline
Q' & \mb 1 & \mb k & \ol{\mb N_f} &0&1 \\
V^+ &\mb 1 & \mb 1 & \mb 1 & \pm 1 & k+1
\end{array}
\ee

These bulk flows are nicely compatible with the first pair of dual boundary conditions in Aharony duality \eqref{Abc}, namely ($\CD$,D,D) in Theory A and ($\CN$,N,N,D,N)$+\Gamma,\eta$ in Theory B. These boundary conditions ensure that no edge modes arise as some of the bulk fields become massive, with positive axial mass as above. The boundary Fermi multiplets $\Gamma,\eta$ are untouched by the flow, and continue to have charges
\be
\begin{array}{c|cc|ccc}
& U(N) & U(k) & SU(N_f) & U(1)_y & U(1)_R \\\hline
\Gamma & \ol{\mb N} & \ol{\mb k} & \mb 1&0&0 \\
\eta & \text{det} & \text{det}^{-1} & \mb 1 &1&0 
\end{array}
\ee
Thus we derive a duality of boundary conditions
\be \label{A-Fbc} (\CD,\overset{Q}{\text{D}}) \quad\leftrightarrow \quad (\CN,\overset{Q'}{\text{N}},\overset{V^+}{\text{N}})+\Gamma,\eta\,. \ee
Similarly, we obtain a duality interface
\be \overset{\text{Theory B}}{(\CN,\text{N},\text{N})}\big|\Gamma,\eta\big|\overset{\text{Theory A}}{(\CN,\text{N})}\, \ee
with superpotential $\int d\theta^+ \Gamma Q Q'$.

Note that the boundary anomalies in the dual pair \eqref{A-Fbc} are guaranteed to match by consistency of the flow. Indeed, the usual computation of anomalies gives
\begin{align}
\CI_{(\CD,\text{D})} &= \CI^A_{\rm bulk} -N_c\Tr(\mb s^2) +(\Tr\,\mb s)^2-\tfrac12 N_c^2\mb r^2
+\tfrac12 \big[ N_c\Tr(\mb x^2)+N_f\Tr(\mb s^2)+2N_f(\Tr\,\mb s)(-\mb r)+N_cN_f(-\mb r)^2\big] \notag \\
& = \CI_{(\CD,\text{D},\text{D})}^{\text{Aharony}}\,,\\
\CI_{(\CN,\text{N},\text{N})} &= \CI^A_{\rm bulk}+k\Tr(\mb s'{}^2) -(\Tr\,\mb s')^2+\tfrac12 k^2\mb r^2
-\tfrac12 \big[ k\Tr(\mb x^2)+N_f\Tr(\mb s'{}^2)\big]  -\tfrac12(\mb y+k'\mb r)^2 \notag \\
&= \CI_{\text{($\CN$,N,N,D,N)}}^{\text{Aharony}}\,,
\end{align}
as required for anomaly matching in the RG flows. Therefore, \eqref{A-match} implies the desired
\be \CI_{(\CD,\text{D})} = \CI_{(\CN,\text{N},\text{N})} + \CI_\Gamma + \CI_\eta\,. \ee

The equality of half-indices corresponding to the pair of b.c. \eqref{A-Fbc} also follows from equality of the half-indices in Aharony duality \eqref{II-AD1}, \eqref{II-AN1}, by applying a limit to fugacities that mimics the limit of real masses in the flow. Specifically, we should redefine
\be s_i\to s_i/a\,,\qquad z_i\to a z_i\,,\qquad y\to a^{N_f}y \ee
and then send $a\to\infty$. Conveniently, the prefactor in the $\II_{(\CD,\text{D,D})}$ index (that depends on the boundary anomaly) remains independent of $a$, thanks to our careful mixing of axial and topological symmetries. Moreover, $a$ only appears in the arguments of quantum dilogarithms with negative exponents, thanks to our careful alignment of N vs. D b.c. and axial charges. Thus the $a\to\infty$ is well defined, and simply sends every quantum dilogarithm containing a negative power of $a$ to `$1$'. This just means that the contributions of all massive bulk fields disappear. Assuming equality of the half-indices \eqref{II-AD1}, \eqref{II-AN1}, we derive an equality
\begin{align} \label{II-AF} \II_{\text{($\CD$,D)}} &= \frac{1}{(q)_\infty^{N}}\sum_{m\in \Z^N} \frac{q^{\frac12\big[km\cdot m+(\sum_i m_i)^2\big]}s^{km}\big(y\,{\textstyle\prod_i s_i}\big)^{\sum_i m_i}}
{\prod_{i\neq j} (q^{1+{m_i-m_j}}s_i/s_j;q)_\infty } \prod_{1\leq i\leq N\atop 1\leq \alpha\leq N_f} \II_{\rm D}(q^{m_i}s_ix_\alpha;q)  \\
&=  \frac{(q)_\infty^{k}}{k!} \oint \prod_{i=1}^{k}\frac{dz_i}{2\pi i z_i}
 \prod_{i\neq j} (qz_i/z_j;q)_\infty \prod_{1\leq i\leq k\atop 1\leq \alpha\leq N_f} \II_{\rm N}\Big(\frac{-q^{\frac12}z_i}{x_\alpha};q\Big) \text{F}\Big(\frac{-q^{\frac12}y\prod_i  s_i}{\prod_j z_j};q\Big) \prod_{1\leq i \leq N\atop 1 \leq j \leq k}\text{F}(-q^{\frac12}s_iz_j;q) \notag \\
 &=\II_{\text{($\CN$,N,N)}+\Gamma,\eta}\,. \notag
\end{align}

Finally, note that we could equally well have triggered the above flows with a large negative axial mass.
This leads to
\be \begin{array}{cc}
\text{Theory A}:\quad & \text{$U(N)_{-k-N+\tfrac{N_f}{2}}$ with $N_f$ fundamentals $Q_{3d}$} \\[.2cm]
\text{Theory B}:\quad & \text{$U(N)_{k+N-\tfrac{N_f}{2}}$ with $N_f$ fundamentals $Q'_{3d}$ + a singlet $V^-_{3d}$}
\end{array}\ee
Now the Aharony-dual boundary conditions that avoid edge modes are  ($\CN$,N,N)$+\Gamma',\eta'$ and ($\CD$,D,D,N,D), which flow to
\be (\CN,\overset{Q}{\text{N}})+\Gamma',\eta' \quad\leftrightarrow \quad (\CD,\overset{Q'}{\text{D}},\overset{V^+}{\text{D}}), \ee
and the opposite interface $\overset{\text{Theory A}}{(\CN,\text{N})}\big|\Gamma',\eta'\big|\overset{\text{Theory B}}{(\CN,\text{N},\text{N})}$.

\subsection{$ SU(N)_{k + N - \frac{N_f}{2}} \leftrightarrow  U(k)_{-k-N + \frac{N_f}{2}, -k + \frac{N_f}{2}} $ with $N_f$ fundamentals}
\label{sec:A-SU}

Just as in level-rank duality, one may obtain a U-SU duality with matter from U-U by gauging the topological $U(1)_y$ symmetry.%
\footnote{At the level of the full superconformal index, such gauging procedures for Seiberg-like dual pairs were discussed in \cite{park2013seiberg}.} %
If we start from the duality with fundamental matter from Section \ref{sec:A-F}, gauging the topological symmetry results in a $SU(N)_{k + N - \frac{N_f}{2}} \leftrightarrow  U(k)_{-k-N + \frac{N_f}{2}, -k + \frac{N_f}{2}}$ duality in the bulk, with $N_f$ fundamental chirals on each side.

On the boundary, the baryonic Fermi multiplet $\eta$ is lost by the same mechanism as in Section \ref{sec:UtoSU}. 
The $N$ fundamental Fermi multiplets $\Gamma$ coupled to Neumann b.c. are sufficient to cancel gauge anomalies. We expect, for example, that
\be \text{$SU(N)_{k + N - \frac{N_f}{2}}$ w/ ($\CD$,D)}\quad\leftrightarrow\quad
\text{$U(k)_{-k-N + \frac{N_f}{2}, -k + \frac{N_f}{2}}$ w/ $(\CN,\text{N})+\Gamma$}\ee
A full match of 't Hooft anomalies proceeds exactly as in U-SU level-rank duality with the additional fundamental contributions.

The half-indexes are
\begin{align} \label{USUF} \II_{\text{($\CD$,D)}} &= \frac{1}{(q)_\infty^{N-1}}\sum_{m\in \Z^{N},\; \sum_i m_i = 0} \frac{q^{\frac12\big[km\cdot m\big]}s^{km}}
{\prod_{\alpha} (q^{1+{m.\alpha}}s^{\alpha};q)_\infty } \prod_{1\leq i\leq N\atop 1\leq \alpha\leq N_f} \II_{\rm D}(q^{m_i}s_ix_\alpha;q)  \\
&=  \frac{(q)_\infty^{k}}{k!} \oint \prod_{i=1}^{k}\frac{dz_i}{2\pi i z_i}
 \prod_{i\neq j} (qz_i/z_j;q)_\infty \prod_{1\leq i\leq k\atop 1\leq \alpha\leq N_f} \II_{\rm N}\Big(\frac{-q^{\frac12}z_i}{x_\alpha};q\Big) \prod_{1\leq i \leq N\atop 1 \leq j \leq k}\text{F}(-q^{\frac12}s_iz_j;q) \notag \\
 &=\II_{\text{($\CN$,N)}+\Gamma}\,. \notag
\end{align} where now the $s$ fugacities parameterize the torus of $SU(N)$ and $m$ are vectors in the cocharacter lattice of $SU(N)$.

Let us present a simple non-abelian example. On the Neumann side we consider $U(2)_{-9/2, -3/2}$ with one fundamental and three boundary Fermi multiplets, and on the Dirichlet side we consider its dual $SU(3)_{9/2}$ with one fundamental.

The Neumann side of the half-index is
\begin{align*}
\II_{(\CN, N)} &= \frac{(q)^2_{\infty}}{2} \oint \prod_{i=1}^2 \frac{dz_i }{ 2\pi i z_i} \prod_{\substack{i \neq j \\ i, j =1}}^2(z_i z_j^{-1}; q)_{\infty} \frac{\prod_{i=1}^2 F(-q^{1/2}z_i s_1)F(-q^{1/2}z_i s_2/s_1) F(-q^{1/2}z_i/s_2) }{ \prod_{i=1}^2(-q^{3/2}a^{-1} z_i; q)_{\infty}} 
\end{align*}
while the Dirichlet side is
\begin{align*}
\II_{(\CD, D)} &= \frac{1 }{ (q)^2_{\infty}}\sum_{\substack{m_1,m_2 \in \Z}}\frac{q^{2 m_1^2 - 2 m_1 m_2 + 2 m_2^2} s_1^{4 m_1 - 2m_2} s_2^{4 m_2 - 2 m_1} }{ (q^{1 + m_1 - 2 m_2} s_1/s_2^2)_{\infty} (q^{1 - m_1 - m_2}/s_1 s_2)_{\infty} (q^{1 + 2 m_1 - m_2} s_1^2/s_2)_{\infty}} \\
 &\times \frac{\II_{\rm D}((a/q) q^{-1 + m_1}s_1)\II_{\rm D} ((a/q) q^{-1 - m_2}/s_2)\II_{\rm D}((a/q) q^{-1 - m_1 + m_2} s_2/s_1) }{ (q^{1 - 2m_1 + m_2} s_2/s_1^2)_{\infty}(q^{1 + m_1 + m_2}s_1 s_2)_{\infty}(q^{1 + m_1 + 2 m+2}s_2^2/s_1)_{\infty}}
\end{align*} where, as in other examples, we have shifted the fugacity for the axial symmetry $a \rightarrow a/q$ for convenience. One can check that these expressions match order by order in $q$. The expansion may be written compactly in terms of $SU(3)$ characters 
\begin{align*}
\II_{\CT_{1}, (\CN, N)} = \II_{\CT_2, (\CD, D)} &= -\chi_{(0, 0)}(c_1, c_2) - q\chi_{(1, 1)}(c_1, c_2) \\
& \hspace{-.3in} + q^2 \big[-\chi_{(0, 0)}(c_1, c_2) + \frac1a\chi_{(0, 1)}(c_1, c_2)- \chi_{(2, 2)}(c_1, c_2) - 2 \chi_{(1, 1)}(c_1, c_2) \big]+ \ldots
\end{align*}
where
\begin{align*}
\chi_{(n_1, n_2)}(t_1, t_2) &= 
   \frac{1 }{ {\left(t_1^2-t_2\right) (t_1 t_2-1) \left(t_1-t_2^2\right)}} \Big[ -t_2^4 t_1^{n_2+1} \left(\frac{t_2}{t_1}\right)^{n_1+n_2}+t_2^3 t_1^{n_1+n_2+3}
   \left(\frac{t_2}{t_1}\right)^{n_2} \\
   & \hspace{-1cm} -\left(\frac{1}{t_2}\right)^{n_2-1}
   t_1^{n_1+n_2+4}+t_1^{n_2+3}
   \left(\frac{1}{t_2}\right)^{n_1+n_2}+\left(\frac{1}{t_2}\right)^{n_2-3}
   \left(\frac{t_2}{t_1}\right)^{n_1+n_2}-t_1 \left(\frac{1}{t_2}\right)^{n_1+n_2-1}
   \left(\frac{t_2}{t_1}\right)^{n_2} \Big].
\end{align*}

\subsection{$U(N)_{k+N-\frac{n_f+n_a}2}\leftrightarrow U(k)_{-k-N+\frac{n_f+n_a}2}$ with $n_f$ fundamentals, $n_a$ anti-fundamentals}
\label{sec:A-AF}

Generalizing the previous example, we may \emph{partially} integrate out some fundamentals and antifundamentals from Aharony duality. If we wish to avoid edge modes, we may use (say) the ($\CD$,D,D) $\leftrightarrow$ ($\CN$,N,N,D,N)$+\Gamma,\eta$ duality of boundary conditions, and make sure that both fundamentals and antifundamentals are integrated out with positive mass in Theory A, and negative mass in Theory B.

Suppose that we wish to integrate out $N_f-n_f$ fundamentals (leaving $n_f$ behind) and $N_f-n_a$ anti-fundamentals (leaving $n_a$ behind), all with positive mass in Theory A. Following \cite{BCC-Seiberg}, this can be accomplished by breaking the flavor symmetry $SU(N_f)\to SU(n_f)\times SU(N_f-n_f)\times U(1)_f$ and  $\wt{SU(N_f)}\to \wt{SU(n_a)}\times \wt{SU(N_f-n_a)}\times \wt{U(1)_f}$, and, roughly speaking, using a combination of axial, $U(1)_f$, and $\wt{U(1)_f}$ masses 
to give the last $N_f-n_f$ fundamentals ($N_f-n_a$ anti-fundamentals) a positive mass while keeping the rest massless. 
As long as $n_a,n_f<N_f$, no scalar, gauge-invariant monopole operators remain.

The corresponding transformation in Theory B gives the last  $N_f-n_f$ fundamentals ($N_f-n_a$ anti-fundamentals) a negative mass. It also gives some of the mesons a positive mass, and (due to mixing with the topological symmetry) it turns out to give the $V^\pm$ singlets a negative mass so long as $n_a,n_f<N_f$. This aligns perfectly with the (N,N,D,N) b.c. in Theory B, so no edge modes are generated.

After integrating out all massive fields in the bulk, we arrive at
\be \label{LR-FA-bulk} \begin{array}{l}
\text{Theory A:\quad $U(N)_{k+N-\frac{n_f+n_a}{2}}$ with $n_f$ fund's $Q_{3d}$, $n_a$ anti-fund's $\tilde Q_{3d}$} \\[.2cm]
\text{Theory B:\quad $U(k)_{-k-N+\frac{n_f+n_a}{2}}$ with $n_f$ fund's $Q_{3d}'$, $n_a$ anti-fund's $\tilde Q'_{3d}$, $n_f\times n_a$ singlets $M_{3d}$} \\[.2cm]
\end{array}
\ee
with a superpotential $W=M_{3d}\tilde Q_{3d}\tilde Q'_{3d}$ in Theory B, and gauge and flavor charges
\be
\begin{array}{c|cc|ccccc}
& U(N) & U(k) & SU(n_f) & \widetilde{SU(n_a)} & U(1)_a& U(1)_y & U(1)_R \\\hline
Q & \mb N & \mb 1 & \mb n_f & \mb 1 & 1&0&0 \\
\tilde Q  & \ol{\mb N} & \mb 1 & \mb 1 & \ol{\mb n_a} & 1 &0&0 \\\hline
Q' & \mb 1 & \mb k & \ol{\mb n_f} & \mb 1 & -1&0&1 \\
\tilde Q' & \mb 1 & \ol{\mb k} & \mb 1 & \mb n_a & -1 &0&1 \\
M & \mb 1 & \mb 1 & \mb n_f & \ol{\mb n_a} & 2 &0&0 
\end{array}
\ee
The full anomaly polynomials encoding the new bulk UV CS levels come out as%
\footnote{To obtain this form, we shifted the global CS levels for both theories A and B by a constant amount.}
\begin{align} \CI^A_{\rm bulk} &= \big(k+N-\tfrac{n_f+n_a}{2}\big)\Tr(\mb s^2)+(\Tr\,\mb s)\big[2\mb y+(n_a-n_f)(\mb a-\mb r)\big] \\
\CI^B_{\rm bulk} &= -\big(k+N-\tfrac{n_f+n_a}{2}\big)\Tr(\mb s'{}^2)+(\Tr\,\mb s')\big[2\mb y+(n_a-n_f)\mb a\big] \\
 &\quad + \tfrac12(k+N-n_a)\Tr\,\mb x^2+\tfrac12(k+N-n_f)\Tr\,\tilde{\mb x}^2 \notag \\
 &\quad + \big[\tfrac12(n_a+n_f)(k+N)-2n_an_f\big]\mb a^2 +\big[-(n_a+n_f)N+2n_an_f\big]\mb a\,\mb r \notag \\
 &\quad -\frac12\big[k^2+(N-n_a)(N-n_f)\big]\mb r^2-\mb y^2 \notag
\end{align}

The Aharony-dual boundary conditions ($\CD$,D,D) $\leftrightarrow$ ($\CN$,N,N,D,N)$+\Gamma,\eta$ now flow to
\be \begin{array}{l@{\quad\leftrightarrow\quad}l}
  (\CD,\overset{Q}{\text{D}},\overset{\tilde Q}{\text{D}})  & (\CN,\overset{Q'}{\text{N}},\overset{\tilde Q'}{\text{N}},\overset{M}{\text{D}}) + \Gamma,\eta\,, 
\end{array}\label{LR-FA}\ee
with the Fermi multiplets $\Gamma,\eta$ remaining untouched --- they have the same bifundamental and bi-det charges as in \eqref{AGe}. Anomaly matching along the RG flow on either side ensures that boundary anomalies will still match perfectly; and an identity of half-indices may be derived from the Aharony half-indices \eqref{II-AD1}--\eqref{II-AN1} by sending appropriate fugacities to infinity, just as in \eqref{II-AF}. Finally, from \eqref{LR-FA} (or by flowing with the Aharony interface \eqref{Aint}), we obtain the duality interface
\be \overset{\text{Theory B}}{(\CN,\text{N,N,D})} \big|\Gamma,\eta\big| \overset{\text{Theory A}}{(\CN,\text{N,N})} \ee
with a superpotential $\int d\theta^+\big[\Gamma Q Q'+\mu Q\tilde Q\big]$ and $E_\Gamma=\tilde Q\tilde Q'$, so as to factorize the bulk superpotential $M\tilde Q\tilde Q'$.

Alternatively, we could have used real masses of the opposite signs, and the Aharony-dual boundary conditions ($\CN$,N,N)$+\Gamma',\eta'$ $\leftrightarrow$ ($\CD$,D,D,N,D), to obtain dual boundary conditions for the bulk theories with opposite Chern-Simons levels, and an opposite interface $(\CN,\text{N,N})\big|\Gamma',\eta'\big|(\CN,\text{N,N,D})$.

\subsection{Other Seiberg-like dualities}
\label{sec:otherS}

There are many more Seiberg-like dualities in the bulk that flow from Aharony duality by integrating out chirals with a combination of positive and negative masses. This generates bulk Chern-Simons levels that are smaller than those in \eqref{LR-FA-bulk}. In the presence of our usual boundary conditions, such a flow will necessarily produce edge modes, which must be incorporated into a duality of boundary conditions in the way that was outlined at the end of Section \ref{sec:A-UU}. This leads to boundary conditions that appear rather complicated, though in principal their analysis is systematic.


\section{Adjoint matter}

We conclude with a simple example of a gauge theory coupled to adjoint matter: $SU(2)_{k}$ plus a single adjoint chiral multiplet $\Phi_{\rm 3d}$. If the Chern-Simons level is equal to $\pm1$, then according to the ``duality appetizer'' of Jafferis and Yin \cite{Jafferis-appetizer} this theory is dual to a single free $\CN=2$ chiral multiplet. We consider the following duality amuse-bouche by finding simple dual boundary conditions. 

We assume that the chiral $\Phi_{\rm 3d}$ has R-charge zero, and charge $+1$ for a $U(1)_x$ flavor symmetry. We also set all the background CS levels in the bulk to zero.

First suppose that the dynamical Chern-Simons level is non-negative, $k\geq 0$, so that Dirichlet b.c. are well behaved. Taking $\CD$ b.c. for the gauge multiplet and D b.c. for the chiral, the anomaly polynomial is
\begin{align} \CI_{\CD,\text{D}} &= k\text{Tr}(\mb f^2)-\big[2\text{Tr}(\mb f^2)+\tfrac32\mb r^2\big] + \big[2\text{Tr}(\mb f^2)+\tfrac32(\mb x-\mb r)^2\big]  \notag \\
 &= k\text{Tr}(\mb f^2) +\tfrac32\mb x^2-3\mb x\mb r\,.
\end{align}
The corresponding half-index is
\begin{equation}
\II_{\CD,\text{D}}(x, y; q) = \frac{1 }{ (q)_{\infty}} \II_{\rm D}(x;q) \sum_{m \in \mathbb{Z}} q^{k m^2} y^{2km} \frac{\II_{\rm D}(q^{2m}x y^2;q) \II_{\rm D}( q^{-2m}x y^{-2} ;q) }{(q^{1 + 2m} y^2; q)_{\infty} (q^{1 - 2m} y^{-2}; q)_{\infty}}\,,
\end{equation}
and converges nicely as long as $k\geq 0$. (A substitution $x\to x/q$ aids computations). Here $x$ is the fugacity for the bulk $U(1)_x$ flavor symmetry and $(y,y^{-1})$ are the fugacities for the $SU(2)_\pd$ boundary flavor symmetry.

For $k=1$ we find
\begin{equation}
\II^{\;k=1}_{\CD,\text{D}}(x, y; q){=} \II_{\rm D}(x^2;q) \,\chi_{0}[SU(2)_1](y)\,,
\end{equation}
\emph{i.e.} the index of  D b.c. for a free chiral $\text{Tr}(\Phi^2)$ (as predicted by the duality appetizer), together with a decoupled $SU(2)_1$ WZW model on the boundary. The WZW model carries the $\text{Tr}(\mb f^2)$ boundary 't Hooft anomaly. In order to get a perfect match of boundary anomalies, the RHS also requires background CS terms $-\tfrac12(\mb x+\mb r)^2$ in the bulk.

Amusingly, at $k=0$ we also find
\be \II^{\;k=0}_{\CD,\text{D}}(x,y;q)= \II_{\rm D}(x^2;q)\II_{\rm N}(x^{-1};q)\,, \label{k0D} \ee
suggesting that the bulk theory is dual to two free chirals of $U(1)_x$ charge $+2,-1$, with N and D b.c., respectively. Indeed, we can identify the $+2$ chiral with $\text{Tr}(\Phi^2)$, and the $-1$ chiral with a monopole operator that hits the unitarity bound during RG flow.
The anomalies match on the nose, since on the RHS $\CI_{\rm D}+\CI_{\rm N}=\frac12(2\mb x-\mb r)^2-\tfrac12(\mb x+\mb r)^2=\tfrac32\mb x^2-3\mb x\mb r$.

We may similarly consider Neumann b.c., assuming that the Chern-Simons level is non-positive, $k\leq 0$. At $k=0$, the anomaly polynomial of $\CN$ b.c. on the gauge multiplet and N b.c. on the adjoint chiral is
\be \CI_{\CN,\text{N}} \overset{k=0}= -\tfrac32 \mb x^2+3\mb x \mb r\,.\ee
Happily, there is no gauge anomaly to cancel.
We suspect by comparison to \eqref{k0D} that in the IR we should find two chirals of $U(1)_x$ charges $+1,-1$, with N and D b.c., respectively. Indeed, the half-index is
\be \II^{\;k=0}_{\CN,\text{N}}(x;q) =  \frac{(q)_{\infty} }{ 2} \oint \frac{ds }{ 2\pi i s} \frac{(s^2; q)_{\infty}( s^{-2}; q)_{\infty} }{ (x; q)_{\infty}(x s^2; q)_{\infty}(x s^{-2}; q)_{\infty}} = \II_{\rm N}(x^2;q)\II_{\rm D}(x^{-1};q)\,.\ee

At $k=-1$, the anomaly polynomial is
\be \CI_{\CN,\text{N}} \overset{k=-1}= -\text{Tr}(\mb f^2)-\tfrac32\mb x^2+3\mb x\mb r\,, \ee
and a gauge anomaly must be cancelled. This can be done by adding two boundary Fermi multiplets $\Gamma_1,\Gamma_2$ of R-charge zero, transforming as a doublet of $SU(2)$. 
These Fermi multiplets also transform with charge +1 (say) under a boundary $U(1)_z$ flavor symmetry.
The total boundary anomaly becomes
\be \CI_{\CN,\text{N}}+\CI_\Gamma = -\text{Tr}(\mb f^2)-\tfrac32\mb x^2+3\mb x\mb r
 + \big[\text{Tr}(\mb f^2)+2\mb z^2\big] = 2\mb z^2 -\tfrac32\mb x^2+3\mb x\mb r\,. \ee

We find that the corresponding half-index at $k=-1$ is
\begin{align}
\II^{\;k=-1}_{\CN,\text{N}+\Gamma}(x,z; q) &= \frac{(q)_{\infty} }{ 2} \oint \frac{ds }{ 2\pi i s} \frac{(s^2; q)_{\infty}( s^{-2}; q)_{\infty} }{ (x; q)_{\infty}(x s^2; q)_{\infty}(x s^{-2}; q)_{\infty}} \text{F}(-q^{\frac12}s z; q) \text{F}(-q^{\frac12}s^{-1} z; q) \notag\\[.1cm]
 &= \II_{\rm N}(x^2;q)\, \chi_{0}[SU(2)_1](z)\,.
\end{align}
This shows the expected $\text{Tr}(\Phi^2)$ chiral with N b.c., in addition to --- surprisingly --- a decoupled $SU(2)_1$ WZW model. The WZW model indicates an enhancement of the $U(1)_z$ boundary flavor symmetry to $SU(2)_z$, and carries the boundary 't Hooft anomaly $2\mb z^2 = \text{Tr}\bsp \mb z&0 \\ 0 &-\mb z\esp^2$.

\section*{Acknowledgements}

We would like to thank Kevin Costello, Sergei Gukov, Pavel Putrov, and Nathan Seiberg for valuable discussions and inspiration.
T.D. is partially supported by ERC Starting Grant no. 335739 ``Quantum fields and knot homologies,'' funded by the European Research Council under the European Union's Seventh Framework Programme.
T.D. and N.P. would like to thank the Perimeter Institute for Theoretical Physics for hospitality and support during some phases of this project.
The work of D.G. was supported by the Perimeter Institute for Theoretical Physics.
Research at the Perimeter Institute is supported by the Government of Canada through Industry Canada and by the Province of Ontario through the Ministry of Economic Development \& Innovation.
N.P. is supported by a Sherman Fairchild Postdoctoral Fellowship. This material is based upon work supported by the U.S. Department of Energy, Office of Science, Office of High Energy Physics, under Award Number DE-SC0011632.





\appendix

\section{Fermionic T-duality}
\label{app:T}

There is a well known symmetry in 2d $\CN=(0,2)$ GLSM's that exchanges a Fermi multiplet and its conjugate $\Gamma\leftrightarrow \Gamma^\dagger$, while at the same time swapping its E and J terms $E\leftrightarrow J$.
More precisely, this is a duality of off-shell Fermi superfields that replaces a Fermi superfield $\Gamma$ constrained to obey $\ol D_+\Gamma=E$ and entering in a superpotential coupling $\int d\theta^+\Gamma J$ (for some specified functions $E,J$ of the chirals) with another Fermi superfield $\Lambda$ constrained to obey $\ol D_+\Lambda=J$ and a coupling $\int d\theta^+\Lambda E$. Going on shell, these superfields are related by
\be \Lambda = \Gamma^\dagger\,. \label{Tdual}\ee
We show here that this can be viewed as a fermionic analogue of T-duality. We thank N. Seiberg for suggesting this perspective to us.

We begin by considering a 2d $\CN=(2,2)$ theory with a free, periodic chiral multiplet $\Phi$, satisfying $\ol D_\pm \Phi=0$. With respect to a $\CN=(0,2)$ subalgebra, the multiplet splits into a chiral and a fermi
\be C := \Phi\big|_{\theta^-=\bar\theta^-=0}\,,\qquad \Gamma := D_-\Phi\big|_{\theta^-=\bar\theta^-=0}\,. \ee
The T-dual of $\Phi$ is a twisted-chiral multiplet $\tilde \Phi$, which also splits into a $\CN=(0,2)$ chiral and a fermi
\be \tilde C := \tilde\Phi\big|_{\theta^-=\bar\theta^-=0}\,,\qquad \Lambda := \ol D_-\tilde\Phi\big|_{\theta^-=\bar\theta^-=0}\,. \ee
Note that $\Gamma$ and $\Lambda$ are ordinary (0,2) Fermi multiplets, both with $E=J=0$, which must capture the same degrees of freedom.
Since the (2,2) chiral and its T-dual are related on shell by
(\emph{cf.} \cite{HoriVafa})
\be \Phi+\Phi^\dagger = \tilde\Phi + \tilde\Phi^\dagger\,, \label{PhiT} \ee
we immediately see that, on shell,
\be D_-(\Phi+\Phi^\dagger)\big|_{\theta^-=\bar\theta^-=0} = D_-(\tilde\Phi + \tilde\Phi^\dagger)\big|_{\theta^-=\bar\theta^-=0}\quad\Rightarrow\quad \Gamma = \Lambda^\dagger\,,
\ee
which is the same as \eqref{Tdual}.

To incorporate general E and J terms, we work directly in an interacting 2d $\CN=(0,2)$ theory. Consider a general 2d $\CN=(0,2)$ theory with a distinguished Fermi multiplet $\Gamma$, which is constrained to satisfy $\ol D_+\Gamma=E_\Gamma(C)$ (a function of the chiral multiplets) and has a superpotential interaction $\int d\theta^+ \Gamma J_\Gamma(C)$. We assume that the kinetic term for $\Gamma$ has been diagonalized,%
\footnote{We assume this only for clarity of presentation. The duality transformation below can be generalized in straightforward way to theories with a general Fermi kinetic term.} %
so that the part of the superspace Lagrangian involving $\Gamma$ takes the form
\be \CL(\Gamma) = \int d\theta^+d \bar\theta^+ h(C,C^\dagger,...)\Gamma\Gamma^\dagger + \int d\theta^+ \Gamma J_\Gamma + c.c. \ee
%
Note that supersymmetry requires the total $E\cdot J$ to vanish (or more generally to be constant, see Footnote \ref{foot:EJ}), but as long as there are other Fermi multiplets in the theory we may assume that $E_\Gamma$ and $J_\Gamma$ are totally generic.

We can rewrite the action in terms of a general, unconstrained superfield $\Gamma$ by introducing a second Fermi multiplet $\Lambda$ to act as a Lagrange multiplier:
\be \CL(\Gamma,\Lambda) = \int d\theta^+d \bar\theta^+ h(C,C^\dagger,...)\Gamma\Gamma^\dagger + \int d\theta^+\big[ \Gamma J_\Gamma - \Lambda (\ol D_+\Gamma-E_\Gamma)\big] + c.c. \ee
A priori, $\ol D_+\Lambda$ could be arbitrary (the chiral constraint $\ol D_+\Gamma=E_\Gamma$ is implemented regardless); but new superpotential term preserves SUSY only if the total `$E\cdot J$' is unchanged (up to a constant). The new effective contribution to $E\cdot J$, replacing $E_\Gamma J_\Gamma$, is
\be  \ol D_+\big[ \Gamma J_\Gamma - \Lambda (\ol D_+\Gamma-E_\Gamma)\big]
 = \ol D_+\Gamma J_\Gamma - \ol D_+\Lambda \ol D_+\Gamma+\ol D_+\Lambda E_\Gamma\,. \ee
Rather nicely, this precisely equals $E_\Gamma J_\Gamma$ if we choose $\boxed{\ol D_+\Lambda = J_\Gamma}$. With this choice, we may also conveniently rewrite the superpotential as
\be   \Gamma J_\Gamma - \Lambda (\ol D_+\Gamma-E_\Gamma) = -\ol D_+(\Gamma\Lambda) + \Lambda E_\Gamma\,, \ee
so that the Lagrangian takes the form
\be \CL(\Gamma,\Lambda) = \int d\theta^+ d\bar\theta^+\big[h \Gamma\Gamma^\dagger - \Gamma\Lambda - \Lambda^\dagger\Gamma^\dagger\big] + \int d\theta^+\Lambda E_\Gamma + c.c. \label{LGL} \ee
Finally, we integrate out the unconstrained superfield $\Gamma$ to obtain a dual Lagrangian in terms of $\Lambda$,
\be \CL(\Lambda) = \int d\theta^+d\bar\theta^+ \frac{1}{h}\Lambda \Lambda^\dagger +  \int d\theta^+\Lambda E_\Gamma + c.c. \ee
As desired, the E and J terms have been swapped. More interestingly, the metric has also been inverted. It is also easy to see from \eqref{LGL} that, on shell,
\be h\, \Gamma^\dagger =  \Lambda \,.\ee

Note that in a gauge theory, if $\Gamma$ has charge $q$ under an abelian (say) gauge symmetry, 
the effective metric in the Lagrangian looks like $h\sim e^{2qA_+}$. Its inverse in the $\Lambda$ kinetic term is $1/h\sim e^{-2qA_+}$, consistent with the fact that $\Gamma$ and $\Lambda$ must have opposite  charges.

\section{Character of the Vacuum Module}
\label{app:WK}

In this section, we show that equation (\ref{eq:ourWK}) is equivalent to the standard Weyl-Kac character formula for the vacuum module. 

The standard Weyl-Kac character formula for an integrable representation of an affine Kac-Moody algebra of highest weight $\hat{\lambda}$ is
\begin{equation}
ch_{\hat{\lambda}}[G_{k}] = {{\sum_{w \in \hat{W}}} \epsilon(w) e^{w(\hat{\lambda} + \hat{\rho})} \over e^{\hat{\rho}} \prod_{\hat{\alpha} >0}(1 - e^{-\hat{\alpha}})^{mult(\hat{\alpha})}}
\end{equation} where we can write the affine weight and Weyl vectors in terms of their finite counterparts as $\hat{\lambda} = (\lambda, k, 0), \hat{\rho} = (\rho, h, 0)$\footnote{The third entry of the vector, called the grade of the representation, is zero for an integrable highest weight.}. In particular, the highest weight vector corresponding to the vacuum module is $\lambda = (0, k, 0)$, where the first entry is understood to be the vector of length $r$ with all components equal to zero. In general, a hatted quantity denotes its affine counterpart. It is convenient to rewrite the `normalized character' following \cite{Difrancesco} as 
\begin{equation}\label{eq:WK}
\chi_{\hat{\lambda}}[G_{k}] = q^{s(\hat{\lambda})}ch_{\hat{\lambda}} = {{\sum_{w \in W}} \epsilon(w) \Theta_{w(\hat{\lambda} + \hat{\rho})} \over {\sum_{w \in W}} \epsilon(w) \Theta_{w \hat{\rho}}} = {{\sum_{w \in W}} \epsilon(w) \Theta_{w(\hat{\lambda} + \hat{\rho})} \over q^{|\rho|^2 \over 2 h}e^{\hat{\rho}} \prod_{\hat{\alpha} >0}(1 - e^{-\hat{\alpha}})^{mult(\hat{\alpha})}}
\end{equation} where the theta functions, defined below, in the numerator and denominator together include the overall factor of the modular anomaly, see footnote \ref{fn:modular}. Notice that the sum in the numerator is now over the finite Weyl group. The theta functions are
\begin{equation}
\Theta_{\hat{\lambda}}(q, x)=\sum_{\alpha^{\vee} \in Q^{\vee}}e^{(\lambda + k \alpha^{\vee}, k, |\lambda|^2/2k - |\lambda + k \alpha^{\vee}|^2/2k)} =: \sum_{m \in Q^{\vee}} x^{k m}x^{\lambda} q^{{k \over 2} |m + \lambda/k|^2} = \sum_{m \in Q^{\vee}} x^{k m}x^{\lambda} q^{{k \over 2}(m, m)}q^{(m, \lambda)}q^{|\lambda|^2/2k}
\end{equation} with the sum taken over the coroot lattice $Q^{\vee}$.

We can use this to rewrite the numerator of the Weyl-Kac formula for the vacuum module as
\begin{align*}
q^{-|\rho|^2 \over 2 h}e^{-\hat{\rho}}\sum_{w \in W} \epsilon(w) \Theta_{w(\hat{\lambda}_{vac} + \hat{\rho})} &=  q^{-|\rho|^2 \over 2 h}x^{-\rho}\sum_{w \in W} \epsilon(w) \Theta_{(w \rho, k + h, 0)} \\
&= q^{-|\rho|^2 \over 2 h}x^{-\rho}\sum_{w \in W}\sum_{m \in Q^{\vee}} \epsilon(w) x^{w \rho} q^{(m, w\rho)} x^{(k + h)m} q^{{1 \over 2}(k + h) (m, m)} q^{{1 \over 2}|\rho|^2/(k + h)} \\
&= q^{-|\rho|^2 \over 2 h} q^{{1 \over 2}|\rho|^2/(k + h)} \sum_{m \in Q^{\vee}} q^{{1 \over 2}(k + h)|m|^2} x^{(k + h)m}q^{(m, \rho)} \sum_{w \in W}\epsilon(w) x^{w \rho - \rho}q^{(m, w \rho - \rho)} \\
&=q^{s(\hat{\lambda}_{vac})} \sum_{m \in Q^{\vee}} q^{{1 \over 2}(k + h)|m|^2} x^{(k + h)m} q^{(m, \rho)}\prod_{\alpha \in \Phi^{+}}(1 - q^{m.\alpha} x^{\alpha}) \\
&= q^{s(\hat{\lambda}_{vac})} \sum_{m \in Q^{\vee}} q^{{1 \over 2}(k + h)|m|^2} x^{(k + h)m} \prod_{\alpha \in \Phi^{+}}(-q^{1/2})^{-m.\alpha}(1 - q^{m.\alpha} x^{\alpha})
\end{align*}
In the second line we used the fact that Weyl translates of $\rho$ have the same length and in the second-to-last line we have used the Weyl denominator formula: $\sum_{w \in W}\epsilon(w)e^{w(\rho)-\rho} = \prod_{\alpha \in \Phi^+}(1 - e^{-\alpha})$. Finally, in the last line, we used the identity $\prod_{\alpha \in \Phi^{+}}(q^{1/2})^{-m.\alpha} = q^{(m, \rho)}$ (after accounting for our particular sign conventions). The extra factor of $(-1)^{m.\alpha}$ is trivial in the case of $G$ simply connected, since $m.\rho \in \mathbb{Z}, m \in \Lambda^{\vee} = Q^{\vee}$ so that $(-1)^{m.\alpha} = 1$. 

This is precisely the numerator of equation (\ref{eq:ourWK})---except for the modular characteristic $q^{{1 \over 2}|\rho|^2/(k + h) - {1 \over 2}|\rho|^2/h}$ which we drop---if the group $G$ is simply connected and hence the coroot lattice can be identified with the cocharacter lattice. 

Manipulating the denominator formula of the affine Weyl-Kac character formula is even easier: 
\begin{align*}
\prod_{\hat{\alpha} >0}(1 - e^{-\hat{\alpha}})^{mult(\hat{\alpha})} &= (q)^r_{\infty}\prod_{\alpha \in \Phi^+}(q x^{\alpha}; q)_{\infty}(q x^{-\alpha}; q)_{\infty}(1 - x^{\alpha})
\end{align*} where we recall that the set of positive affine roots $\hat{\alpha} = (\alpha, n, 0)$ includes roots with positive, negative, and zero $\alpha$ with positive $n$, as well as roots of the form $\hat{\alpha} = (\alpha, 0, 0)$ with $\alpha \in \Phi^{+}$.  

For non-simply connected $G$ our half-index formulae suggest that one should just replace the coroot lattice with the cocharacter lattice in the definition of our theta function, as well as honestly include the $(-1)^{m}$ factor, which may be nontrivial in general. The former replacement indeed occurs in the work of \cite{wendt}, who proposes a non-simply connected generalization of the Weyl-Kac formula in Theorem 5.5. It would be interesting to precisely match our formula to the literature in the case of simple but not simply connected $G$ (or else refine our conjecture).

\bibliographystyle{JHEP_TD}
\bibliography{halfindex}

\end{document}